\documentclass[sn-basic]{sn-jnl}
\theoremstyle{thmstyleone}%

\theoremstyle{thmstyletwo}%

\theoremstyle{thmstylethree}%

\usepackage{hyperref}
\hypersetup{hypertex=true,
	colorlinks=true,
	citecolor=blue}
\usepackage{graphicx}
\usepackage{amsmath}
\usepackage{amsfonts}
\usepackage[linesnumbered,ruled,vlined]{algorithm2e}
\usepackage{subfigure}
\usepackage[section]{placeins}
\usepackage{float}
\usepackage{chemarrow}
\usepackage{color}

\renewcommand{\vec}[1]{\boldsymbol{#1}}

\begin{document}

\title[Epigenetic inheritance in DTP-mediated resistance]{Epigenetic state inheritance drivers drug-tolerant persister-induced resistance in solid tumors: A stochastic agent-based model}

\author{\fnm{Xiyin} \sur{Liang}}\email{liangxiyin@tiangong.edu.cn}
\author*{\fnm{Jinzhi} \sur{Lei}}\email{jzlei@tiangong.edu.cn}

\affil{\orgdiv{School of Mathematical Sciences}, \orgname{Center for Applied
		Mathematics}, \orgname{Tiangong University}, \orgaddress{\city{Tianjin},  \postcode{300387}, \country{China}}}


\abstract{The efficacy of anti-cancer therapies is severely limited by the emergence of drug resistance. While genetic drivers are well-characterized, growing evidence suggests that non-genetic mechanisms, particularly those involving drug-tolerant persisters (DTPs), play a pivotal role in solid tumor relapse. To elucidate the evolutionary dynamics of DTP-induced resistance, we develop a stochastic agent-based model (ABM) of solid tumor evolution that couples macroscopic population dynamics with microscopic epigenetic state inheritance during the cell cycle. Our simulations accurately reproduce the temporal progression of relapse observed in experimental studies, capturing the dynamic transition from sensitive cells to DTPs, and ultimately to stable resistant phenotypes under prolonged therapy. By explicitly modeling the epigenetic plasticity of individual cells, our model bridges the gap between cellular heterogeneity and population-level tumor evolution. Furthermore, we performed \textit{in silico} clinical trials using virtual patient cohorts to evaluate therapeutic outcomes, demonstrating that optimized adaptive treatment strategies can significantly delay tumor relapse compared to standard dosing. This study provides a quantitative framework for dissecting DTP-driven resistance mechanisms and designing more effective, biologically informed therapeutic strategies. 
}
	

\maketitle

\section{Introduction}
\label{sec1}

Over the past decades, targeted therapies have significantly improved the survival of cancer patients. However, drug resistance remains the greatest challenge to achieving durable remission or cure. While classical evolutionary theories have predominantly focused on genetic drivers (e.g., pre-existing or de novo mutations), growing evidence indicates that non-genetic mechanisms play a critical role  in mediating drug resistance in solid tumors, particularly during the initial phase of treatment  \citep{Boumahdi20greatescape,McDonald:2024aa,Calderon:2024aa}. Understanding how these non-genetic adaptive strategies drive tumor relapse is therefore essential for designing more effective therapeutic schedules.

A key contributor to non-genetic resistance is the emergence of drug-tolerant persister cells (DTPs) \citep{Bell20Principles,McDonald:2024aa}. The phenomenon of drug tolerance, originally characterized in bacterial populations as `persistence' \citep{Balaban:2019aa}, implies a transient state of multidrug tolerance that is not encoded by genetic mutations. DTPs have been observed across multiple cancer types, including non-small cell lung cancer (NSCLC) \citep{Sharma:2010aa}, melanoma \citep{Rambow:2018aa}, colorectal cancer (CRC) \citep{Cell21Colorectal}, and glioblastoma  \citep{Liau17Glioblastoma}. Importantly, DTPs exhibit distinct dynamics under short- versus long-term treatments. For example, in NSCLC PC9 cells treated with erlotinib, a small fraction of cells survive in a slow-cycling state and revert to drug sensitivity upon drug withdrawal \citep{Sharma:2010aa}. However, under prolonged exposure, these cells eventually evolve into stably resistant populations.

The molecular basis of DTPs involves complex intracellular rewiring.  Multiple processes, including chromatin remodeling, metabolic reprogramming, and activation of alternative signaling pathways, have been implicated in the emergence and maintenance of DTPs. For instance, cancer cells exposed to stress can generate a reversible DTP population characterized by upregulation of IGF-1R, CD133, and KDM5A \citep{Sharma:2010aa}. IGF-1R signaling is required for the establishment of drug tolerance, which is associated with distinct chromatin states that can be disrupted pharmacologically. Several signaling pathways, such as Wnt$/\beta$-catenin and Notch, have also been shown to contribute to DTP formation \citep{Reyes:2025aa,Natcom18Notch}. In NSCLC, enhanced acetylcholine (ACh) metabolism promotes DTP formation by activating Wnt$/\beta$-catenin signaling \citep{Huz22JCI}. Moreover, stromal cells in the tumor microenvironment (TME), such as cancer-associated fibroblasts (CAFs) and tumor-associated macrophages (TAMs), secrete cytokines that further support DTP survival \citep{Straussman12CAF}. These findings suggest that the cell state is not a simple binary switch (sensitive vs. resistant) but rather a high-dimensional continuous phenotype modulated by microenvironment cues and epigenetic regulators.

Mathematical modelling has proven effective in decoding these complex evolutionary dynamics. Several approaches have provided insights into the trade-offs between cell kill and tolerance induction. Classical ordinary differential equation (ODE) models typically stratify cell populations into discrete compartments (e.g., sensitive, tolerant, resistant) to study transition rates and dosage optimization \citep{KuosmanenPlosCB2025,Gevertz2025npjSBA}. More recent works have employed optimal control theory to balance toxicity and resistance suppression \citep{FischerJTB2024,Gunnarsson2025npjSBA}. To capture spatial heterogeneity, partial differential equation (PDE) models have been employed to study the role of the tumor microenvironment in resistance \citep{SunX18MCT}.  Furthermore, agent-based lineage models have been developed to infer resistance dynamics by integrating single-cell data, offering higher resolution on the timing of DTP emergence \citep{IyerPlosCB2025,Whiting2025Natcom,Wang:2025aa}. Notably, Chisholm et al. proposed coupled individual-based and integro-differential equation models to study reversible phenotypic evolution under cytotoxic drug exposure, demonstrating how adaptation within non-genetically unstable populations drives the outgrowth of proliferative resistant clones \citep{CR15Emergence}.

However, most existing frameworks rely on mean-field approximations or discrete-state assumptions, which may overlook a defining feature of DTPs: the ``resistance continuum'' \citep{Feinberg2023Science}. Experimental data from single-cell RNA sequencing reveal that acquired resistance involves a series of gradual cell-state transitions driven by intracellular epigenetic variations \citep{France2024Nature}. Crucially, these epigenetic states are inherited with noise during cell division, creating a ``memory'' effect that standard ODEs fail to capture \citep{Probst2009Epigenetic}. Therefore, a framework that explicitly integrates microscopic epigenetic inheritance with macroscopic population dynamics is necessary to elucidate how continuous phenotypic drift drives relapse.

In this study, we develop a stochastic agent-based model (ABM) specifically calibrated for NSCLC treatment. Theoretically grounded in the stochastic inheritance framework proposed by Lei \citep{LeiJTB20framework}, our model instantiates the cell state as a continuous vector coupling drug sensitivity with stemness. By defining an inheritance probability kernel, we explicitly model how epigenetic phenotypes are passed from mother to daughter cells with varying degrees of fidelity. This multiscale approach allows us to: (1) reproduce the temporal progression from reversible tolerance to irreversible resistance; (2) quantify how epigenetic plasticity induces DTP-mediated resistance; and (3) conduct \textit{in silico} clinical trials using virtual patient cohorts to identify adaptive therapy strategies that significantly prolong progression-free survival.

\section{Models and Methods}
\label{sec2}

Tumors are characterized by uncontrolled cellular proliferation. Here, we propose a multi-scale mathematical model,  based on stem cell regeneration dynamics, to quantify the evolutionary dynamics of tumor relapse under targeted therapy. While the population dynamics are formally governed by an integro-differential equation (IDE) (derived in Appendix A), we permit a high-dimensional representation of cell states by implementing the model as a stochastic agent-based model (ABM). In this section, we first outline the biological rules and structure of the ABM, followed by a detailed definition of the epigenetic state variables and kinetic rates. Finally, we provide the parameter estimation strategy based on experimental benchmarks.

\subsection{Model description}
\label{sec:2.1}

\begin{figure}[t]
	\centering
	\includegraphics[width=13cm]{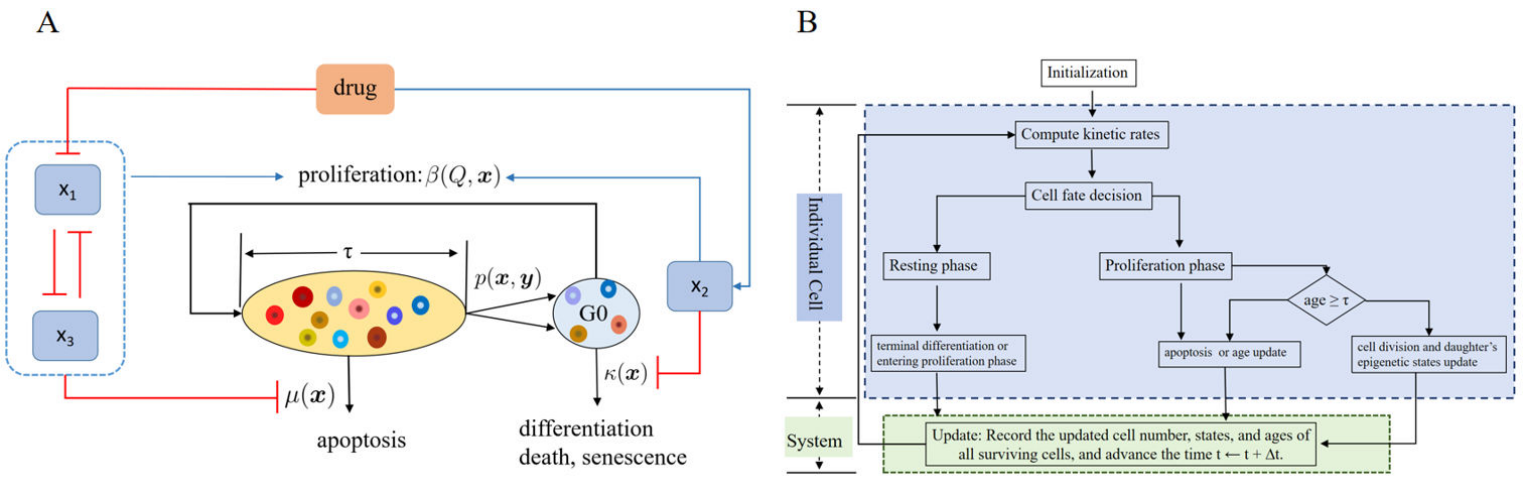}
	\caption{\textbf{Schematic of the multi-scale theoretical framework and the agent-based simulation flow.} \textbf{A} The heterogeneous stem cell regeneration model. Tumor dynamics follow a G0 cell cycle structure where kinetic rates depend on the cell's epigenetic state $\vec{x} = (x_1, x_2, x_3)$. $Q(t, \vec{x})$ represents the density of G0 cells with state $\vec{x}$. The orange region denotes the proliferative phase (duration $\tau$). Cellular plasticity is mathematically captured by the inheritance probability kernel $p(\vec{x}, \vec{y})$, describing how epigenetic states are transmitted with variation from mother ($\vec{y})$ to daughter ($\vec{x}$). Colored circles within blue and orange regions represent cells with varying phenotypes. The lines with red bars denote inhibition, while blue arrows indicate promotion, and black arrows represent cell transition. \textbf{B} Flowchart of the agent-based stochastic simulation. The system is initialized with all cells in the resting phase. At each time step, the probabilities of cell fate decisions (division, death, differentiation) are computed for each agent based on its state $\vec{x}$. Detailed implementations are provided in Section \ref{sec:ICBSS}.}
	\label{Fig-Model}
\end{figure}

To capture the interplay between cell cycle progression and phenotypic plasticity, our model is built upon the heterogeneous stem cell regeneration framework established by Lei \citep{Lei20evolutionary,LeiJTB20framework}. As illustrated in Fig. \ref{Fig-Model}A, the cell cycle is partitioned into two distinct phases: a resting phase (G0) and a proliferative phase. The life cycle of an individual cell is governed by the following stochastic rules:
\begin{itemize}
\item Resting phase (G0): Cells in the G0 phase can either re-enter the cell cycle at a proliferation rate $\beta$ or exit the population at a removal rate $\kappa$, which aggregates the effects of terminal differentiation, senescence, and death.
\item Proliferative phase: Once in the cell cycle, cells undergo a fixed duration $\tau$ of preparation. During the period, they are subject to apoptosis at a rate $\mu$. Surviving cells divide into two daughter cells, which immediately return to the G0 phase to begin a new cycle.
\end{itemize}

We consider target therapy using EGFR-TKI1, such as erlotinib and osimertinib, which are primarily cytostatic agents rather than cytotoxic ones. They work by blocking the growth signaling pathway and interfering with the inheritance of epigenetic states, rather than directly increasing apoptosis.

To elucidate the molecular mechanism of DTP-induced resistance, we characterized each cell by a multidimensional epigenetic state vector $\vec{x}$, rather than discrete phenotypic labels. As shown in Fig. \ref{Fig-Model}A, the variation in $\vec{x}$ dictates cellular heterogeneity, directly modulating the kinetic rates $\beta$, $\kappa$, and $\mu$. Crucially, phenotypic plasticity arises from the stochastic inheritance of $\vec{x}$ during mitosis. Due to the molecular complexity of chromatin remodeling, we do not model specific pathways but instead describe this ``memory loss'' or ``rewiring'' via a phenomenological inheritance probability kernel, $p(\vec{x}, \vec{y})$. This kernel defines the probability density that a mother cell with state $\vec{y}$ produces a daughter cell with state $\vec{x}$, thereby coupling microscopic epigenetic drift with macroscopic population evolution.

The schematic diagram of the stochastic simulation algorithm is presented in Fig. \ref{Fig-Model}B. This ABM approach allows us to track the lineage and trajectory of rare persister cells, which is computationally prohibitive in grid-based numerical methods for high-dimensional IDE. Specific details of the simulation algorithm are provided in Section \ref{sec:ICBSS}.

\subsubsection{Epigenetic states and cell phenotypes}

Biological evidence suggests that acquired resistance is not a binary switch but a dynamic remodeling process driven by transcriptional plasticity. In an experimental work by \citet{Sharma:2010aa}, PC9 NSCLC cells treated with erlotinib exhibited a transient drug-tolerant state (DTPs) associated with the loss of EGFR signaling and the upregulation of stemness markers (e.g., CD133, CD44) and IGF-1R. Under prolonged exposure, these cells eventually evolved into stably resistant colonies (DRCs), often characterized by the activation of alternative pro-survival pathways (e.g., increased acetylcholine metabolism or Wnt signaling) to bypass EGFR inhibition \citep{Huz22JCI, Reyes:2025aa}.

\paragraph{Epigenetic states}
To formulate the biological complexities into a tractable mathematical framework, we reduce the high-dimensional molecular profile of a cell as a three-dimensional epigenetic state vector $\vec{x} = (x_1, x_2, x_3) \in \Omega \subset \mathbb{R}^3$. These three components correspond to the distinct functional axes of resistance evolution:
\begin{enumerate}
\item Drug sensitivity ($x_1$): Represents the activity of the primary oncogenic drivers (e.g., EGFR pathway and downstream ERK/AKT signaling) targeted by the drug.
\item Stemness/Plasticity ($x_2$): Quantifies the differentiation potential. High levels of $x_2$ indicate a slow-cycling, stem-like state characteristic of DTPs. Biologically, $x_2$ serves as a proxy for stem cell markers such as CD133, CD44, and Oct4. 
\item Alternative adaptation ($x_3$): Represents the activation of bypass signaling pathways that restore proliferative capacity in the presence of the drug. These signaling pathways include IGF-1R \citep{Sharma:2010aa}. Specifically for the data used in this study \citep{Huz22JCI}, $x_3$ represents the Acetylcholine (ACh) signaling axis (measured by ACh levels or ChAT expression).
\end{enumerate}
Table \ref{tab-Epistate3} provides a summary of the biological interpretation of these variables.

Given the state vector, $Q(t, \vec{x})$ represents the density of G0 cells with state $\vec{x}$, and $Q(t) = \int_\Omega Q(t, \vec{x})$ give the total number of G0 cells at time $t$. 

\begin{table}[htbp]
	\centering
	\caption{Microscopic epigenetic state variables and their biological interpretation.}
	\label{tab-Epistate3}
	\begin{tabular}{@{}ll@{}}
		\toprule
		Variable & Biological interpretation \\  
		\midrule  
		\vspace{1ex}
		$x_1$ & \parbox[c]{10cm}{Activity of the drug-targeted pathway. Higher $x_1$ implies a higher growth rate but greater susceptibility to therapy.} \\ 
		\vspace{1ex}
		$x_2$ & \parbox[c]{10cm}{Potential of self-renewal and plasticity. Higher $x_2$ induces quiescence (G0 arrest) and promotes survival (DTP state).} \\ 
		$x_3$ & \parbox[c]{10cm}{Activity of alternative pathways. Higher $x_3$ allows cells to re-enter the cell cycle despite drug inhibition.} \\
		\botrule
	\end{tabular}
\end{table}

\begin{figure}[htbp]
\centering
\includegraphics[width=0.8\textwidth]{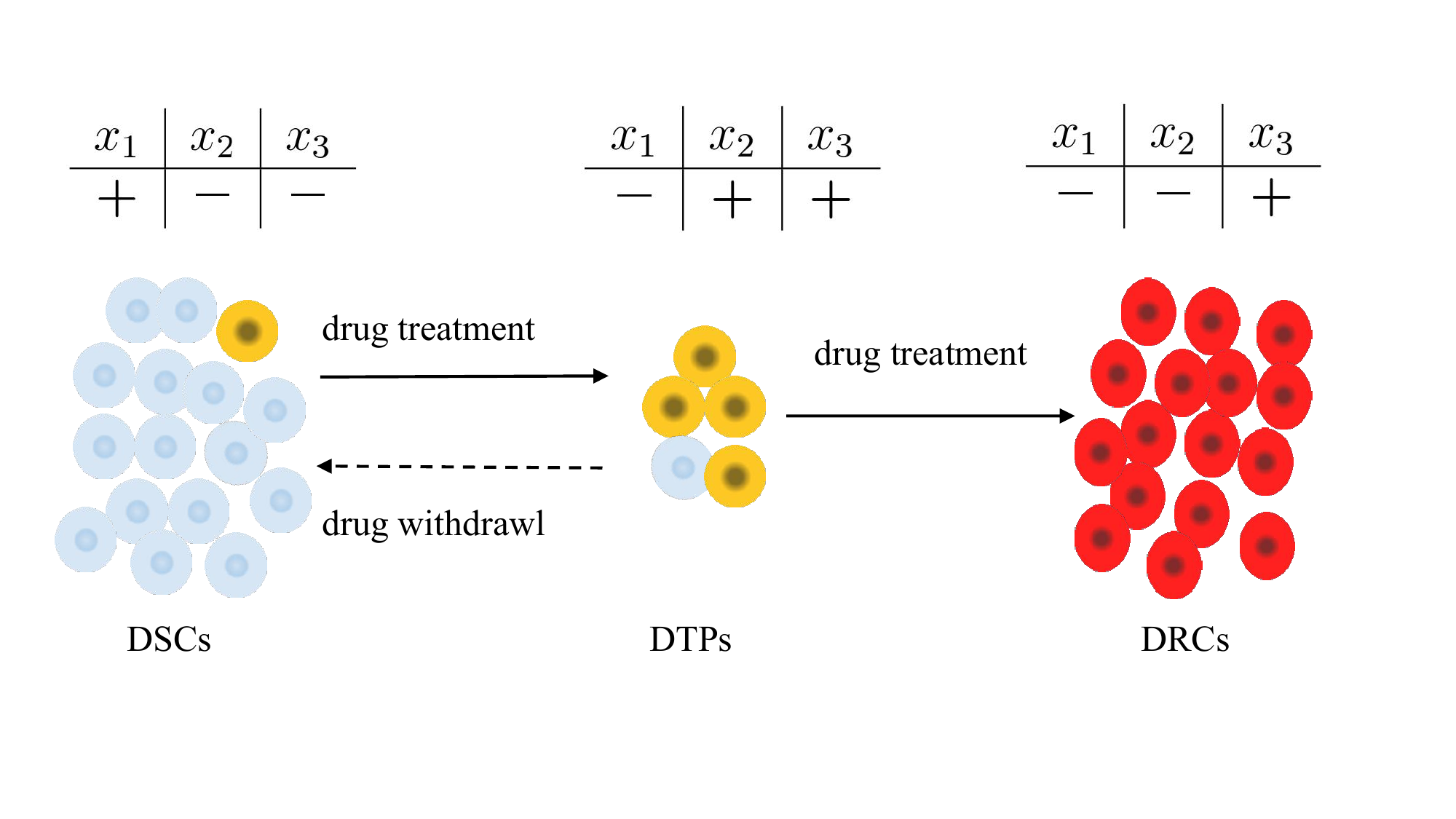}
\caption{\textbf{Phenotypic transitions mapped on the epigenetic landscape.} Before therapy, the population is dominated by DSCs (high $x_1$). Drug treatment suppresses $x_1$, selecting for DTPs (high $x_2$) which survive in a dormant state. Under continuous pressure, stochastic epigenetic drift eventually accesses the high-$x_3$ basin, leading to the outgrowth of proliferative DRCs.}
\label{Fig-Model-phenotype}
\end{figure}

\paragraph{Cell phenotype identification}
While the underlying epigenetic state $\vec{x}$ is continuous---reflecting the ``resistance continuum'' \citep{France2024Nature}---there are clinically observed distinct cell types. To identify cell phenotypes from the epigenetic state, we classify cells into three canonical phenotypes based on expression thresholds $x_s^1$, $x_s^2$, and $x_s^3$ (Table \ref{tab-phenotype-epi}):
\begin{itemize}
\item DSCs (Drug-Sensitive Cells): high target expression ($x_1 > x_s^1$) and low stemness ($x_2 < x_s^2$).
\item DTPs (Drug-Tolerant Persisters): low target expression ($x_1 \le x_s^1$) but high stemness ($x_2 > x_s^2$), corresponding to a reversible, slow-cycling state.
\item DRCs (Drug-Resistant Cells): low target expression ($x_1 \le x_s^1$) and low stemness ($x_2 \le x_s^2$), but high adaptation ($x_3 > x_s^3$), regaining proliferative capacity.
\end{itemize}
This classification allows us to map continuous simulation trajectories onto clinically recognizable stages of relapse (Fig. \ref{Fig-Model-phenotype} and Table \ref{tab-phenotype-epi}).

\begin{table}[t]
	\caption{Classification criteria for tumor cell phenotypes. }
	\label{tab-phenotype-epi}
	\begin{tabular}{lll}
		\toprule
		Phenotype & Active Markers & Criterion (Notation)\\ 
		\midrule  
		DSCs & $x_1^{\text{high}}, x_2^{\text{low}}$ & $(x_1,x_2,x_3)\in (x^1_s, +\infty)\times(0, x^2_s]\times(0,x^3_s]$ 
		\\ 
		DTPs & $x_1^{\text{low}}, x_2^{\text{high}}$ & $(x_1,x_2,x_3)\in (0, x^1_{s}]\times(x^2_s, +\infty)\times(x^3_s, +\infty)$ 
		\\   
		DRCs & $x_1^{\text{low}}, x_2^{\text{low}}, x_3^{\text{high}}$ & $(x_1,x_2,x_3)\in (0, x^1_{s}]\times(0, x^2_s]\times(x^3_s, +\infty)$ \\
		\botrule
	\end{tabular}
\end{table}

\subsubsection{Kinetic rates}
\label{sec:2.2.1}

Standard constant-rate models cannot capture the adaptive landscape of resistance. Here, we assume the kinetic rates ($\beta, \mu$, $\kappa$) are functionally dependent on the epigenetic state $\vec{x}$, governed by the following biological logic (illustrated in Fig.~\ref{Fig-Model}A):
\begin{enumerate}
\item Proliferation is driven by both the primary target ($x_1$) and alternative pathways ($x_3$), but is suppressed by stemness ($x_2$). This creates a non-monotonic dependence of proliferation on $x_2$, where intermediate stemness favors expansion while extreme stemness favors dormancy.
\item There exists a reciprocal inhibition between the dominant sensitive pathway ($x_1$) and the adaptive pathway ($x_3$), reflecting the metabolic cost of switching signaling dependencies.
\item Therapeutic agents inhibit the function of $x_1$, thereby reducing proliferation. However, this stress selectively favors cells with high $x_2$ (DTPs) or high $x_3$ (DRCs).
\end{enumerate}
The precise mathematical formulations of these dependencies are detailed below.

\paragraph{Proliferation and differentiation rates}
The proliferation rate $\beta$ incorporates population-level feedback regulation to maintain tissue homeostasis. Biologically, the self-renewal capacity of stem cells is constrained by microenvironmental signals and available niche space. Following established theoretical frameworks \citep{Bernard:2003aa,LeiJTB20framework}, we model this feedback using a phenomenological Hill function dependent on the total resting phase population $Q(t)$:
$$
\beta = \beta_0(\vec{x}) \dfrac{\theta(\vec{x})^m}{\theta(\vec{x})^m + Q(t)^m},
$$
where $\beta_0(\vec{x})$ represents the intrinsic proliferation potential at low cell density, and $\theta(\vec{x})$ denotes the effective carrying capacity (or characteristic cell number) at which the proliferation rate is halved.

The intrinsic proliferation rate $\beta_0$ and terminal differentation rate $\kappa$ are modulated by the stemness state $x_2$. We construct these functions to reflect the biological logic of stem cells:
\begin{itemize}
\item High $x_2$ corresponds to quiescent stem cells (low proliferation, negligible differentiation); 
\item Intermediate $x_2$ represents transient amplifying cells (maximal proliferation); 
\item Low $x_2$ indicates differentiated cells (reduced proliferation, high differentiation rate).
\end{itemize}

\begin{figure}[htbp]
\centering
\includegraphics[width=6cm]{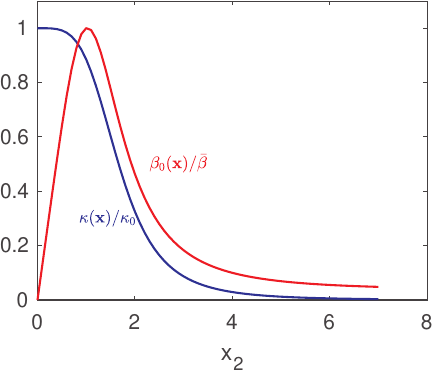}
\caption{\textbf{Functional dependence of kinetic rates on the stemness $x_2$.} The proliferation rate $\beta_0(\vec{x})$ (red curve, normalized by $\bar{\beta}$) exhibits a unimodal profile, peaking at intermediate $x_2$ values (progenitor state) and decreasing at high $x_2$ (quiescent stem state). The differentiation rate $\kappa(\vec{x})$ (blue curve, normalized by $\kappa_0$) decreases monotonically with increasing stemness.}
\label{Fig-beta-kappa}
\end{figure}

To capture this behavior, we define $\beta_0(\vec{x})$ as a unimodal function of $x_2$, while the differentiation rate $\kappa(\vec{x})$ as a monotonically decreasing function:
\begin{equation}
\label{beta0}
\beta_0(\vec{x})=\bar{\beta}\ \frac{a_1x_2+(a_2x_2)^4}{1+(a_3x_2)^4},
\end{equation}
\begin{equation}
\label{kappa}
\kappa(\vec{x})=\kappa_0\frac{1}{1+(\kappa_1 x_2)^{n_{\kappa}}},
\end{equation}
where $\bar{\beta}$ and $\kappa_0$ characterize the maximum rates of proliferation and differentiation, respectively. The coefficients $a_i$, $\kappa_1$, and Hill exponent $n_{\kappa}$ shape the respnse curves, as illustrated in Fig.~\ref{Fig-beta-kappa}.

The parameter $\theta$ is regulated by drug sensitivity ($x_1$) and adaptive resistance ($x_3$). We assume these two pathways act explicitly to promote survival and growth signaling. The effective carrying capacity parameter $\theta$ is defined as: 
\begin{equation}
\label{theta}
\theta(\vec{x}) = \theta_0 \left(1 + a_5 \dfrac{(a_4 (x_1 + \alpha x_3))^6}{1+(a_4 (x_1 + \alpha x_3))^6} \right),
\end{equation}
where $\theta_0$ is the basal level, and the term in the parentheses describes the saturation effect of oncogenic signaling $(x_1 + \alpha x_3)$. Here, $\alpha$ weights the relative contribution of the adaptive pathway $x_3$ compared to the primary target $x_1$.

\paragraph{Apoptosis rate}
Similar to the parameter $\theta$, the apoptosis rate $\mu$ is also jointly regulated by drug sensitivity ($x_1$) and adaptive resistance ($x_3$) through the oncogenic signaling $(x_1 + \alpha x_3)$. Thus, we model $\mu$ as an exponentially decaying function of the combined survival signals:
\begin{equation}
\label{mu}
\mu(x_1, x_3)=\mu_0 e^{\mu_1  \big(\gamma_1-(x_1+\alpha x_3)\big)},
\end{equation}
where $\mu_0$ is the baseline apoptosos rate, and $\mu_1, \gamma_1$ determine the sensitivity of cell death to signal withdrawal. 

The formulations \eqref{mu} and \eqref{beta0} together ensure that cells with low $x_1$ (due to drug inhibition) and low $x_3$ (lack of adaptation) undergo high rates of apoptosis, whereas DTPs and DRCs survive through elevated $x_2$ (quiescence, avoiding $\mu$ in proliferative phase) or elevated $x_3$.

\paragraph{Impact of drug treatment}
The targeted therapy exerts selective pressure by suppressing the growth signal. We model this by reducing the basal capacity $\theta_0$, which recovers only if cells upregulate the adaptive state $x_3$. To describe this dynamic adaptation, we introduce a drug-dependent modulation factor $\eta(x_3)$:
\begin{equation}
\label{theta0}
\theta_0(\vec{x})=\begin{cases}
\bar{\theta}_0,\quad &\mbox{Control (No drug)},\\
\bar{\theta}_0\cdot \eta(x_3), & \mbox{Treatment (With drug)},
\end{cases}
\end{equation}
where the recovery factor $\eta(x_3)$ is govened by a sigmoid function $\psi(x_3)$:
\begin{equation}
\label{eq:psi}
\eta(x_3) = 1 - (1-\eta_0) \psi(x_3),\quad \mbox{with}\quad \psi(x_3) =\dfrac{1}{1+e^{k(x_3 - \bar{x}_3)}}.
\end{equation}
Biologically, immediately upon drug administration, $\theta_0$ drops to $\eta_0 \bar{\theta}_0$, resulting in the suppression of the cell population. However, as cells evolve toward the DRC phenotype (increasing $x_3$), $\psi(x_3)$ approaches $0$, restoring $\eta(x_3) \approx 1$. This captures the emergence of resistance where cells regain proliferative fitness despite the presence of the drug.

\subsubsection{The inheritance probability kernel $p(\vec{x}, \vec{y})$}

To efficiently capture the phenotypic plasticity without resolving detailed intracellular networks, we introduce a phenomenological inheritance probability kernel $p(\vec{x}, \vec{y})$. This kernel defines the conditional probability density that a mother cell with state $\vec{y}$ generates a daughter with state $\vec{x}$, satisfying the normalization condition: 
\begin{equation}
\int_\Omega p(\vec{x}, \vec{y}) d \vec{x} = 1,\quad \forall\vec{y} \in \Omega.
\end{equation}
Biologically, this kernel aggregates molecular noise (e.g., chromatin remodeling) into a stochastic memory process, balancing high-fidelity inheritance with the epigenetic drift required for evolution. Mathematically, it acts as the integral transition operator in the population dynamics equation (see Appendix A), linking microscopic fluctuation to macroscopic population shifts \citep{LeiJTB20framework}.

\paragraph{Mathematical formulation of the probability kernel}
Although the biological traits $x_1, x_2$, and $x_3$ are functionally coupled (e.g., via mutual inhibition pathways), we assume that the stochastic fluctuations affecting each trait during cell division are statistically independent. This simplifies the high-dimensional joint probability distribution into a product of marginal distributions:
$$p(\vec{x}, \vec{y})=\prod^3_{i=1} p_i(x_i, \vec{y}),$$ 
where $p_i(x_i, \vec{y})$ is the inheritance function for the $i$-th component. Crucially, this factorization implies independence only in the noise structure, not in the phenotypes themselves. The biological correlations among traits are rigorously preserved and encoded within the conditional expectation function of these distributions (detailed below), ensuring that the simulated cells exhibit realistic phenotypic coupling.

Since the state variables $x_i$ reflect the expression levels of marker proteins or mRNA, which typically follow right-skewed, non-negative distributions \citep{Friedman06linking, Shahrezaei08PNAS}, we model $p_i(x_i, \vec{y})$ using a Gamma distribution:
\begin{equation}
\label{eq:gamma}
\begin{aligned}
p_i(x_i, \vec{y}) &= \mathrm{Gamma}\big(x_i; a_i(\vec{y}), b_i(\vec{y})\big)\\
& =
\left\{
\begin{array}{ll}
\dfrac{b_i(\vec{y})^{-a_i(\vec{y})}}{\Gamma(a_i(\vec{y}))} x_i^{a_i(\vec{y}) - 1} e^{-x_i/b_i(\vec{y})},\quad & x_i \geq 0, \vspace{0.3em}\\
\ 0,& x_i  < 0,
\end{array}
\right.
\end{aligned}
\end{equation}
where $\Gamma(\cdot)$ is the Gamma function. The parameters $a_i(\vec{y})$ (shape) and $b_i(\vec{y})$ (scale) depend on the mother cell state $\vec{y}$.

The conditional expectation and variance of $x_i$ are related to the Gamma parameters by 
\begin{equation}
\mathrm{E}(x_i \vert \vec{y}) = a_i(\vec{y}) b_i(\vec{y}),\quad \mathrm{Var}(x_i \vert \vec{y}) = a_i(\vec{y}) b_i^2(\vec{y}). 
\end{equation}
Hence, 
$$
a_i(\vec{y}) = \dfrac{\mathrm{E}(x_i \vert \vec{y})^2}{\mathrm{Var}(x_i \vert \vec{y})},\quad b_i(\vec{y}) = \varphi_i(\vec{y})/a_i,
$$
where $\varphi_i(\vec{y}) = \mathrm{E}(x_i \vert \vec{y})$ denotes the conditional expectation of $x_i$. In practice, we often treat the dimensionless parameter $a_i$ as a fixed constant, while $b_i(\vec{y})$ is determined by $\varphi_i(\vec{y})$. Thus, through the predefined $a_i$ and $\varphi_i(\vec{y})$, the scale parameter $b_i(\vec{y})$ and the inheritance probability $p_i(x_i, \vec{y})$ are determiend subsequently.

The conditional expectation $\varphi_i(\vec{y})$ is crucial for $p_i(x_i, \vec{y})$, as it defines the expected epigenetic state of the daughter cell inherited from the mother cell. We construct this function based on three biological considerations:
\begin{enumerate}
\item \textbf{Epigenetic memory:} High expression levels in the mother cell ($y_i$) generally bias daughter cells toward high expression ($x_i$), enforcing a monotonic dependence of $\varphi_i(\vec{y})$ on $y_i$.
\item \textbf{Saturation constraints:} Biological transcription and translation rates are physically limited by cellular resources, preventing unbounded accumulation of state variables.
\item \textbf{Nonlinear cooperativity:} Epigenetic maintenance often involves cooperative enzyme kinetics (e.g., histone methylation writers effectively recruiting more writers), creating switch-like or sigmoidal responses essential for distinct phenotypic stabilization \citep{LeiJTB20framework}.
\end{enumerate}
Consequently, we model $\varphi_i(\vec{y})$ using a generalized Hill function, which naturally captures the transition from linear inheritance memory to saturation:
\begin{equation}
\label{varphifun}
\varphi_i(\vec{y}) = \varphi_{i,0}(\vec{y}) + \varphi_{i,1}(\vec{y})\frac{y_i^{n_i}}{K_i(\vec{y})^{n_i}+y_i^{n_i}},
\end{equation}
Here, baseline coefficient $\varphi_{i,0}$ and maximum capacity $\varphi_{i,1}$ determine the bounds of the state variable, while the half-saturation threshold $K_i(\vec{y})$ and Hill coefficient $n_i$ determine the sensitivity of the switch. Crucially, by allowing these coefficients to depend on other state components (coupling), we can embed the structure of the underlying gene regulatory network into the inheritance kernel \citep{Li:2025bd}.

\paragraph{Parameter identifiability and drug modulation}
The functional form in \eqref{varphifun} provides a direct link between model parameters and experimental observables. In a population at dynamic equilibrium, the population-averaged expression level of marker $x_i$ roughly corresponds to the fixed point of the conditional expectation map, governed by the equation $x_i \approx \mathrm{E}[x_i \vert \vec{x}] = \varphi_i(\vec{x})$. This relationship allows us to calibrate the coefficients (e.g., $\varphi_{i,0}, \varphi_{i,1}$, and $K_i$) using mean fluorescence intensity (MFI) data from flow cytometry of bulk RNA-seq data.

Furthermore, therapeutic agents do not merely select for pre-existing states but actively remodel the epigenetic landscape. We posit that drug treatment alters the inheritance kernel by modulating the shape coefficients---specifically the baseline coefficients $\varphi_{i, 0}$, the maximum capacity $\varphi_{i, 1}$, and the sensitivity threshold $K_i$. This reflects drug-induced transcriptional reprogramming.

In the following, we specify the explicit forms of $\varphi_i(\vec{y})$ for each epigenetic state variable $x_i$. Unless otherwise specified, all parameters are positive constants. The drug concentration is denoted by a normalized variable $D$, where $0\le D \le 1$.

\vspace{0.25cm}
\noindent\textbf{1. Regulation of drug sensitivity ($x_1$).}

The inheritance of sensitivity markers (e.g., EGFR) is disrupted by drug exposure and modulated by the adaptive state ($x_3$). We assume: 
\begin{itemize}
\item \textbf{Drug inhibition:} The drug dosage $D$ reduces the expression of $x_1$ in progeny, reflecting the selection against sensitive phenotypes of drug-induced transcriptional repression.
\item \textbf{Protective adaptation:} High levels of the adaptive state ($x_3$) can buffer this drug effect.
\end{itemize}

These mechanisms are formulated as:
\begin{equation}
\label{varphi-x1}
\begin{aligned}
\varphi_{1,0}(\vec{y}) & = \varphi_{10} - D c_{10} \psi(y_3),\\
\varphi_{1,1}(\vec{y}) & = \varphi_{11} - D c_{11} \psi(y_3),\\
K_1(\vec{y}) &= k_{10} + k_{11}\dfrac{y_3^{n_{K1}}}{(K_{10} - r D)^{n_{K_1}}+y_3^{n_{K1}}}.
\end{aligned}
\end{equation}
Here, $\psi(y_3)$ (defined in \eqref{eq:psi}) corresponds to the drug adaptation that approaches $0$ for high $y_3$, mitigating the drug-induced domulation ($-D c \psi(y_3)$). Moreover, the drug increases the saturation threshold $K_1$ to reduce the expression of $x_1$ in progeny.   

\vspace{0.25cm}
\noindent\textbf{2. Stress-induced stemness ($x_2$).}

Consistent with the concept of adaptability-driven reprogramming \citep{Huz22JCI}, we assume that drug treatment actively promotes the inheritance of the stem-like state $x_2$:
\begin{equation}
\label{varphi-x2}
\varphi_{2,0}(\vec{y}) = \varphi_{20} + D c_{20} \psi(y_3),\ \varphi_{2,1}(\vec{y}) =  \varphi_{21} + D c_{21} \psi(y_3),\ K_2(\vec{y}) = K_2.
\end{equation}
Here, the terms $+Dc\psi(y_3)$ imply that under drug pressure (and before full resistance $x_3$ is acquired), daughter cells tend to have higher stemness, facilitating the transition to DTPs.

\vspace{0.25cm}
\noindent\textbf{3.  Adaptation pathway ($x_3$).}

The adaptive state $x_3$ (e.g., bypass signaling) typically exhibits a see-saw relationship with the primary target $x_1$. We model this by making the half-saturation constant $K_3$ dependent on $y_1$:
\begin{equation}
\label{varphi-x3}
\varphi_{3,0}(\vec{y}) = \varphi_{30}, \varphi_{3,1}(\vec{y}) = \varphi_{31},
\end{equation}
with a Hill-type dependence of $K_3(\vec{y})$ on $y_1$:
\begin{equation}
\label{k3-x3-fun}
K_3(\vec{y}) = k_{30} + k_{31}\dfrac{y_1^{n_{K3}}}{K_{30}^{n_{K_3}}+y_1^{n_{K3}}}.
\end{equation}
A high level of $y_1$ increases $K_3(\vec{y})$, which decreases the expectation response  $\varphi_3(\vec{y})$, effectively suppressing the spontaneous emergence of the resistant state $x_3$ in sensitive populations.

In summary, while specific parameters are estimated phenomenologically, the structure of $p(\vec{x}, \vec{y})$ provides a mechanistic hypothesis: drug resistance emerges not just from selection, but from stress-induced biases in the stochastic inheritance of lagile epigenetic states.

\subsection{Agent-based stochastic simulation}
\label{sec:ICBSS}

While the population dynamics of our system are rigorously governed by the IDE derived from stem cell regeneration dynamics (detailed in Appendix A), numerical solution of this equation in a high-dimensional epigenetic space suffers from the curse of dimensionality. To overcome this computational bottleneck and, crucially, to capture the stochastic extinct events that deterministic models miss, we implement the model using an agent-based stochastic simulation. This approach serves as a Monte Carlo realization of the underlying IDE, where the continuous population density $Q(t, \vec{x})$ is approximated by an ensemble of discrete agents.

The computational procedure follows a fixed time-step evolution, outlined in Algorithm \ref{alg:ICB}.

\vspace{0.25cm}
\begin{algorithm}[H]
	\caption{Agent-based stochastic simulation}
	\label{alg:ICB}
	
	{\bf Initialization:} Set time $t=0$, population size $N = N_0$, and drug schedule. Construct the initial population $\Sigma=\{C_i(\vec{x}, a)\}_{i=1}^N$. Each cell $C_i(\vec{x}, a)$ is characterized by its epigenetic state vector $\vec{x} \in \mathbb{R}^3_+$ and age $a$. Initially, all cells are synchronized in the resting phase with $a=0$, and therefore the resting phase cell number $Q = N$.
	
	\For{$t = 0$ to $T$ with step $\Delta t$}{
		\For{each cell $C_i$ in $\Sigma$}{
			\tcc{1. Update kinetic rates based on current state $\vec{x}$} Calculate $\beta(\vec{x}, Q)$, $\kappa(\vec{x})$, and $\mu(\vec{x})$.
			
			\tcc{2. Determine cell fate within the time interval $(t, t + \Delta t)$ (Bernoulli trials)} Generate a random number $r \sim \text{Uniform}(0, 1)$.
			
			\If{cell is in the resting phase}{
				\If{$r<\beta\Delta t$}{Transition to proliferation phase; set age $a = 0$. State unchanged.}
				\ElseIf{$r<(\beta + \kappa)\Delta t$}{Cell undergoes terminal differentiation (removed from $\Sigma$).}} 
				
			\ElseIf{cell is in the proliferation phase}{
				\If{$a_i < \tau$}{
				\If{$r < \mu \Delta t$}{Cell undergoes apoptosis (removed from $\Sigma$).}
				\Else{Update age: $a \leftarrow a + \Delta t$.}}
				\ElseIf{$a \geq \tau$}{
					\tcc{3. Mitosis and inheritance} Cell divides into two daughter cells. For each daughter, sample new state $\vec{x}_{\mathrm{new}}$ from the kernel $p(\vec{x}_{\mathrm{new}}, \vec{x})$ (Eq. \eqref{eq:gamma}). Reset ages to $0$; add daughters to $\Sigma$ as resting phase cells.}
			}}	
		{\bf Update:}  Update the total population count $N(t)$ and the resting phase cells $Q(t)$; record system state; advance $t\leftarrow t+\Delta t$.
	}
\end{algorithm}
\vspace{0.25cm}

The simulation discretizes time into small intervals $\Delta t$ (set to ensure that transition probabilities $\ll 1$). At each step, cell fate decisions (differentiation, division, apoptosis) are determined via Bernoulli trials comparing the calculated probabilities (e.g., $\beta\Delta t, \kappa \Delta t, \mu \Delta t$) against uniform random numbers. A critical step occurs at mitosis: when a cell divides, the epigenetic states of the two daughter cells are independently sampled from the multivariate Gamma distribution defined by the inheritance kernel $p(\vec{x}, \vec{y})$. This step introduces the necessary epigenetic noise that fuels phenotypic evolution.

Since the stochastic process converges to the deterministic mean-field limit as the system size $Q\to \infty$ (a theoretical consistency we validate in Appendix B), our simulation strategy depends on the population scale:
\begin{enumerate}
\item Deterministic limit: For simulations in Section \ref{sec:3.1} and \ref{sec:3.2}, stochastic fluctuations are negligible relative to the mean trend. Thus, single representative runs are sufficient to characterize the population dynamics.
\item Stochastic regime: For scenarios involving tumor extinction or the initial emergence of rare resistant clones (Fig. \ref{Fig_intermitten} and Section \ref{sec:3.4}), stochastic effects differ significantly between runs. In these cases, we perform ensembles of independent simulations ($n= 3$ for illustrative traces, $n=10$ for robustness analysis) to capture the variance and survival probabilities.
\end{enumerate}

\subsection{Model Parameterization and Calibration}

Parameterizing a high-dimensional agent-based model requires balancing biological plausibility with quantitative accuracy. To ensure the model faithfully reproduces tumor evolutionary dynamics without overfitting, we employed a hierarchical estimation strategy combining literature-derived biological constraints with data-driven calibration.

First, for the fundamental cellular kinetic parameters that govern cell fate decisions---specifically the maximum proliferation rate $\bar{\beta}$, terminal differentiation rate $\kappa_0$, and apoptosis rate $\mu_0$---we restricted our search space to biologically plausible ranges reported in prior quantitative oncology studies \citep{Lima2021Bayesian, PlosCB21Inferring, McClatchy20Modeling,Kolokotroni2016PlosCB}. These physiological constraints (summarized in the Range column of Table \ref{tab-modelpara}) ensure that the simulated cellular turnover rates remain consistent with known cell cycle biological limits.

\begin{figure}[htbp]
	\centering
	\includegraphics[width=13cm]{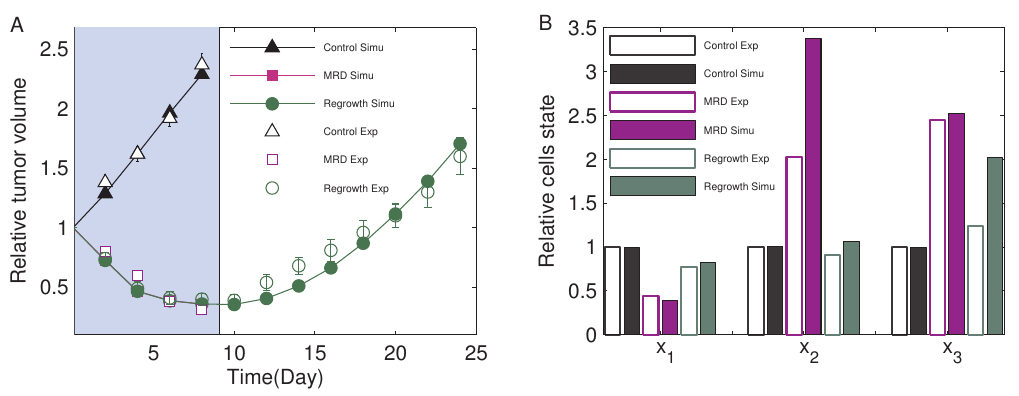}
	\caption{\textbf{Model calibration against in vivo data.} \textbf{A} Fitting of macroscopic tumor volume dynamics. The model (solid lines) accurately captures both the unperturbed growth of Control tumors ($D = 0$) and the regression-relapse trajectory of osimertinib-treated tumors ($D = 1$). Experimental data (symbols, mean $\pm$SEM) derived from PC9 xenografts in \citet{Huz22JCI}. \textbf{B} Validation of microscopic phenotypic heterogeneity. The simulated proportions of microscopic phenotypic heterogeneity. The simulated proportions of epigenetic states correspond to the relative expression levels observed in distinct tumor phases (Control, MRD, Regrowth), confirming the model's ability to reproduce drug-induced phenotypic plasticity. Experimental benchmarks derived from \citet{Huz22JCI} and \citet{Aissa20Natcom}.
}
	\label{Fig-SimuExp}
\end{figure}

For the phenomenological parameters governing the epigenetic inheritance kernel and drug interactions (e.g., shape parameters and Hill coefficients), we performed a sequential calibration using longitudinal data from osimertinib PC9 xenograft models \citep{Huz22JCI}. In \cite{Huz22JCI},  PC9 xenograft mice were treated with the EGFR-TKI osimertinib for 9 consecutive days to establish osimertinib-regressed minimal residual disease (MRD) tumors. The drug was then withdrawn for 14 days, resulting in regrown tumors. As illustrated in Fig. \ref{Fig-SimuExp}, the calibration proceeded in three stages: 
\begin{enumerate}
\item \textbf{Baseline Growth ($D = 0$):} We first disabled the drug term and turned the intrinsic growth parameters to match the tumor volume evolution of the control group (Fig. \ref{Fig-SimuExp}A, black triangles).
\item \textbf{Therapeutic Response ($D = 1$):} Fixing the baseline growth parameters, we then introduced the drug effect to calibrate the sensitivity parameters ($c_{i,j}, r$), fitting the rapid regression and subsequent regrowth phases observed in the experimental group (Fig. \ref{Fig-SimuExp}A, colored markers).
\item \textbf{Phenotypic Distributions:} To validate the microscopic realism of the model, we fine-tuned the inheritance kernel parameters so that the simulated proportions of sensitive ($x_1$), stem-like ($x_2$), and resistant ($x_3$) cells matched the transcriptomic signatures of control, MRD, and regrown tumors reported in \citet{Huz22JCI} (Fig. \ref{Fig-SimuExp}B).
\end{enumerate}

To critically evaluate the influence of parameter uncertainty, we conducted a systematic local sensitivity analysis (detailed in Section \ref{sec:3.4.3}). This analysis identified that tumor recurrence dynamics are primarily driven by kinetic rates and competition coefficients, while being less sensitive to minor variations in specific kernel shape parameters. Consequently, less sensitive parameters were fixed at their calibrated baseline values to maintain model identifiability, while key sensitive parameters were later varied to construct virtual patients for exploring treatment heterogeneity.

A complete summary of the baseline parameter values, their units, and the sources of their derivations is provided in Table \ref{tab-modelpara}. Unless otherwise specified in the figure legends, all subsequent simulations typically utilize these baseline parameter values.

\renewcommand{\arraystretch}{1.2}
\begin{table}[htbp]
	\centering
	\caption{\footnotesize Summary of key system parameters used in the simulations.}
	\label{tab-modelpara}
	\begin{tabular}{@{}lllll@{}}
		\toprule
		Parameter & Description  & Value  & Unit  & Source/Method\\  
		\midrule
		$\bar{\beta}$ & Maximum proliferation rate & $0.118^*$ & $\mathrm{h}^{-1}$  & Range 0.01--0.4 (1-4) \\
		$\bar{\theta}_0$ & Proliferation saturation capacity & $5.9\times 10^2$ & cells & Cali. Control\\
		$m$ & Hill coefficient (proliferation) & $1$ & - & Assumed (Standard) \\
		$\eta_0$ & Drug effect on $\bar{\theta}_0$ & $0.66$ & - & Cali. Treated\\
		$\kappa_0$ & Maximum differentiation rate & $0.051^*$ & $\mathrm{h}^{-1}$ & Range $10^{-4}$--$10^{-1}$ (1-4)\\
		$\tau$ & Proliferartive phase duration & $18$ & $h$ & Range $12 - 36$ (1-4)\\
		$\mu_0$ & Maximum apoptosis rate & $0.0071^*$ & $\mathrm{h}^{-1}$ & Range $10^{-4}$--$10^{-1}$ (1-4)\\
		$\mu_1$ & Effective coefficient of $x_1$ and $x_3$ to apoptosis  & $0.65$ & $\mathrm{h}^{-1}$ & Cal. Treated\\
		$\alpha$ & Relative weight of $x_3$ of joint effect & $0.95^*$ & - & Estimated\\
		$\bar{x}_3$ & Adaption threshold of $x_3$ & $4.2$ & - & Cali. Phenotype\\
		$\gamma_1$ & Baseline combined effect of $x_1$ and $x_3$ & $5.5$ & - & Estimated\\
		$\varphi_{10}$ &  Hill coefficient for $x_1$ & $0.7$ &  -  & Cali. Phenotype\\
		$\varphi_{11}$ &  Hill coefficient for $x_1$ & $4.9$ & - & Cali. Phenotype\\
		$c_{10}$ & Drug effect coefficient on $x_1$ & $0.18$ & - & Cali. Treated\\
		$c_{11}$ & Drug effect coefficient on $x_1$ & $2.45$ & - & Cali. Treated\\
		$k_{10}$ & Effect of $x_3$ on $x_1$ & $0.7$ & - & Cali. Treated\\
		$k_{11}$ & Effect of $x_3$ on $x_1$ & $3.45$ & - & Cali. Treated\\
		$\varphi_{20}$ &  Hill coefficient for $x_2$ & $0.65$ & - & Cali. Phenotype\\
		$\varphi_{21}$ &  Hill coefficient for $x_2$ & $4.5$ & - & Cali. Phenotype\\
		$c_{20}$ & Drug effect coefficient on $x_2$ & $0.2$ & - & Cali. Treated\\
		$c_{21}$ & Drug effect coefficient on $x_2$ & $2.0$ & - & Cali. Treated\\
		$K_{10}$ & Half-effective coefficient of $K_1$ & $2.6$ & - & Estimated\\
		$K_{2}$ & Half-effective coefficient of $x_2$  & $3.45$ & - & Estimated\\
		$K_{30}$ & Half-effective coefficient of $K_3$ & $2.3$ & - & Estimated\\
		$r$ & Drug effect coefficient on $K_{10}$ & $0.3$ & - & Cali. Treated\\
		$\varphi_{30}$ &  Hill coefficient for $x_3$ & $0.7$ & - & Cali. Phenotype\\
		$\varphi_{31}$ &  Hill coefficient for $x_3$ & $4.9$ & - & Cali. Phenotype\\
		$k_{30}$ & Effective coefficient of $x_1$ to $x_3$ & $0.7$ & - & Estimated\\
		$k_{31}$ & Effective coefficient of $x_1$ to $x_3$ & $3.45$ & - & Estimated\\
		$n_i$ & Hill coefficient for $x_i$ ($i=1,2,3$) & $2$ &  - &Estimated\\
		$n_{K_i}$ & Hill coefficient for $x_i$ ($i=1, 3$) & $4$ & - &Estimated\\
		$a$ & Shape parameter of Gamma distribution & $50$ & - & Range $30$--$120$ (5)\\
		$a_1$ & Coefficient in \eqref{beta0} & $1.3$ & - & Cali. Control\\
		$a_2$ & Coefficient in \eqref{beta0} & $0.33$ & -& Cali. Control \\
		$a_3$ & Coefficient in \eqref{beta0} & $0.75$ & -& Cali. Control\\
		$a_4$ & Coefficient in \eqref{beta0} & $0.23$ & - & Cali. Control\\
		$a_5$ & Coefficient in \eqref{beta0} & $10.0$ & - & Cali. Control\\
		$\kappa_1$ & Coefficient in \eqref{kappa}& $0.6$ & - & Cali. Control\\
		$k$ & Coefficient in \eqref{eq:psi} & $ 5$ & - & Cali. Treated\\
		\botrule
		\multicolumn{5}{p{12cm}}{\small Methods: ``Range''  indicates values constrained by literature-constrained biological priors; ``Cali.'' indicates values calibrated by fitting simulation results to data in Fig. \ref{Fig-SimuExp} (Control, Treated, or Phenotype heterogeneity); ``Estimated'' indicates parameters with low sensitivity and estimated based on biological priors.}\\		\multicolumn{5}{p{12cm}}{\small Sources: 1=\cite{Lima2021Bayesian}, 2=\cite{PlosCB21Inferring}, 
			3= \cite{McClatchy20Modeling}, 4=\cite{Kolokotroni2016PlosCB}, 5=\cite{LeiJTB20framework}.}\\
\multicolumn{5}{p{12cm}}{\small $^*$: Parameters with high sensitivity and are varied to generate virtual patient cohorts.}
	\end{tabular}
\end{table}

\section{Results}
\label{sec3}

\subsection{Tumor evolution dynamics in the absence of treatment}
\label{sec:3.1}

\begin{figure}[htbp]
	\centering
	\includegraphics[width=12cm]{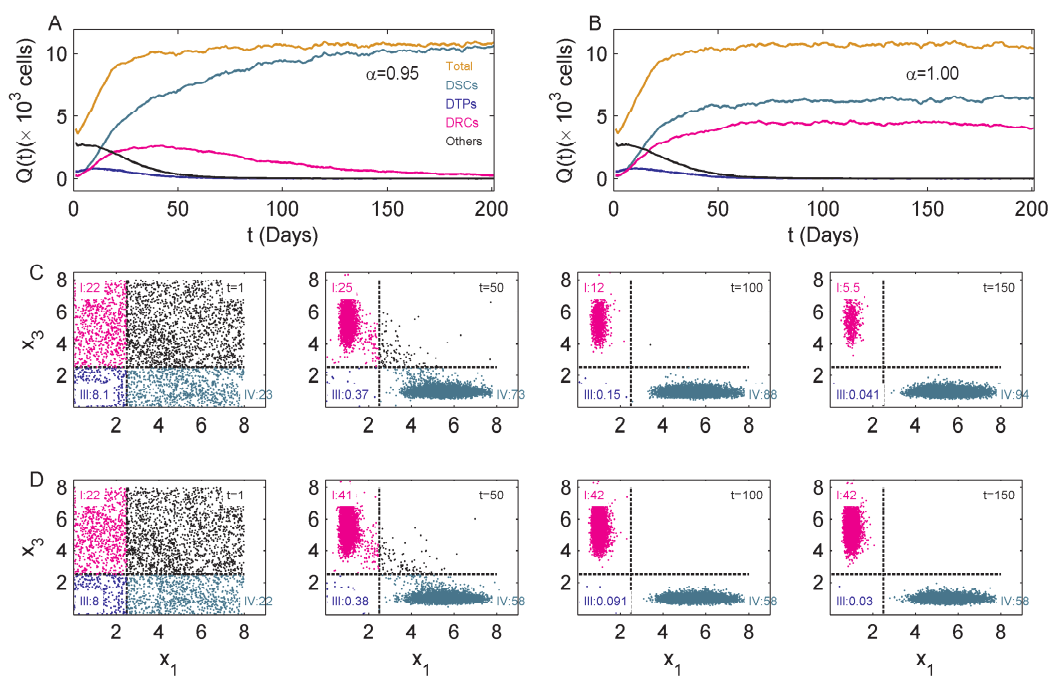}
	\caption{\textbf{Intrindic tumor evolution driven by different competition parameter $\alpha$.} \textbf{A} and \textbf{B} Temporal evolution of tumor subpopulations starting from a uniform epigenetic state distribution, with \textbf{A} biased competition ($\alpha=0.95$), and \textbf{B} balanced competition ($\alpha=1.0$). Subpopulations are color-coded: Total tumor cells (orange), DSCs (pale teal), DTPs (blue), DRCs (magenta), and other phenotypes (black). \textbf{C} and \textbf{D} Evolution of the epigenetic landscape $(x_1, x_3)$ at days $1$, $50$, $100$, and $150$,  with \textbf{C} $\alpha  = 0.95$, and \textbf{D} $\alpha=1.0$. Black dashed lines indicate the phenotypic threshold (see Table \ref{tab-phenotype-epi}), partitioning the state space into four distinct quadrants: I(Non-viable/Othersm black), II(DRCs, magenta), III(DTPs, blue), IV(DSCs, pale teal). Percentages denote the cell fraction within each quadrant.}
	\label{Fig-tumorgrowth}
\end{figure}

We first investigated the intrinsic tumor evolution dynamics in the absence of drug administration ($D(t) = 0$). To examine tumor cell plasticity during tumor expansion, we initialized the system with a naive population of $4\times 10^3$ cells, assigning initial epigenetic states $x_i$ uniformly distributed on $[0, 8]$. The parameter $\alpha$ governs the competition between the complementary pathways associated with drug-sensitive ($x_1$) and resistant ($x_3$). We analyzed two distinct scenarios, $\alpha = 0.95$ and $\alpha = 1.0$, tracking both the macroscopic population growth and microscopic phenotypic drift (Fig. \ref{Fig-tumorgrowth}).

For the cast of $\alpha = 0.95$, the total tumor burden increased rapidly before reaching saturation around day 50 (Fig. \ref{Fig-tumorgrowth}A). Interestingly, while the macroscopic volume stabilized, the internal population structure continued to evolve. The fraction of drug-resistant cells (DRCs) spontaneously declined from 20\% to 5.5\%, while drug-sensitive cells (DSCs) expanded from 73\% to 94\% between days $50$ to $150$ (Fig. \ref{Fig-tumorgrowth}C). This indicates that in a drug-free environment, the system exhibits a natural selection pressure favoring the highly proliferative sensitive phenotype.

In contrast, when $\alpha=1.0$, representing an equal fitness potential for adaptive pathways, the macroscopic growth trajectory remained similar, but the steady-state composition differed markedly (Fig. \ref{Fig-tumorgrowth}B). The population settled into a heterogeneous equilibrium with substantial proportions of both DRCs ($\sim 42\%$) and DSCs ($\sim 58\%$) (Fig. \ref{Fig-tumorgrowth}D). These results demonstrate that the competition coefficient $\alpha$ is a critical determinant of the intrinsic heterogeneity level within the tumor. 

Since clinically observable EGFR-mutant tumors are typically dominated by drug-sensitive clones prior to TKI treatment (reflecting the fitness cost often associated with resistance mechanisms absent drug pressure), we selected $\alpha = 0.95$ for the subsequent simulations to replicate this clinically relevant baseline.

\subsection{Dynamics of DTPs-mediated drug resistance}
\label{sec:3.2}

To investigate how drug-tolerant persister (DTP) cells drive drug resistance, we simulated the evolution of epigenetic states under short-term EGFR-TKI treatment calibrated to PC9 xenograft mice \citep{Huz22JCI}. Experimental observations show that tumor volume markedly decreases during drug treatment (days 1--9), but relapses after drug withdrawal, eventually exceeding 150\% of its pre-treatment size by day 24 (Fig. \ref{Fig-SimuExp}A).  

\begin{figure}[htbp]
	\centering
	\includegraphics[width=12cm]{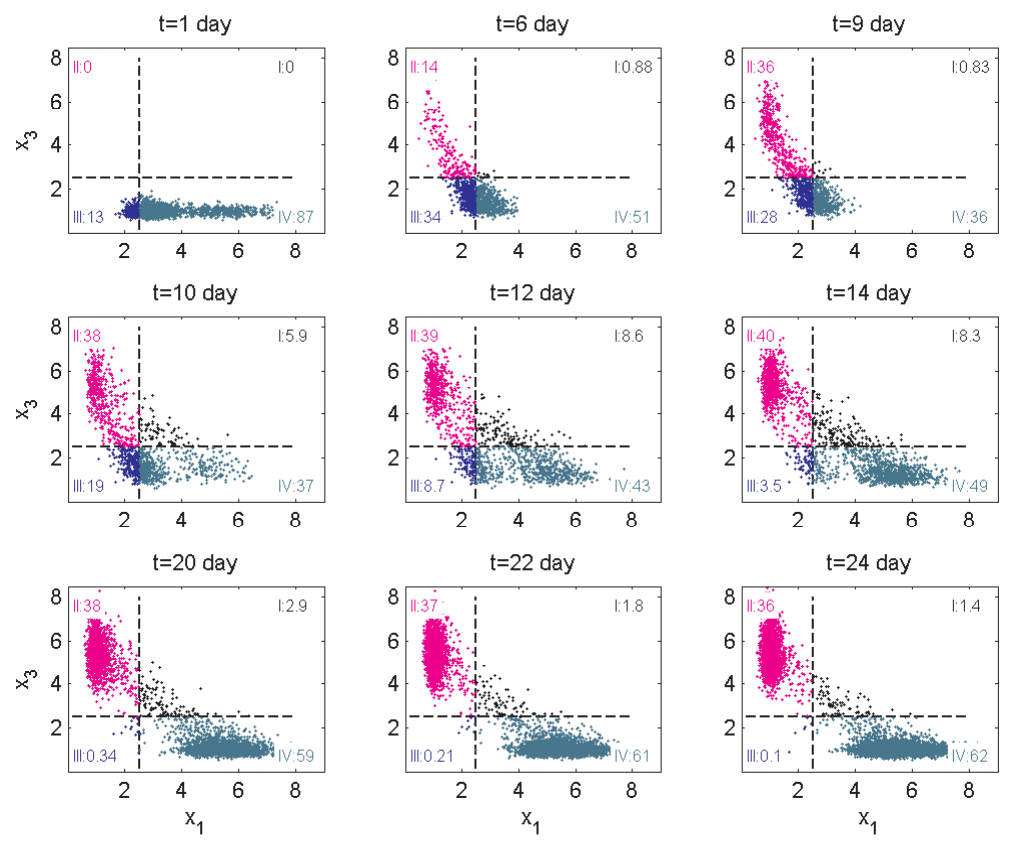}
	\caption{\textbf{Evolution of epigenetic states $(x_1, x_3)$ following drug treatment.} Each point corresponds to an individual tumor cell in the $(x_1, x_3)$ plane. The plane is partitioned into four quadrants defined by the phenotype threshold in Table \ref{tab-phenotype-epi}, with colors indicating the respective regions. The percentage values denote the fraction of cells residing in each quadrant. Parameter settings are identical to those used in Fig. \ref{Fig-SimuExp}.}
	\label{Fig-epistatex1x3-regrowth}
\end{figure}

Figure \ref{Fig-epistatex1x3-regrowth} depicts the evolution of drug sensitivity ($x_1$) and adaptation ($x_3$) states during treatment and tumor relapse. During remission (days 1--9), DSCs with high $x_1$ were depleted, while DTPs with elevated $x_3$ emerged. Between days 10 and 14, following drug withdrawal, both drug-sensitive and drug-tolerant populations expanded, whereas the fraction of transition cells (with simultaneously low $x_1$ and $x_3$) declined. By day 14, a stable drug-resistant phenotype characterized by high $x_3$ and low $x_1$ was established. At later stages (days 20--24), cells segregated into distinct phenotypic clusters with well-separated $x_1$ and $x_3$ states. Notably, by day 24, about 62\% of cells had reverted to a drug-sensitive state, indicating that a substantial fraction of the population can re-establish sensitivity once treatment is discontinued.

\begin{figure}[htbp]
	\centering
	\includegraphics[width=12cm]{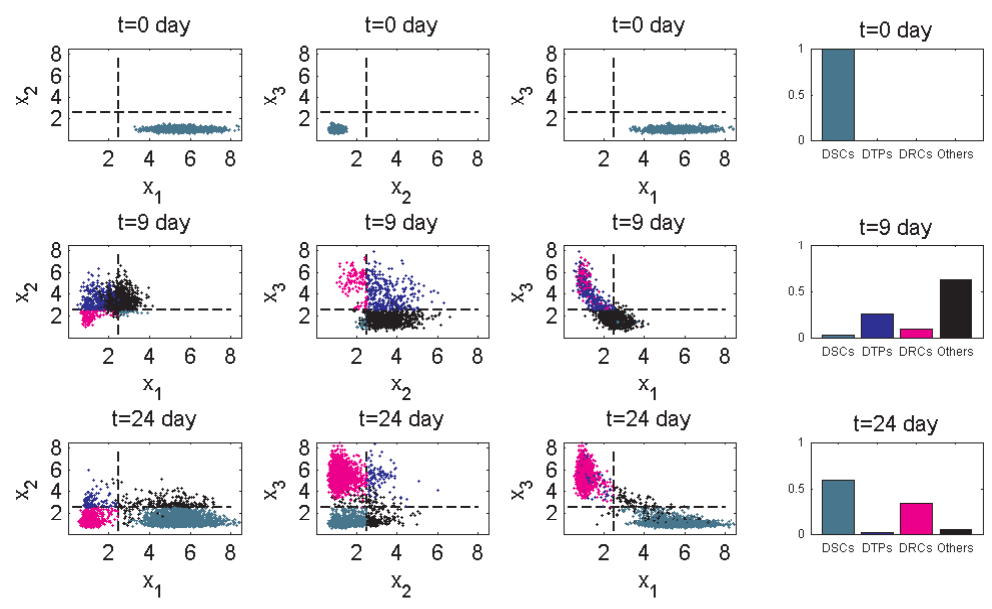}
	\caption{\textbf{Epigenetic state dynamics during tumor regrowth.} Scatter plots show the distributions of $(x_1, x_2)$, $(x_2, x_3)$, and $(x_3, x_1)$ at days 0, 9, and 24 after treatment. Each point represents an individual tumor cell. Colors denote cell phenotypes: DSCs (pale teal), DTPs (blue), DRCs (magenta), and other phenotypes (black). Black dashed lines mark the threshold values $x^i_s$ that define phenotype boundaries (see Table \ref{tab-phenotype-epi}). The rightmost column summarizes the phenotype proportions at each time point. }
	\label{Fig-epistate-regrowth}
\end{figure}

To further characterize tumor cell plasticity and heterogeneity during relapse, we analyzed the evolution of phenotypic fractions (Fig. \ref{Fig-epistate-regrowth}). Phenotypes were defined according to Table \ref{tab-phenotype-epi}. Prior to treatment (day $0$), the majority of cells were DSCs, characterized by high $x_1$ and low $x_2$, $x_3$. During treatment, the DSC fraction dropped sharply, and surviving cells transitioned into DTPs (day 9), showing upregulated $x_2$ and $x_3$ along with reduced $x_1$. At this time, many cells displayed mixed features corresponding to intermediate transitional states. After treatment withdrawal, the DSC fraction progressively increased as $x_1$ was restored. DTPs largely disappeared by day $24$, whereas DRCs rose to $36\%$. Prolonged simulation predicted that the DRC fraction would eventually decline to zero by day $150$ (Fig. \ref{Fig-num-x1x3-drug}A). Together, these results highlight the molecular state transitions underlying DTP-induced resistance and demonstrate that our model captures experimentally observed tumor relapse dynamics. 

We next examined how cell plasticity influences relapse dynamics by varying the shape parameter $a$ in the inheritance function $p(\vec{x}, \vec{y})$. The parameter $a$ associates with the shape parameter of the Gamma distribution \eqref{eq:gamma}. A larger $a$ corresponds to a narrow distribution (less variance) for fixed means, and increasing $a$ effectively reduces the stochastic phenotypic drift (plasticity) during cell division.

\begin{figure}[htbp]
	\centering
	\includegraphics[width=13cm]{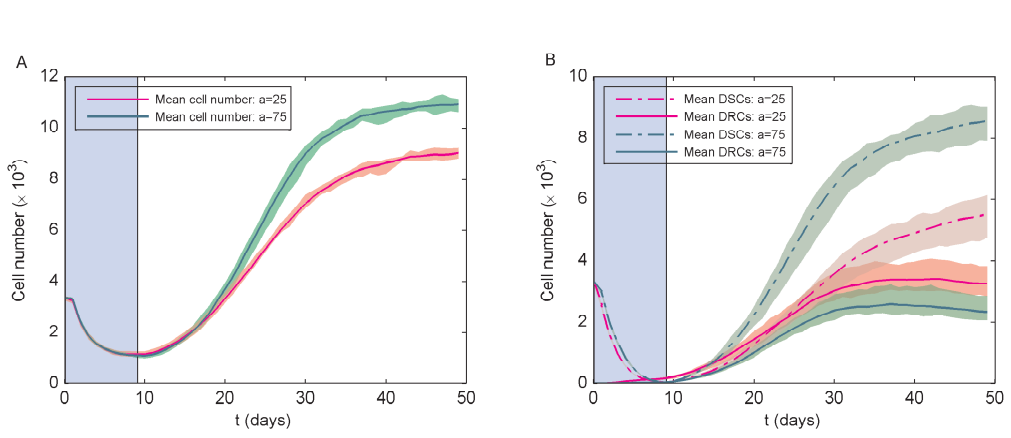}
	\caption{\textbf{Relapse dynamics under different plasticity parameters $a$.} Simulations followed the same protocol as in Fig. \ref{Fig-SimuExp}, with continuous drug treatment applied for 9 days. Model parameters are given in Table \ref{tab-modelpara}. For each $a$, ten independent simulations with different random seeds were performed; solid lines represent mean trajectories, and shaded regions indicate the range across replicates. \textbf{A} Total tumor cell numbers for $a = 25$ (magenta) and $a=75$ (pale teal). \textbf{B} Dynamics of DSC and DRC subpopulations for $a=25$ (magenta) and $a=75$ (pale teal). Solid lines correspond to DSCs, and dash-dotted lines to DRCs. Increased plasticity ($a = 75$) accelerates the expansion of DRCs and promotes earlier relapse compared to low plasticity $a = 25$. Lines represent the mean of 3 independent simulations, and the shaded regions indicate the Mix-to-Max range.}
	\label{Fig-plasticity}
\end{figure}

Simulations with $a = 25$ and $a = 75$ revealed that higher $a$ accelerated tumor relapse (Fig. \ref{Fig-plasticity}A). Analyzing DSC and DRC populations separately showed that a larger $a$ yielded a smaller DRC fraction and a larger DSC fraction (Fig. \ref{Fig-plasticity}B). These results suggest that reduced plasticity diminishes the generation of DRCs and favors the faithful restoration of DSCs. Because sensitive cells proliferate naturally faster than resistant ones, higher inheritance fidelity (reduced plasticity) ultimately accelerates regrowth, leading to a higher tumor burden after treatment withdrawal. 

These simulations demonstrate that DTPs act as a transient reservoir enabling tumor survival during treatment and shaping relapse dynamics through their interplay with both drug-sensitive and drug-resistant populations.

\subsection{Formation of irreversible drug-resistant cells}
\label{subsectionTumorevolution}

Experimental studies have demonstrated that DTPs are transient and reversible: they can survive short-term drug exposure but typically revert to a drug-sensitive phenotype once treatment is withdrawn \citep{Sharma:2010aa}. However, prolonged drug administration may drive a transition from DTPs to irreversible DRCs. Unlike DTPs, DRCs maintain resistance even after drug withdrawal and recover proliferative capacity comparable to the original DSCs \citep{Marine:2020aa}. Here, we utilized our model to investigate the emergence of irreversible resistance and tumor progression under prolonged drug treatment.

\begin{figure}[htbp]
	\centering
	\includegraphics[width=13cm]{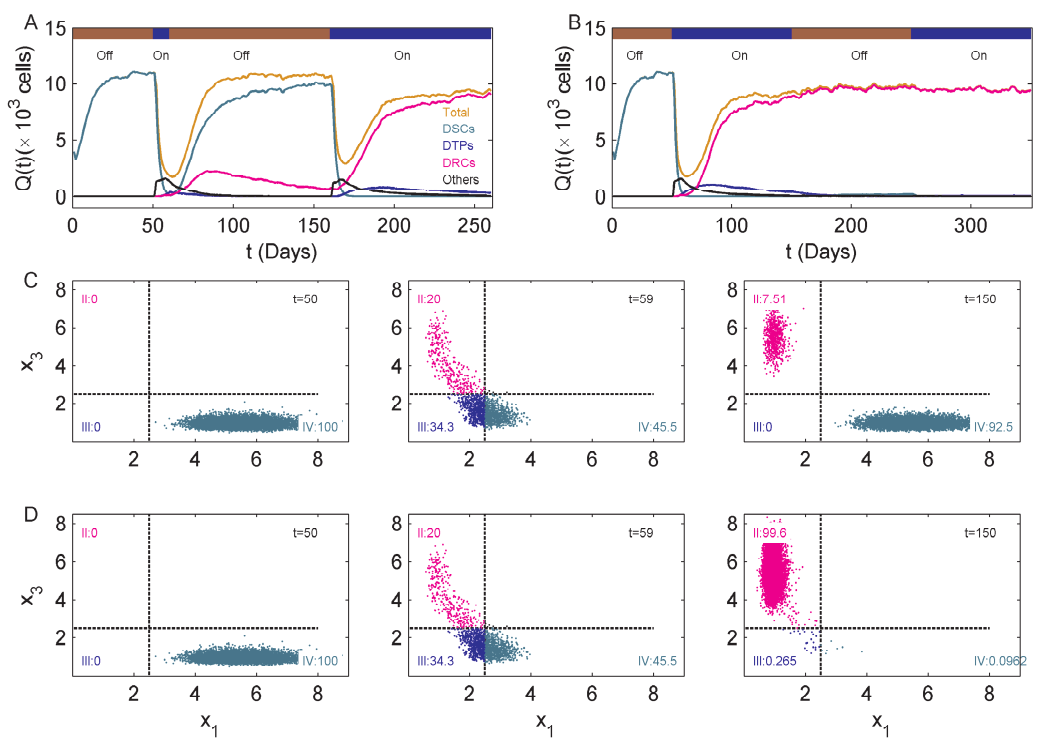}
	\caption{\textbf{Tumor evolution under short- and long-term drug treatments.} \textbf{A} Cell number dynamics and \textbf{C} epigenetic states $(x_1, x_3)$ at different time points following short-term treatment. \textbf{B} Cell number dynamics and \textbf{D} epigenetic states $(x_1, x_3)$ at different time points following long-term treatment. All tumors were initialized with DSCs. In A and C, the drug was applied during $50 \leq t \leq 60$ and $160 \leq t \leq 260$. In B and D, the drug was applied during $50 \leq t \leq 150$ and $250 \leq t \leq 350$. In C and D, the four quadrants of $(x_1,x_3)\in [0, 8]\times [0, 8]$ are color-coded: I(black), II(magenta), III(blue), IV(pale teal). Numbers indicate the proportions of cells in each quadrant. Short-term treatment suppresses growth transiently without altering epigenetic diversity, whereas long-term treatment profoundly shifts cell states and promotes resistant subpopulations.}
	\label{Fig-num-x1x3-drug}
\end{figure}

To compare tumor responses to short- and long-term treatment, we simulated two protocols: (i) a short $10$-day treatment followed by a $100$-day drug-free period before retreatment, and (ii) a long $100$-day treatment followed by the same drug-free period and retreatment (Fig. \ref{Fig-num-x1x3-drug}). 

During the short-term treatment, tumor cell numbers declined sharply upon drug administration, accompanied by a transient rise in DTPs and DRCs (Fig. \ref{Fig-num-x1x3-drug}A). In the epigenetic state space, DSCs with high $x_1$ shifted toward regions of reduced $x_1$ and elevated $x_3$, indicating a rapid transition into DTPs and DRCs (middle panel of Fig. \ref{Fig-num-x1x3-drug}C). After drug withdrawal, DSCs expanded and gradually regained dominance, while DRCs persisted at low levels due to competitive suppression by faster-growing DSCs. By day 160, approximately 93\% of cells had restored their drug-sensitive state (right panel of Fig. \ref{Fig-num-x1x3-drug}C). Upon retreatment, most cells were again eliminated, though the tumor nadir (minimum) was higher than during the first treatment cycle due to the residual resistant subpopulation. Nevertheless, these dynamics are consistent with experimental observations that DTPs are reversible under short-term therapy \citep{Sharma:2010aa, Ramirez16Diverse}.

In contrast, under long-term continuous treatment (Fig. \ref{Fig-num-x1x3-drug}B), DRCs progressively adapted to the drug environment and became the dominant population by day $150$, while DSCs declined to negligible levels. DTPs initially increased during the first 20 days of treatment but subsequently decreased, indicating a continuous flux from transient tolerance to stable resistance. The corresponding changes in $x_1$ and $x_3$ reveal a clear phenotypic inversion, with most cells adopting a drug-resistant phenotype by day $150$ (right panel of Fig. \ref{Fig-num-x1x3-drug}D). After drug withdrawal, DRCs remained the predominant population, and retreatment at day $250$ elicited little response, confirming the establishment of irreversible resistance. 

The above simulations demonstrate that our model recapitulates the experimentally observed distinction between short-term reversible tolerance and long-term irreversible resistance, and identifies the DTP-to-DRC transition as a key mechanism driving stable therapeutic failure. 

\subsection{Dynamic therapies}
\label{sec:3.4}

Continuous treatment promotes the emergence of irreversible DRCs and tumor recurrence, whereas short-term treatment tends to generate reversible DTPs. To restore drug sensitivity and delay resistance, dynamic treatment strategies involving drug discontinuation (``drug holidays'') have been proposed. Two widely studied approaches are intermittent therapy and adaptive therapy. In both cases, the timing of drug administration and withdrawal is critical for therapeutic efficacy. We used our model to simulate and compare the effects of these strategies.

\subsubsection{Intermittent therapy}

Intermittent therapy applies drugs in periodic on/off cycles. In our simulations, treatment was initiated at day $50$, with the drug administered for $T_{\mathrm{on}}$ days followed by a drug-free interval of $T_{\mathrm{off}}$ days. Each $(T_{\mathrm{on}}, T_{\mathrm{off}})$ combination defines a therapy protocol. The cycle was repeated until tumor recurrence, defined as the tumor size exceeding $8\times 10^3$ cells. We quantify treatment efficacy using the tumor relapse time, defined as the duration from the start of treatment to recurrence:
 \begin{equation}
 \mbox{tumor relapse time} = \min_{t > 50} \{t \vert N(t) > 8000\} - 50.
 \end{equation}

\begin{figure}[htbp]
	\centering
	\includegraphics[width=13cm]{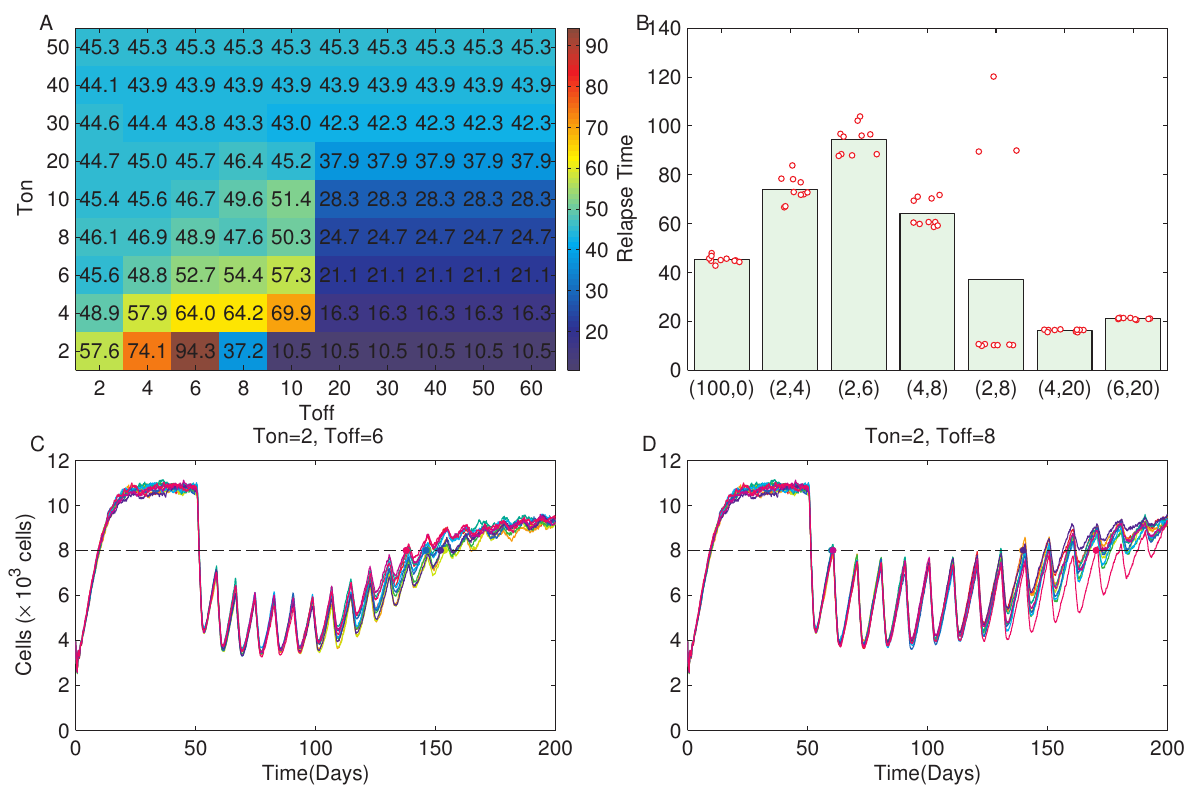}
	\caption{\textbf{Tumor relapse time under intermittent therapy.} \textbf{A} Heatmap of tumor relapse time across different combinations of $T_{\mathrm{on}}$ (duration of drug on) and $T_{\mathrm{off}}$ (duration of drug off). \textbf{B} Tumor relapse time from independent simulation runs for different $(T_{\mathrm{on}}, T_{\mathrm{off}})$ combinations. Bars represent the average relapse time, while circles denote the relapse time from 10 independent runs. The first column $(100, 0)$ corresponds to continuous treatment, and subsequent columns represent intermittent therapy with specified $(T_{\mathrm{on}}, T_{\mathrm{off}})$.  \textbf{C} and \textbf{D} Tumor dynamics for 10 independent runs for selected $(T_{\mathrm{on}}, T_{\mathrm{off}})$ combinations. Each curve shows tumor cell number over time. Black dotted lines indicate the relapse threshold, and filled circles mark the time points when tumors cross this threshold. Intermittent therapy prolongs relapse time compared to continuous treatment, but the outcome strongly depends on the choice of $T_{\mathrm{on}}$ and $T_{\mathrm{off}}$.}
	\label{Fig_intermitten}
\end{figure}

We tested various $(T_{\mathrm{on}}, T_{\mathrm{off}})$ combinations, performing 10 independent simulations per case. The average relapse times are summarized in Fig. \ref{Fig_intermitten}A. Results indicate that proper coordination of drug exposure and holidays is essential to maximize relapse time. The longest delay ($94$ days) occurred with $(T_{\mathrm{on}}, T_{\mathrm{off}}) = (2,6)$, whereas poorly timed drug holidays (e.g., $T_{\mathrm{off}}\geq 10$) led to rapid relapse within $10$ days. Increasing $T_{\mathrm{on}}$ beyond $20$ days provided no further benefit, with relapse times plateauing near $45$ days regardless of $T_{\mathrm{off}}$. 

To assess robustness, we compared relapse times across independent runs (Fig. \ref{Fig_intermitten}B). Most combinations yielded consistent outcomes, except for $(T_{\mathrm{on}}, T_{\mathrm{off}}) = (2,8)$, where some runs resulted in early relapse ($10$ days) and others in late relapse ($90$ days). This variability suggests that for certain protocols, outcomes are highly sensitive to intrinsic stochastic in tumor evolution, leading to unpredictable responses.

We next examined tumor dynamics under selected schedules (Fig. \ref{Fig_intermitten}C--D). With short $T_{\mathrm{on}}$, tumor burden declined during treatment but rebounded during holidays. When subsequent treatments were applied before the tumor reached the relapse threshold, long-term control was achievable (Fig. \ref{Fig_intermitten}C). However, prolonged drug-free intervals allowed tumors to cross the threshold prematurely, shortening relapse time (Fig. \ref{Fig_intermitten}D). These results highlight that fixed-interval schedules may not be optimal, emphasizing the critical importance of treatment timing. 

\subsubsection{Adaptive therapy}

Adaptive therapy dynamically adjusts treatment according to tumor burden \citep{Gatenby09Adaptive}. The rationale is to maintain a stable tumor size by preserving a population of drug-sensitive cells that competitively suppress resistant subpopulations. In our simulations, therapy was turned on when tumor size exceeded a threshold $P_{\mathrm{on}}$ and turned off when it dropped below a lower threshold $P_{\mathrm{off}}$ (with $P_{\mathrm{on}} \ge P_{\mathrm{off}}$).  

\begin{figure}[htbp]
\centering
\includegraphics[width=13cm]{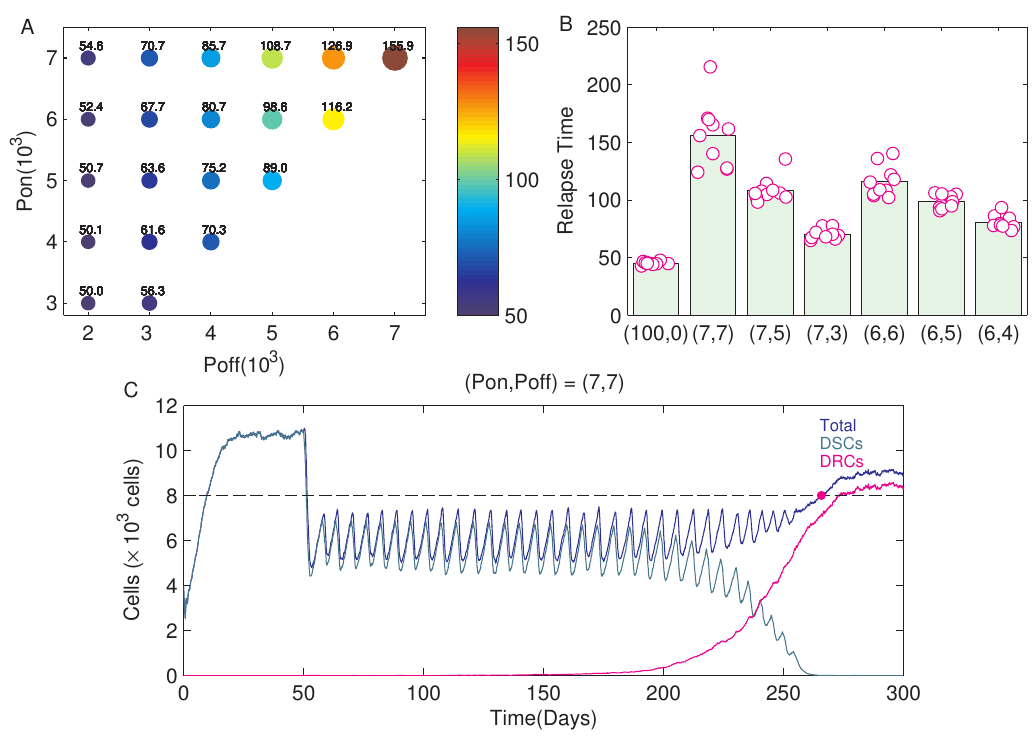}
\caption{\textbf{Tumor relapse time under adaptive therapy.} \textbf{A} Heatmap of relapse time across different combinations of drug on/off thresholds. $P_{\mathrm{on}}$ and $P_{\mathrm{off}}$ denote the thresholds for drug initiation and withdrawal, respectively. \textbf{B} Comparison of continuous treatment with adaptive therapy. The first column shows the results of continuous treatment, while subsequent columns represent adaptive therapy at specified $(P_{\mathrm{on}}, P_{\mathrm{off}})$ values. Bars indicate the average relapse time, and magenta circles correspond to individual stochastic simulation runs. \textbf{C} Tumor evolution dynamics under adaptive therapy with $(P_{\mathrm{on}}, P_{\mathrm{off}}) = (7,7)$. Curves represent the trajectories of total tumor cells (blue), DSCs (pale teal), and DRCs (magenta). The black dotted line marks the relapse threshold, and the magenta dot indicates the time point when the tumor crosses this threshold. Adaptive therapy prolongs relapse compared to continuous treatment, with outcomes depending on the choice of threshold values.}
\label{Fig-Adaptive}
\end{figure}

Fig. \ref{Fig-Adaptive}A shows the average relapse times for different $(P_{\mathrm{on}}, P_{\mathrm{off}})$ strategies, based on 10 independent simulations per case. Adaptive therapy markedly extended relapse times compared with intermittent therapy. Increasing either threshold generally postponed relapse. Moreover, results across runs were highly consistent (Fig. \ref{Fig-Adaptive}B), indicating robust therapeutic outcomes. By contrast, continuous therapy yielded an average relapse time of only $50$ days, underscoring the advantage of adaptive control.

To illustrate the dynamics, Fig. \ref{Fig-Adaptive}C shows tumor composition for $(P_{\mathrm{on}}, P_{\mathrm{off}}) = (7, 7)$. Here, DRCs remained consistently suppressed by DSCs throughout repeated ON/OFF cycles. Unlike long continuous therapy, which drives rapid DRC expansion (cf. Fig. \ref{Fig-num-x1x3-drug}B), adaptive therapy maintained DRCs at low levels until their eventual takeover triggered relapse.

\subsection{Virtual clinical trials and patient stratification}

\subsubsection{Generation and validation of the virtual patient cohort}
\label{sec:3.4.3}

In-silico clinical trials using virtual patients (VPs) provide a powerful framework to assess the therapeutic strategies in the presence of interpatient heterogeneity. Here, a virtual patient is defined as a specific instance of the mathematical model, characterized by a unique vector of physiological parameters sampled from biological distributions. This approach allows us to explore how individual variations in model parameters influence treatment outcomes without the logistical constraints of physical trials.

To construct a representative VP cohort, we first identified the most critical parameters driving treatment response via local sensitivity analysis (LSA). We examined the sensitivity of the relapse time $R_t$ (defined as the first time the tumor burden exceeds $8,000$ cells) to small perturbations in the definition parameters list in Table \ref{tab-modelpara}, under a standard treatment window $t \in [50, 150]$. The local sensitivity index for the $i$-th parameter $\theta_i$ is calculated as:
\begin{equation*}
	LSA_i=\dfrac{\partial R_t}{\partial \theta_i}.
\end{equation*}

\begin{figure}[htbp]
	\centering
	\includegraphics[width=12cm]{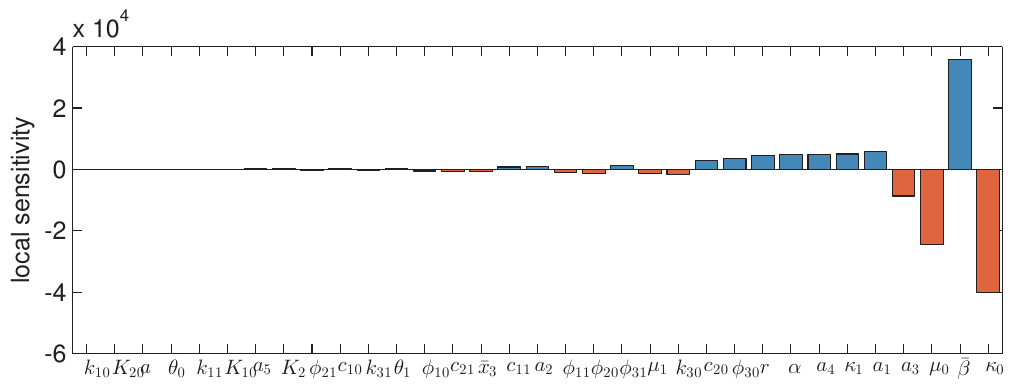}
	\caption{\textbf{Local sensitivity analysis (LSA) for parameter selection.} Sensitivity of tumor relapse time ($R_t$) to model parameters under a treatment window of $t\in [50,150]$. Parameters are ranked by the absolute magnitude of their sensitivity. Light blue bars indicate a positive correlation (increasing the parameter delays relapse), while bright orange bars indicate a negative correlation. }
	\label{Fig-relativesensitivity}
\end{figure}

As shown in Fig. \ref{Fig-relativesensitivity}, parameters governing cellular kinetics---specifically the proliferation rate $\bar{\beta}$, maximum differentiation rate $\kappa_0$, and basal apoptosis rate $\mu_0$---exhibit the highest impact on relapse time. Additionally, the competition coefficient $\alpha$, which indicates the fitness cost of resistance, was included due to its direct role in shaping the evolutionary landscape. Consequently, each virtual patient is represented by a parameter vector $(\bar{\beta}, \kappa_0, \alpha, \mu_0)$, while other less sensitive parameters were held constant. 

The virtual cohort was generated using a Gaussian Mixture Model (GMM) to capture the realistic distribution of these four parameters (see Appendix \ref{sec:supp-VP} for sampling details). To validate the physiological relevance of our cohort, we compared the simulated progression-free survival (PFS) under continuous therapy against clinical data from the EURTAC trial, a pivotal Phase 3 study comparing erlotinib with chemotherapy in EGFR-mutant NSCLC  \citep{Rosell:2012aa}. Our model accurately reproduced the clinical Kaplan-Meier curves (Figure \ref{Fig-PFS-VP-Adaptive_Ppara}A, blue vs. pale teal), confirming that the generated virtual population captures the essential heterogeneity observed in real-world patients.

\subsubsection{Mechanisms determining therapeutic benefit}

\begin{figure}[htbp]
	\centering
	\includegraphics[width=13cm]{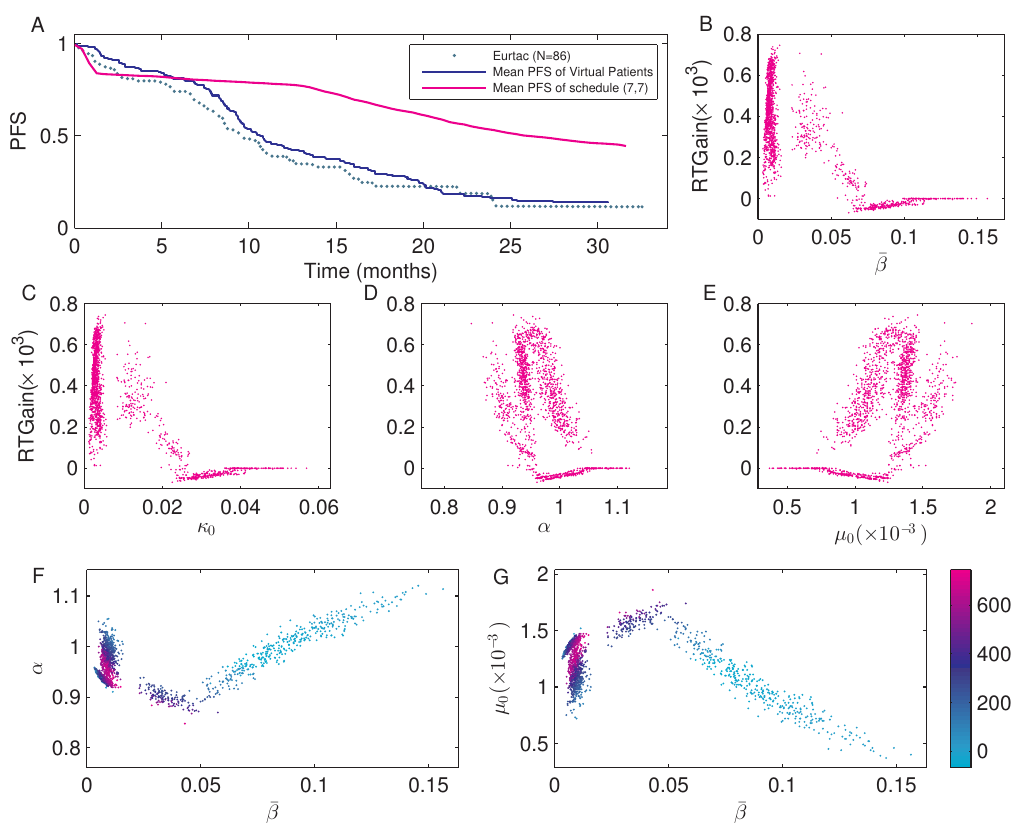}
	\caption{\textbf{Stratification of therapeutic outcomes in virtual patients.} \textbf{A}  Progression-free survival (PFS) Kaplan-Meier curves. The virtual cohort (blue line) reproduces the clinical EURTAC trial data (pale teal dotted line) under continuous therapy. Adaptive therapy (magenta line) significantly extends overall PFS. \textbf{B}--\textbf{E} Dependencies of Relapse Time Gain (RTGain) on the four key parameters across the cohort. \textbf{F}--\textbf{G}  Two-dimensional stratification of patient response. Scatter plots showing RTGain (color-coded) as a function of proliferation rate $\bar{\beta}$ vs. competition $\alpha$ (F) and apoptosis $\mu_0$ (G). Three distinct zones (High Proliferation/Failure, Ecological Control/Success, and Intermediate) reveal the mechanistic constraints of adaptive therapy.}
	\label{Fig-PFS-VP-Adaptive_Ppara}
\end{figure}

We next evaluated the efficacy of adaptive therapy $(P_{\mathrm{on}}, P_{\mathrm{off}}) = (7, 7)$ within this validated cohort. While the overall population showed a clear survival benefit compared to continuous treatment (Fig. \ref{Fig-PFS-VP-Adaptive_Ppara}A, magenta line), the degree of benefit varied significantly across individuals. To quantify this, we defined the Relapse Time Gain (RTGain) for each patient as:
\begin{equation}
\mbox{RTGain}  =  T_{\text{relapse}}^{\text{adaptive}}  - T_{\text{relapse}}^{\text{continuous}}.
\end{equation}

By projecting RTGain onto the parameter space (Fig. \ref{Fig-PFS-VP-Adaptive_Ppara}B-G), we identified three distinct mechanistic zones that explain why some patients benefit from adaptive therapy while others do not:
\begin{enumerate}
\item High Proliferation Zeon (Negative Gain): Patients with high proliferation rates ($\bar{\beta}) > 0.07$, Fig. \ref{Fig-PFS-VP-Adaptive_Ppara}B) exhibited negative RTgain. In these highly aggressive tumors, the rapid regrowth during drug holidays outpaces the suppression exerted by sensitive cells. Consequently, the tumor rebounds prematurely before the next treatment cycle can be effective. For such patients, continuous maximum-tolerated dose (MTD) therapy remains the superior strategy. 
\item Ecological Control Zone (High Gain): The most significant benefits ($\mathrm{RTGain} > 0$) were observed in patients with low proliferation rates ($\bar{\beta} < 0.02$, Fig. \ref{Fig-PFS-VP-Adaptive_Ppara}B), particularly when combined with high intrinsic apoptosis ($\mu_0 > 1.2 \times 10^{-3}$) and lower resistant fitness ($\alpha < 0.95$) (Fig. \ref{Fig-PFS-VP-Adaptive_Ppara}F-G). In this regime, the slower growth and higher turnover of the tumor allow for effective ecological control. During drug holidays, the restored drug-sensitive population effectively outcompetes resistant clones, suppressing their expansion and prolonging the time to resistance.
\item Intermediate Compensation Zone: In the transition region where proliferation is moderate, positive outcomes are still achievable through a compensatory mechanism. Even if $\bar{\beta}$ is relatively high, a combination of stronger apoptosis pressure (high $\mu_0$) and intense interclonal competition (low $\alpha$) can preserve the efficacy of adaptive cycles. 
\end{enumerate}

These findings suggest that adaptive therapy is not a ``one-size-fits-all solution''. Its success relies on the specific balance of tumor kinetics, where the cost of resistance can be leveraged. High-frequency turnover and moderate growth rates favor adaptive protocols, whereas hyper-proliferative tumors require continuous suppression.

\section{Conclusion}

Non-genetic resistance mechanisms play a critical role in tumor recurrence. A prominent example is resistance mediated by drug-tolerant persister cells (DTPs), a transient phenotype that allows tumor cells to survive treatment through reversible changes in gene expression, epigenetic modifications, or metabolic reprogramming. Over prolonged exposure, these DTPs may further evolve into stable drug-resistant cells (DRCs). However, the quantitative description of tumor recurrence dynamics and the link between micro-level epigenetic regulation and macro-level tumor population behavior remain poorly understood.

In this study, we developed an agent-based computational framework, combining a mathematical model with a stochastic simulation algorithm, to capture the evolution of cellular epigenetic states and tumor population dynamics during recurrence. To characterize cellular heterogeneity, we introduced three quantitative indicators---sensitivity, proliferative differentiation potential (stemness), and adaptation (resistance)---that directly modulate cell-specific kinetic rates and fitness. To bridge intracellular regulation with population-level outcomes, we defined an inheritance probability function describing the likelihood of daughter cells adopting epigenetic states distinct from their mothers. This framework captures both phenotype transitions and heterogeneity arising from cellular plasticity.

While foundational studies utilizing ordinary differential equations (ODEs) \citep{KuosmanenPlosCB2025,Gevertz2025npjSBA} have elegantly demonstrated that phenotypic plasticity alone can drive resistance, they fundamentally rely on mean-field approximations. These macro-scale models assume cells within a compartment (e.g., a tolerant state) are homogeneous. However, drug resistance often originates from the tails of the population distribution---rare subpopulations with extreme epigenetic states \citep{Boumahdi20greatescape,Sharma:2010aa,Cell21Colorectal}. Our agent-based model (ABM) complements these macroscopic continuous frameworks by providing microscopic resolution. By tracking individual cells, our model captures the stochastic emergence of these rate outlier cells, which are often averaged out in deterministic ODE formulations but are critical for predicting the exact timing of relapse.

Crucially, recent single-cell lineage tracing by Whiting et al. \citep{Whiting2025Natcom} has experimentally validated the existence of a continuous transcriptional trajectory during drug adaptation. Our model aligns with this biological reality by defining the cell state as a continuous vector $\vec{x}= (x_1, x_2, x_3)$ rather than discrete bins. This allows us to simulate the gradual epigenetic drift along the fitness landscape, offering a mechanistic explanation for the intermediate phenotypes (partial resistance) observed in the clinic.

Parallel efforts in the field have focused on quantifying the dynamics of heterogeneity, for instance, by applying Shannon entropy measures to non-genetic information \citep{IyerPlosCB2025}. While providing valuable theoretical metrics for monitoring diversity, our work aims to bridge the gap towards translational application. By explicitly modeling biologically interpretable exes (e.g., $x_2$ representing stemness/differentiation potential), we extend biological mechanisms into clinical simulation. This structure enables the generation of virtual patient cohorts to test specific dynamic treatment protocols (e.g., tumor-burden-triggered adaptive therapy), demonstrating how continuous epigenetic feedback can be exploited to delay relapse in a personalized manner.

The model was applied to fit experimental data from NSCLC mouse xenografts under EGFR-TKI treatment. Simulations recapitulated tumor growth trajectories and qualitative gene expression changes observed in vivo, while also revealing cell-cell epigenetic dynamics and phenotype proportions that are experimentally inaccessible. In particular, the inheritance function parameters were shown to strongly influence plasticity, thereby shaping epigenetic state distributions, phenotype transitions, and tumor growth.

The model further reproduced the reversible nature of DTPs under short-term treatment and the emergence of irreversible DRCs during long-term treatment. This transition was associated with a decline in epigenetic state $x_1$, an increase in state $x_3$, and the rising fitness of DRCs.

Based on xenograft-fitted parameters, we simulated intermittent and adaptive therapy regimens. Adaptive therapy consistently outperformed continuous dosing, prolonging progression-free survival (PFS) in most virtual patients. However, for patients with intrinsically short PFS, adaptive therapy yielded limited or even adverse outcomes, highlighting the need for personalized treatment optimization.

The critical feature of our multi-scale framework is the explicit mapping between mathematical parameters and biologically actionable mechanisms, which allows for the direct simulation of specific therapeutic interventions. First, the kinetic rate parameters---specifically the maximum proliferation rate $\bar{\beta}$, basal apoptosis rate $\mu_0$, and differentiation rate $\kappa_0$---serve as quantifiable proxies for cell fate decisions. These rates are not arbitrary variables but can be routinely measured \textit{in vitro} using standard assays such as Ki-67 staining, Annexin V/PI flow cytometry, or live-cell imaging \citep{MacKey:2001aa,Darzynkiewicz:1994aa,Kim:2015aa}. In a clinical context, these parameters represent the primary targets of standard-of-care treatment; for instance, cytotoxic agents like cisplatin and paclitaxel or cytostatic drugs directly modulate $\bar{\beta}$ and $\mu_0$ to induce tumor regression \citep{Shirmanova:2017aa,Drew:2025aa,Gupta:2025aa}.

Second, beyond macroscopic growth kinetics, our model captures the molecular regulation of cellular plasticity through the epigenetic inheritance function $\varphi_i$. This function governs the probability of phenotypic switching during cell division, a process biologically regulated by epigenetic ``writers'' and ``erasers'' such as DNA methyltransferases and histone modifiers. Consequently, the parameter $\varphi_i$ provides a specific interface for modeling epigenetic therapies. It can be experimentally perturbed using remodeling agents (e.g., DNMT inhibitors like 5-azacytidine or EZH2 inhibitors) \citep{Guler:2017aa,Ramaiah:2021aa} or via precise epigenetic editing technologies utilizing CRISPR-Cas9-based epi-editors \citep{Wei:2025aa,Villiger:2024aa}. By integrating these distinct layers of regulation, our model offers a mechanistic platform to evaluate the efficacy of combination strategies---such as coupling chemotherapy (targeting kinetic rates) with epigenetic therapy (targeting phenotypic plasticity)---to overcome drug resistance.

While the proposed framework offers valuable insights into DTP-mediated resistance, it has limitations. Detailed intracellular signaling networks were abstracted into coarse-grained indices of epigenetic state, which may limit accuracy in reproducing specific datasets. Moreover, the influence of the tumor microenvironment, including macrophages and cancer-associated fibroblasts (CAFs), was not considered. Future work should integrate signaling pathways and microenvironmental interactions to dissect DTP-driven resistance in greater depth and to refine the design of optimal, patient-specific therapeutic strategies.

\backmatter

\bmhead{Acknowledgements}

This work was funded by the National Natural Science Foundation of China (Grant No. 12331018).

\bmhead{Data and Code Availability} The experimental data are available from references. The source code that supports the findings of this study is available from GitHub: \texttt{https://github.com/jinzhilei/DTP-Resistance}.

\begin{appendices}
	
\section{Mathematical formulations}

In this section, we provide a detailed mathematical derivation of the delay integro-differential equation (DIDE) model, which is derived from the stochastic process described in the main text. This formulation is based on the general framework for heteroeneous stem cell regeneration originally established by \citet{LeiJTB20framework}.

Classical cell cycle models, such as the G0 model \citep{Burns70onthe}, typically assume cell populations are homogeneous. However, single-cell RNA sequencing data have revealed significant heterogeneity in stem cells, characterized by continuous epigenetic states (e.g., transcriptomes, DNA methylation, and histone modifications). Crucially, these states are not perfectly inherited during cell division, leading to dynamic heterogeneity in the progeny. To capture this mathematically, we model the cell population density $Q(t, \vec{x})$ using a population balance equation framework. Unlike standard ODEs, this approach accounts for the continuous distribution of phenotypes. The core component is the inheritance probability kernel $p(\vec{x}, \vec{y})$, which defines the conditional probability that a mother cell with epigenetic state $\vec{y}$ produces a daughter cell with state $\vec{x}$. This formulation ensures that population evolution is driven by the interplay between state-dependent proliferation and stochastic inheritance noise.

As illustrated in Fig. \ref{Fig-Model}A, let $s(t, a, \vec{x})$ denote the density of proliferating cells at time $t$ with epigenetic state $\vec{x}$ and age $a$, and let $Q(t, \vec{x})$ represent the density of resting (G0) cells with state $\vec{x}$. Adopting the notation from the framework in \cite{Lei20evolutionary,LeiJTB20framework}, the system is governed by the following partial differential equations:
\begin{eqnarray}
	\label{sQ-stax}
	\frac{\partial s(t, a, \vec{x})}{\partial t} + \frac{\partial s(t, a, \vec{x})}{\partial a} & =& -\mu(\vec{x}) s(t, a, \vec{x}), \quad t>0, 0<a<\tau(\vec{x}),\\
	\label{sQ-Qtx}
	\frac{\partial Q(t, \vec{x})}{\partial t} & =& - \Big(\beta\big(Q(t), \vec{x})+\kappa(\vec{x}\big)\Big) Q(t, \vec{x}) \nonumber\\
	&&{}+ 2\int_{\Omega} s\big(t,\tau(\vec{y}), \vec{y}\big) p(\vec{x}, \vec{y}) d \vec{y},
\end{eqnarray}
with the boundary condition for cells entering the proliferative phase:
\begin{equation}
	\label{eq:bc}
	s(t,0, \vec{x}) = \beta\big(Q(t), \vec{x}\big)Q(t,\vec{x}).
\end{equation}
Here, the total number of cells in the resting phase at time $t$ is 
\begin{equation}
	Q(t)=\int_{\Omega} Q(t, \vec{x}) d \vec{x}.
\end{equation}
Note that the operator in Eq. \eqref{sQ-stax} represents the transport of clls through cell cycle age $a$. The total tumor cell population is given by
\begin{equation}
	N(t)  = Q(t) + \int_{\Omega}\int_0^{\tau(\vec{x})} s(t, a, \vec{x})\,d a\, d  \vec{x}. 
\end{equation}

Equation \eqref{sQ-stax}, coupled with the boundary condition \eqref{eq:bc}, can be solved using the interaction of characteristic lines (method of characteristics), yielding:
\begin{equation}
	\label{eq:sx}
	s\big(t,\tau(\vec{x}), \vec{x}\big)=\beta\big(Q_{\tau(\vec{x})}(t), \vec{x}\big)Q_{\tau(\vec{x})}(t, \vec{x})e^{-\mu(\vec{x})\tau(\vec{x})},
\end{equation}
where we define the time-delayed tomers $Q_{\tau(\vec{x})}(t) = Q\big(t - \tau(\vec{x})\big)$ and $Q_{\tau(\vec{x})}(t, \vec{x})=Q\big(t-\tau(\vec{x}), \vec{x}\big)$. Substituting \eqref{eq:sx} into \eqref{sQ-Qtx}, we eliminate the age variable $a$ and obtain the following closed delay integro-differential equation for $Q(t, \vec{x})$:
\begin{equation}
	\label{Qtx}
	\begin{aligned}
		\frac{\partial Q(t, \vec{x})}{\partial t} = &-Q(t, \vec{x})\Big(\beta\big(Q(t), \vec{x}\big) + \kappa(\vec{x})\Big) \\
		&\ {}+ 2\int_{\Omega} \beta\big(Q_{\tau(\vec{y})}(t), \vec{y}\big)Q_{\tau(\vec{y})}(t, \vec{y})e^{-\mu(\vec{y})\tau(\vec{y})}p(\vec{x}, \vec{y}) d \vec{y}.
	\end{aligned}
\end{equation}

Equation \eqref{Qtx} represents the deterministic continuum limit corresponding to the stochastic simulation framework used in this study. This type of equation, verified in \cite{LeiJTB20framework} to describe plastic regeneration, allows for the quantitative analysis of tumor relapse dynamics driven by continuous phenotype transitions \citep{Zhang:2021gd,Ma:2023aa,Wang:2025aa}.  The specific functional forms of the kinetic rates $\beta, \kappa, \mu$, the inheritance kernel $p(\vec{x}, \vec{y})$, and the drug-induced perturbations are detailed in Section \ref{sec:2.1} of the main text.

\section{Reduction to one-dimensional epigenetic state and validation of the agent-based model}

To establish the theoretical equivalence between the deterministic framework and the agent-based stochastic simulation, we first simplify the model to a single epigenetic state $x_1$. This reduction eliminates dimensionality challenges while retaining the core mechanisms of phenotypic plasticity---specifically, the dependence of proliferation ($\beta$) and apoptosis ($\mu_1$) rates on $x_1$, and the stochastic inheritance of state during division. The simplified model allows direct numerical integration for comparison with stochastic simulations.

State $x_1$ affects only the proliferation and apoptosis rates, so Eq. \eqref{Qtx} is rewritten as (we write $x_1$ as $x$ for simplicity)
\begin{equation}
\label{Qtx-1d}
 \frac{\partial Q(t,x)}{\partial t}=-Q(t,x)\big(\beta(Q,x)+\kappa\big)+2\int_{\Omega} \beta\big(Q(t-\tau),y\big)Q(t-\tau,y)e^{-\mu(y)\tau}p(x,y) dy,
\end{equation}
where $Q(t, x)$ denotes the population density of resting cells with epigenetic state $x$ at time $t$, $\tau = 15 \mathrm{h}$ is the constant cell cycle duration, and $\kappa_0$ is the state-independent terminal differentiation rate.

The corresponding kinetic rates are defined as
\begin{equation*}
\beta(Q, x)=\bar{\beta}\frac{\theta(x)}{\theta(x)+Q}, \ \ \kappa=\kappa_0,
\end{equation*}
where $\theta(x)$ modulates density dependence:
$$
\theta(x)=\bar{\theta}\left (1+a_5\frac{(a_4 x)^6}{1+(a_4 x)^6}\right),
$$ 
and apoptosis increases with $x$ via:
\begin{equation*}
\mu(x)=\mu_0\exp \big( \mu_1  (\gamma_1 - x) \big).
\end{equation*}

The inheritance probability kernel $p(x, y)$ follows a Gamma distributions:
\begin{equation*}
p(x,y) = \dfrac{b(y)^{-a}}{\Gamma(a)} x^{a-1} e^{-x/b(y)}, \mbox{ with } b(y)=\dfrac{\varphi(y)}{a},
\end{equation*}
where $\varphi(y)$ describes drug-dependent inheritance variance as:
\begin{equation}
\label{phiy1_drug}
\varphi(y)=(\varphi_{10} - D c_{10}) + (\varphi_{11} - D c_{11})\frac{y^{n_1}}{K_1^{n_1}+y^{n_1}}.
\end{equation}

To validate the agent-based model, we compared deterministic and stochastic outcomes for Eq. \eqref{Qtx-1d}. The deterministic solution was obtained by discretizing $\Omega = [0, 8]$ into 100 bins and solving via finite differences ($\Delta t = 0.1 h$). Agent-based stochastic simulations (Algorithm \ref{alg:ICB}) were repeated 3 times with $N(0) = 400$ cells, initialized with $x \sim \mathcal{U}(0, 8)$. Drug was administered from day $50$ to $70$ ($D = 1$ in Eq. \ref{phiy1_drug}). We set the parameter values as
\begin{equation}
\label{para-result0}
\begin{aligned}
&\bar{\beta}=0.10h^{-1},\ \bar{\theta}=1\times 10^3,\ \kappa_0=0.01h^{-1}, a_4 =  0.25,\ a_5=10, \vspace{1em}\\
&\mu_0=9.0\times 10^{-3}h^{-1},\ \mu_1=0.9,\ \tau =15h,\gamma_1= 4.5, a=50, \vspace{1em}\\
&\varphi_{10}=0.7,\ c_{10}=0.15,\ \varphi_{11}=4.35,\ c_{11}=1.5,\ K_1=2.2,\ n_1=2.0,
\end{aligned}
\end{equation}
in the simulation.
 
\begin{figure}[htbp]
	\centering
	\includegraphics[width=13cm]{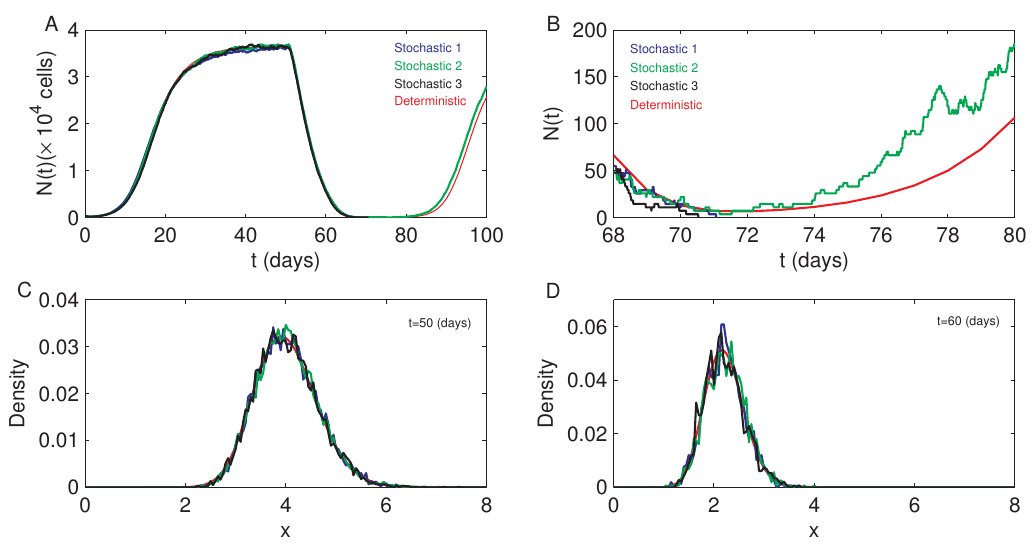}
	\caption{\textbf{Validation of agent-based simulation (blue, green, and black) against deterministic IDE solutions (red).}  \textbf{A} Evolution of the total cell number $N(t)$. \textbf{B} Zoom-in of $N(t)$ near extinction/relapse. Stochastic runs show divergent outcomes (extinction in 2/3 cases and relapse in 1/3 cases), while the deterministic model predicts recovery. \textbf{C-D} Epigenetic state distributions at day 50 (pre-drug) and 60 (on-drug), showing strong agreement between different runs. Parameters are given in \eqref{para-result0}; drug administered from day $50$ to $70$. }
	\label{Fig-result0}
\end{figure}

Simulation results showed excellent agreement in mean $N(t)$ (Fig. \ref{Fig-result0}A) and state distribution (Fig. \ref{Fig-result0}C-D), confirming the ABM accurately reproduces deterministic dynamics for large populations. However, stochasticity drives divergence when $N(t) < 100$ cells (Fig. \ref{Fig-result0}B): extinction occurs in 2/3 runs due to demographic noise, whereas the deterministic model (lacking such noise) predicts inevitable relapse. This highlights the ABM's unique value in capturing rare extinction events that are critical for predicting clinical outcomes.

\section{Virtual patients and PFS data generation}
\label{sec:supp-VP}

To capture inter-patient heterogeneity within the cohort, we define a ``virtual patient'' (VP) as a unique instance of the tumor model characterized by a specific parameter set. Based on the local sensitivity analysis presented in Section \ref{sec:3.4.3}, we identified four key parameters---$\bar{\beta}$, $\kappa_0$, $\alpha$, and $\mu_0$---that most significantly influence tumor relapse time. Consequently, a VP is represented by the vector $\vec{v} = (\bar{\beta}, \kappa_0, \alpha, \mu_0)$, while other parameters remain fixed at their baseline values (Table \ref{tab-modelpara}).

The generation of the virtual cohort involves a rejection sampling approach coupled with a Gaussian Mixture Model (GMM), summarized in the following steps:
\begin{enumerate}
\item \textbf{Baseline Estimation:} We first estimated the baseline parameter vector by fitting the model to the median Progression-Free Survival (PFS) observed in clinical data (EURTAC trial): 
$$\bar{\beta}_{\text{base}} = 0.011, \kappa_{0, \text{base}} = 0.004, \mu_{0, \text{base}} = 0.0013, \alpha_{\text{base}} = 0.95.$$
	\item \textbf{Parameter Space Construction:} To account for patient variability and biological trade-offs, we introduced two scalar tuning factors, $R_1$ and $R_2$, to perturb the baseline parameters. The candidate parameter space is defined as:
	\begin{equation}
	\label{eq:R1R2}
	\begin{aligned}
	\bar{\beta} & = \bar{\beta}_{\text{base}}\times R_1\\
	\kappa_0 & = \kappa_{0, \text{base}} \times R_1\\
	\alpha &= \alpha_{\text{base}} \times (0.9 + 0.2 R_2),\\
	\mu_0 &= \mu_{0, \text{base}} \times (1.4 - 0.8 R_2).
	\end{aligned}
	\end{equation}
Here, $R_1$ modulates the overall turnover rate (proliferation and differentiation), while $R_2$ introduces an inverse correlation between competition strength ($\alpha$) and intrinsic apoptosis ($\mu_0$), reflecting the cost-of-resistance trade-off.
	\item \textbf{Sampling and Filtering (Rejection Sampling):} We discretized the domains of $R_1$ and $R_2$ to generate $2,000$ candidate pairs. Each pair was mapped to a parameter vector $\vec{v}$ via Eq. \eqref{eq:R1R2}. We then performed stochastic simulations for each candidate vector. A subset of parameters was selected (accepted) only if their simulated tumor recurrence times fell within the confidence interval of the clinical observations.
		\item \textbf{Distribution Approximation:} The accepted parameter vectors served as training data to fit a multivariate Gaussian Mixture Model (GMM). This continuous probability distribution, denoted as $\mathcal{P}(R_1, R_2)$, characterizes the population-level heterogeneity of the specific cancer subtype.
		\item \textbf{Virtual Cohot Generation:} Finally, we sampled new cohorts of virtual patients from the learned distribution $\mathcal{P}(R_1, R_2)$. Stochastic simulations were performed on these cohorts under various adaptive therapy protocols to generate the corresponding PFS distributions and evaluate treatment efficacy.
\end{enumerate}




\end{appendices}


\bibliography{NSCLC_DTP}

@string{jtb = {J. Theor. Biol.}}

@article{Villiger:2024aa,
	address = {McGovern Institute for Brain Research, Massachusetts Institute of Technology Cambridge, Cambridge, MA, USA.; Whitehead Institute for Biomedical Research, Cambridge, MA, USA.; Department of Biology, Massachusetts Institute of Technology, Cambridge, MA, USA.; Whitehead Institute for Biomedical Research, Cambridge, MA, USA.; Department of Biology, Massachusetts Institute of Technology, Cambridge, MA, USA.; Whitehead Institute for Biomedical Research, Cambridge, MA, USA.; Department of Biology, Massachusetts Institute of Technology, Cambridge, MA, USA.; Howard Hughes Medical Institute, Massachusetts Institute of Technology, Cambridge, MA, USA.; David H. Koch Institute for Integrative Cancer Research, Massachusetts Institute of Technology, Cambridge, MA, USA.; McGovern Institute for Brain Research, Massachusetts Institute of Technology Cambridge, Cambridge, MA, USA. omar@abudayyeh.science.; McGovern Institute for Brain Research, Massachusetts Institute of Technology Cambridge, Cambridge, MA, USA. jgoot@mit.edu.},
	auid = {ORCID: 0000-0003-2445-670X; ORCID: 0000-0002-7979-3220; ORCID: 0000-0003-2757-2175},
	author = {Villiger, Lukas and Joung, Julia and Koblan, Luke and Weissman, Jonathan and Abudayyeh, Omar O and Gootenberg, Jonathan S},
	copyright = {{\copyright}2024. Springer Nature Limited.},
	crdt = {2024/02/02 23:44},
	date = {2024 Jun},
	dcom = {20240528},
	dep = {20240202},
	doi = {10.1038/s41580-023-00697-6},
	edat = {2024/02/03 00:42},
	ein = {Nat Rev Mol Cell Biol. 2024 Jun;25(6):510. doi: 10.1038/s41580-024-00745-9. PMID: 38740926},
	issn = {1471-0080 (Electronic); 1471-0072 (Linking)},
	jid = {100962782},
	journal = {Nat Rev Mol Cell Biol},
	jt = {Nature reviews. Molecular cell biology},
	language = {eng},
	lid = {10.1038/s41580-023-00697-6 {$[$}doi{$]$}},
	lr = {20240529},
	mh = {Humans; *Gene Editing/methods; *CRISPR-Cas Systems/genetics; *Epigenome/genetics; Animals; *Transcriptome/genetics; Clustered Regularly Interspaced Short Palindromic Repeats/genetics; Genome/genetics},
	mhda = {2024/05/29 00:49},
	month = {Jun},
	number = {6},
	own = {NLM},
	pages = {464--487},
	phst = {2023/12/18 00:00 {$[$}accepted{$]$}; 2024/05/29 00:49 {$[$}medline{$]$}; 2024/02/03 00:42 {$[$}pubmed{$]$}; 2024/02/02 23:44 {$[$}entrez{$]$}},
	pii = {10.1038/s41580-023-00697-6},
	pl = {England},
	pmid = {38308006},
	pst = {ppublish},
	pt = {Journal Article; Review},
	sb = {IM},
	status = {MEDLINE},
	title = {{CRISPR technologies for genome, epigenome and transcriptome editing.}},
	volume = {25},
	year = {2024}}

@article{Wei:2025aa,
	address = {School of Pharmacy, Nanjing University of Chinese Medicine, Nanjing City 210023, China.; National Resource Center for Mutant Mice, GemPharmatech Co., Ltd, Nanjing City 210023, China.; School of Pharmacy, Nanjing University of Chinese Medicine, Nanjing City 210023, China.; State Key Laboratory of Functions and Applications of Medicinal Plants, Guizhou Medical University, Guiyang City 550014, China.},
	author = {Wei, Yixiao and Sun, Jia and Zhu, Ruigong},
	cois = {The authors declare that they have no known competing financial interests or personal relationships that could have appeared to influence the work reported in this paper.},
	copyright = {{\copyright}2025 The Authors.},
	crdt = {2025/11/10 05:58},
	date = {2025},
	dcom = {20251112},
	dep = {20251019},
	doi = {10.1016/j.csbj.2025.10.031},
	edat = {2025/11/10 06:29},
	issn = {2001-0370 (Print); 2001-0370 (Electronic); 2001-0370 (Linking)},
	jid = {101585369},
	journal = {Comput Struct Biotechnol J},
	jt = {Computational and structural biotechnology journal},
	language = {eng},
	lid = {10.1016/j.csbj.2025.10.031 {$[$}doi{$]$}},
	lr = {20251112},
	mhda = {2025/11/10 06:30},
	oto = {NOTNLM},
	own = {NLM},
	pages = {4496--4504},
	phst = {2025/07/21 00:00 {$[$}received{$]$}; 2025/10/15 00:00 {$[$}revised{$]$}; 2025/10/16 00:00 {$[$}accepted{$]$}; 2025/11/10 06:30 {$[$}medline{$]$}; 2025/11/10 06:29 {$[$}pubmed{$]$}; 2025/11/10 05:58 {$[$}entrez{$]$}; 2025/10/19 00:00 {$[$}pmc-release{$]$}},
	pii = {S2001-0370(25)00432-5},
	pl = {Netherlands},
	pmc = {PMC12595283},
	pmcr = {2025/10/19},
	pmid = {41209346},
	pst = {epublish},
	pt = {Journal Article; Review},
	status = {PubMed-not-MEDLINE},
	title = {{CRISPR-epigenetic crosstalk: From bidirectional regulation to therapeutic potential.}},
	volume = {27},
	year = {2025}}

@article{Ramaiah:2021aa,
	address = {Laboratory of Functional genomics and Disease Biology, School of Chemical and Biotechnology, SASTRA Deemed University, Thanjavur 613401, Tamil Nadu, India. Electronic address: janakiramaiah@scbt.sastra.edu.; Department of Applied Biology, CSIR-Indian Institute of Chemical Technology (CSIR-IICT), Hyderabad 500 007, Telangana, India.; Department of Computer Science and Engineering, Koneru Lakshmaiah Education Foundation, Vaddeswaram, Andhra Pradesh, India.},
	author = {Ramaiah, M Janaki and Tangutur, Anjana Devi and Manyam, Rajasekhar Reddy},
	copyright = {Copyright {\copyright}2021 Elsevier Inc. All rights reserved.},
	crdt = {2021/04/19 20:12},
	date = {2021 Jul 15},
	dcom = {20210622},
	dep = {20210416},
	doi = {10.1016/j.lfs.2021.119504},
	edat = {2021/04/20 06:00},
	issn = {1879-0631 (Electronic); 0024-3205 (Linking)},
	jid = {0375521},
	journal = {Life Sci},
	jt = {Life sciences},
	language = {eng},
	lid = {S0024-3205(21)00489-6 {$[$}pii{$]$}; 10.1016/j.lfs.2021.119504 {$[$}doi{$]$}},
	lr = {20210622},
	mh = {Acetylation; Apoptosis/drug effects; Cell Cycle/drug effects; Chromatin/metabolism; Epigenesis, Genetic/drug effects/genetics; Epigenomics/methods; Gene Expression/drug effects; Histone Acetyltransferases/metabolism; Histone Deacetylase Inhibitors/metabolism/*therapeutic use; Histone Deacetylases/metabolism; Histones/metabolism; Humans; Neoplasms/*drug therapy/*genetics/metabolism; Protein Processing, Post-Translational/drug effects},
	mhda = {2021/06/23 06:00},
	month = {Jul},
	oto = {NOTNLM},
	own = {NLM},
	pages = {119504},
	phst = {2021/01/03 00:00 {$[$}received{$]$}; 2021/03/20 00:00 {$[$}revised{$]$}; 2021/04/09 00:00 {$[$}accepted{$]$}; 2021/04/20 06:00 {$[$}pubmed{$]$}; 2021/06/23 06:00 {$[$}medline{$]$}; 2021/04/19 20:12 {$[$}entrez{$]$}},
	pii = {S0024-3205(21)00489-6},
	pl = {Netherlands},
	pmid = {33872660},
	pst = {ppublish},
	pt = {Journal Article; Review},
	rn = {0 (Chromatin); 0 (Histone Deacetylase Inhibitors); 0 (Histones); EC 2.3.1.48 (Histone Acetyltransferases); EC 3.5.1.98 (Histone Deacetylases)},
	sb = {IM},
	status = {MEDLINE},
	title = {{Epigenetic modulation and understanding of HDAC inhibitors in cancer therapy.}},
	volume = {277},
	year = {2021}}

@article{Guler:2017aa,
	address = {Molecular Oncology, Genentech Inc., 1 DNA Way, South San Francisco, CA 94080, USA.; Molecular Oncology, Genentech Inc., 1 DNA Way, South San Francisco, CA 94080, USA.; Molecular Oncology, Genentech Inc., 1 DNA Way, South San Francisco, CA 94080, USA.; Molecular Oncology, Genentech Inc., 1 DNA Way, South San Francisco, CA 94080, USA.; Protein Chemistry, Genentech Inc., South San Francisco, CA, USA.; Protein Chemistry, Genentech Inc., South San Francisco, CA, USA.; Molecular Oncology, Genentech Inc., 1 DNA Way, South San Francisco, CA 94080, USA.; LS Biomarker Development, Genentech Inc., South San Francisco, CA, USA.; LS Biomarker Development, Genentech Inc., South San Francisco, CA, USA.; Molecular Biology, Genentech Inc., South San Francisco, CA, USA.; Molecular Biology, Genentech Inc., South San Francisco, CA, USA.; Pathology, Genentech Inc., South San Francisco, CA, USA.; Molecular Oncology, Genentech Inc., 1 DNA Way, South San Francisco, CA 94080, USA.; Molecular Oncology, Genentech Inc., 1 DNA Way, South San Francisco, CA 94080, USA.; Molecular Oncology, Genentech Inc., 1 DNA Way, South San Francisco, CA 94080, USA.; Epinomics, Menlo Park, CA, USA.; Epinomics, Menlo Park, CA, USA.; Active Motif, Carlsbad, CA, USA.; Bioinformatics, Genentech Inc., South San Francisco, CA, USA.; Bioinformatics, Genentech Inc., South San Francisco, CA, USA.; Molecular Oncology, Genentech Inc., 1 DNA Way, South San Francisco, CA 94080, USA.; Protein Chemistry, Genentech Inc., South San Francisco, CA, USA.; Protein Chemistry, Genentech Inc., South San Francisco, CA, USA.; Molecular Oncology, Genentech Inc., 1 DNA Way, South San Francisco, CA 94080, USA. Electronic address: classon.marie@gmail.com.},
	author = {Guler, Gulfem Dilek and Tindell, Charles Albert and Pitti, Robert and Wilson, Catherine and Nichols, Katrina and KaiWai Cheung, Tommy and Kim, Hyo-Jin and Wongchenko, Matthew and Yan, Yibing and Haley, Benjamin and Cuellar, Trinna and Webster, Joshua and Alag, Navneet and Hegde, Ganapati and Jackson, Erica and Nance, Tracy Leah and Giresi, Paul Garrett and Chen, Kuan-Bei and Liu, Jinfeng and Jhunjhunwala, Suchit and Settleman, Jeff and Stephan, Jean-Philippe and Arnott, David and Classon, Marie},
	copyright = {Copyright {\copyright}2017 Elsevier Inc. All rights reserved.},
	crdt = {2017/08/08 06:00},
	date = {2017 Aug 14},
	dcom = {20170911},
	dep = {20170803},
	doi = {10.1016/j.ccell.2017.07.002},
	edat = {2017/08/07 06:00},
	issn = {1878-3686 (Electronic); 1535-6108 (Linking)},
	jid = {101130617},
	journal = {Cancer Cell},
	jt = {Cancer cell},
	language = {eng},
	lid = {S1535-6108(17)30294-5 {$[$}pii{$]$}; 10.1016/j.ccell.2017.07.002 {$[$}doi{$]$}},
	lr = {20180723},
	mh = {Animals; Antineoplastic Agents/pharmacology; Chromatin/*genetics; Drug Resistance, Neoplasm/*genetics; Epigenetic Repression/*drug effects; Gene Expression Regulation, Neoplastic; Genomic Instability/drug effects; Histone-Lysine N-Methyltransferase/metabolism; Histones/metabolism; Humans; Long Interspersed Nucleotide Elements/*genetics; Methylation; Mice; Mice, Nude; Mice, SCID; Neoplasms/drug therapy/genetics/*pathology; Stress, Physiological; Tumor Cells, Cultured; Xenograft Model Antitumor Assays},
	mhda = {2017/09/12 06:00},
	month = {Aug},
	number = {2},
	oto = {NOTNLM},
	own = {NLM},
	pages = {221--237},
	phst = {2016/08/24 00:00 {$[$}received{$]$}; 2017/05/02 00:00 {$[$}revised{$]$}; 2017/07/05 00:00 {$[$}accepted{$]$}; 2017/08/07 06:00 {$[$}pubmed{$]$}; 2017/09/12 06:00 {$[$}medline{$]$}; 2017/08/08 06:00 {$[$}entrez{$]$}},
	pii = {S1535-6108(17)30294-5},
	pl = {United States},
	pmid = {28781121},
	pst = {ppublish},
	pt = {Journal Article},
	rn = {0 (Antineoplastic Agents); 0 (Chromatin); 0 (Histones); EC 2.1.1.43 (Histone-Lysine N-Methyltransferase)},
	sb = {IM},
	status = {MEDLINE},
	title = {{Repression of stress-induced LINE-1 expression protects cancer cell subpopulations from lethal drug exposure.}},
	volume = {32},
	year = {2017}}

@article{Kim:2015aa,
	address = {Department of Molecular and Cellular Biology, Baylor College of Medicine, Houston, Texas.; Cytometry and Cell Sorting Core, Baylor College of Medicine, Houston, Texas.},
	author = {Kim, Kang Ho and Sederstrom, Joel M},
	copyright = {Copyright {\copyright}2015 John Wiley \& Sons, Inc.},
	crdt = {2015/07/02 06:00},
	date = {2015 Jul 1},
	dcom = {20160907},
	dep = {20150701},
	doi = {10.1002/0471142727.mb2806s111},
	edat = {2015/07/02 06:00},
	gr = {R01 DK046546/DK/NIDDK NIH HHS/United States; P30AI036211/AI/NIAID NIH HHS/United States; S10RR024574/RR/NCRR NIH HHS/United States; P30CA125123/CA/NCI NIH HHS/United States; P30 CA125123/CA/NCI NIH HHS/United States; P30 AI036211/AI/NIAID NIH HHS/United States; S10 RR024574/RR/NCRR NIH HHS/United States; R01DK46546/DK/NIDDK NIH HHS/United States},
	issn = {1934-3647 (Electronic); 1934-3639 (Print); 1934-3647 (Linking)},
	jid = {8908160},
	journal = {Curr Protoc Mol Biol},
	jt = {Current protocols in molecular biology},
	language = {eng},
	lid = {10.1002/0471142727.mb2806s111 {$[$}doi{$]$}},
	lr = {20191127},
	mh = {Animals; *Cell Cycle; Cytological Techniques/*methods; DNA/analysis; Flow Cytometry/*methods; Humans; Ki-67 Antigen/analysis; RNA/analysis; Staining and Labeling/*methods},
	mhda = {2016/09/08 06:00},
	mid = {NIHMS706484},
	month = {Jul},
	oto = {NOTNLM},
	own = {NLM},
	pages = {28.6.1-28.6.11},
	phst = {2015/07/02 06:00 {$[$}entrez{$]$}; 2015/07/02 06:00 {$[$}pubmed{$]$}; 2016/09/08 06:00 {$[$}medline{$]$}; 2016/07/01 00:00 {$[$}pmc-release{$]$}},
	pl = {United States},
	pmc = {PMC4516267},
	pmcr = {2016/07/01},
	pmid = {26131851},
	pst = {epublish},
	pt = {Journal Article; Research Support, N.I.H., Extramural},
	rn = {0 (Ki-67 Antigen); 63231-63-0 (RNA); 9007-49-2 (DNA)},
	sb = {IM},
	status = {MEDLINE},
	title = {{Assaying cell cycle status using flow cytometry.}},
	volume = {111},
	year = {2015}}

@article{Darzynkiewicz:1994aa,
	address = {Cancer Research Institute, New York Medical College, Valhalla 19595.},
	author = {Darzynkiewicz, Z and Gong, J and Traganos, F},
	crdt = {1994/01/01 00:00},
	date = {1994},
	dcom = {19950322},
	doi = {10.1016/s0091-679x(08)61732-x},
	edat = {1994/01/01 00:00},
	gr = {CA 28704/CA/NCI NIH HHS/United States},
	issn = {0091-679X (Print); 0091-679X (Linking)},
	jid = {0373334},
	journal = {Methods Cell Biol},
	jt = {Methods in cell biology},
	language = {eng},
	lr = {20190913},
	mh = {Animals; *Cell Cycle; Cell Division; Cyclins/*analysis; DNA/*analysis; Flow Cytometry; Humans; *Lymphocyte Activation; Neoplasms/pathology; *Ploidies},
	mhda = {1994/01/01 00:01},
	own = {NLM},
	pages = {421--435},
	phst = {1994/01/01 00:00 {$[$}pubmed{$]$}; 1994/01/01 00:01 {$[$}medline{$]$}; 1994/01/01 00:00 {$[$}entrez{$]$}},
	pl = {United States},
	pmid = {7861973},
	pst = {ppublish},
	pt = {Journal Article; Research Support, Non-U.S. Gov't; Research Support, U.S. Gov't, P.H.S.; Review},
	rf = {24},
	rn = {0 (Cyclins); 9007-49-2 (DNA)},
	sb = {IM},
	status = {MEDLINE},
	title = {{Analysis of DNA content and cyclin protein expression in studies of DNA ploidy, growth fraction, lymphocyte stimulation, and the cell cycle.}},
	volume = {41},
	year = {1994}}

@article{MacKey:2001aa,
	address = {Departments of Physiology, Physics, \& Mathematics, Centre for Nonlinear Dynamics in Physiology & Medicine, McGill University, Canada. mackey@cnd.mcgill.ca},
	author = {Mackey, M C},
	crdt = {2001/05/12 10:00},
	date = {2001 Apr},
	dcom = {20010524},
	doi = {10.1046/j.1365-2184.2001.00195.x},
	edat = {2001/05/12 10:00},
	issn = {0960-7722 (Print); 1365-2184 (Electronic); 0960-7722 (Linking)},
	jid = {9105195},
	journal = {Cell Prolif},
	jt = {Cell proliferation},
	language = {eng},
	lr = {20220309},
	mh = {Animals; Cell Division; Hematopoietic Stem Cells/*cytology; Humans; Kinetics; Mice; *Models, Biological},
	mhda = {2001/05/26 10:01},
	month = {Apr},
	number = {2},
	own = {NLM},
	pages = {71--83},
	phst = {2001/05/12 10:00 {$[$}pubmed{$]$}; 2001/05/26 10:01 {$[$}medline{$]$}; 2001/05/12 10:00 {$[$}entrez{$]$}; 2008/07/07 00:00 {$[$}pmc-release{$]$}},
	pii = {195; CPR195},
	pl = {England},
	pmc = {PMC6495991},
	pmcr = {2008/07/07},
	pmid = {11348423},
	pst = {ppublish},
	pt = {Journal Article; Research Support, Non-U.S. Gov't},
	sb = {IM},
	status = {MEDLINE},
	title = {{Cell kinetic status of haematopoietic stem cells.}},
	volume = {34},
	year = {2001}}

@article{Gupta:2025aa,
	address = {Institute of Pharmaceutical Research, GLA University, Mathura, UP, 281406, India.; Medical Laboratory Techniques Department, College of Health and Medical Technology, University of Al-Maarif, Anbar, Iraq. bahaaibrahimsaeed@gmail.com.; Department of Microbiology, Faculty of Science, Marwadi University Research Center, Marwadi University, Rajkot, Gujarat, 360003, India.; Department of Clinical Laboratory Sciences, College of Applied Medical Sciences, King Khalid University, Abha, Saudi Arabia.; Department of Agronomy, Navoi State University of Mining and Technologies, Navoiy, Uzbekistan.; Department of Pharmacy Practice, NIMS Institute of Pharmacy, NIMS University Rajasthan, Jaipur, India.; Department of Basic Science and Humanities, Raghu Engineering College, Visakhapatnam, India.; Centre for Research Impact and Outcome, Chitkara University Institute of Engineering and Technology, Chitkara University, Rajpura, Punjab, 140401, India.; Nursing Department, College of Applied Medical Sciences, Jouf University, Al-Jouf, Saudi Arabia.; Department of Pharmaceutical Chemistry, College of Pharmacy, University of Mosul, Mosul, 41001, Iraq.},
	auid = {ORCID: 0000-0002-8453-0142; ORCID: 0009-0006-7776-1950; ORCID: 0009-0005-7895-8133; ORCID: 0000-0003-0603-6947},
	author = {Gupta, Jitendra and Saeed, Bahaa Ibrahim and Bishoyi, Ashok Kumar and Alkhathami, Ali G and Asliddin, Shodiyev and Nathiya, Deepak and Ravi Kumar, M and Bhanot, Deepak and Rashed, Amera Bekhatroh and Mustafa, Yasser Fakri},
	cois = {Declarations. Conflict of interest: There is no conflict of interest.},
	copyright = {{\copyright}2025. The Author(s), under exclusive licence to Springer Science+Business Media, LLC, part of Springer Nature.},
	crdt = {2025/08/09 11:15},
	date = {2025 Aug 9},
	dcom = {20250809},
	dep = {20250809},
	doi = {10.1007/s12032-025-02973-1},
	edat = {2025/08/09 13:48},
	gr = {RGP.02/534/44/King Khalid University/},
	issn = {1559-131X (Electronic); 1357-0560 (Linking)},
	jid = {9435512},
	journal = {Med Oncol},
	jt = {Medical oncology (Northwood, London, England)},
	language = {eng},
	lid = {10.1007/s12032-025-02973-1 {$[$}doi{$]$}},
	lr = {20250829},
	mh = {Humans; *CDC2 Protein Kinase/metabolism; *Neoplasms/pathology/drug therapy/metabolism/therapy; *Cyclin-Dependent Kinase 2/metabolism; *Carcinogenesis; Animals; *Cell Cycle Checkpoints/physiology; Cell Cycle},
	mhda = {2025/08/09 13:49},
	month = {Aug},
	number = {9},
	oto = {NOTNLM},
	own = {NLM},
	pages = {422},
	phst = {2025/06/04 00:00 {$[$}received{$]$}; 2025/07/28 00:00 {$[$}accepted{$]$}; 2025/08/09 13:49 {$[$}medline{$]$}; 2025/08/09 13:48 {$[$}pubmed{$]$}; 2025/08/09 11:15 {$[$}entrez{$]$}},
	pii = {10.1007/s12032-025-02973-1},
	pl = {United States},
	pmid = {40782258},
	pst = {epublish},
	pt = {Journal Article; Review},
	rn = {EC 2.7.11.22 (CDC2 Protein Kinase); EC 2.7.11.22 (Cyclin-Dependent Kinase 2); EC 2.7.11.22 (CDK1 protein, human); EC 2.7.11.22 (CDK2 protein, human)},
	sb = {IM},
	status = {MEDLINE},
	title = {{From cell cycle control to cancer therapy: exploring the role of CDK1 and CDK2 in tumorigenesis.}},
	volume = {42},
	year = {2025}}

@article{Drew:2025aa,
	address = {BC Cancer Vancouver Centre and Faculty of Medicine, University of British Columbia, Vancouver, British Columbia, Canada.; Research Unit Oncology, EMD Serono, Billerica, MA, USA.; Faculty of Medical Sciences, Newcastle University, Newcastle upon Tyne, UK. nicola.curtin@ncl.ac.uk.},
	auid = {ORCID: 0000-0002-2226-3755; ORCID: 0000-0003-1369-1843},
	author = {Drew, Yvette and Zenke, Frank T and Curtin, Nicola J},
	cois = {Competing interests: Y.D. has participated in advisory boards for Clovis Oncology, AstraZeneca, Merck, Tesaro Inc (now GlaxoSmithKline), GlaxoSmithKline and Genmab. Y.D. has received research grant funding from Clovis Oncology and was involved in the preclinical and clinical development of rucaparib. Y.D. has received royalties for her involvement in the development of rucaparib. Y.D. has received research grant funding from AstraZeneca, Tesaro Inc and Roche. F.T.Z. was an employee at EMD Serono, Billerica, MA an affiliate of Merck KGaA, Darmstadt, Germany. Merck KGaA, Darmstadt, Germany, and/or its affiliates have certain rights in patents, patent applications pertaining to DNA Damage Response inhibitors, and F. Z. is named inventor on several patents. N.J.C. was involved in the development of rucaparib, receiving grant funding from Agouron Pharmaceuticals and Pfizer Oncology. She has received grant funding from BioMarin for translational studies with talazoparib. She was involved in the development of DNA-PK inhibitors and ATM inhibitors, receiving grant funding from Kudos (subsequently acquired by AstraZeneca). She was involved in the preclinical evaluation of ATR inhibitors, receiving grant funding from Vertex and Merck. She is inventor on several patents concerning PARP, DNA-PK, ATM and ATR inhibitors. She has received grant funding from Breakpoint Pharmaceuticals and has consulted for Abbvie and Tesaro on PARP inhibitors and Sierra Oncology on CHK1 inhibitors, she currently consults for Duke Street Bio. She diverts her royalty payments to charity.},
	copyright = {{\copyright}2024. Springer Nature Limited.},
	crdt = {2024/11/12 23:54},
	date = {2025 Jan},
	dcom = {20250106},
	dep = {20241112},
	doi = {10.1038/s41573-024-01060-w},
	edat = {2024/11/13 13:58},
	issn = {1474-1784 (Electronic); 1474-1776 (Linking)},
	jid = {101124171},
	journal = {Nat Rev Drug Discov},
	jt = {Nature reviews. Drug discovery},
	language = {eng},
	lid = {10.1038/s41573-024-01060-w {$[$}doi{$]$}},
	lr = {20250110},
	mh = {Animals; Humans; *Antineoplastic Agents/therapeutic use/pharmacology; DNA Damage; *DNA Repair/drug effects; Drug Development/methods/trends; *Neoplasms/drug therapy/genetics; *Poly(ADP-ribose) Polymerase Inhibitors/therapeutic use/pharmacology},
	mhda = {2025/01/07 00:21},
	month = {Jan},
	number = {1},
	own = {NLM},
	pages = {19--39},
	phst = {2024/09/17 00:00 {$[$}accepted{$]$}; 2025/01/07 00:21 {$[$}medline{$]$}; 2024/11/13 13:58 {$[$}pubmed{$]$}; 2024/11/12 23:54 {$[$}entrez{$]$}},
	pii = {10.1038/s41573-024-01060-w},
	pl = {England},
	pmid = {39533099},
	pst = {ppublish},
	pt = {Journal Article; Review},
	rn = {0 (Antineoplastic Agents); 0 (Poly(ADP-ribose) Polymerase Inhibitors)},
	sb = {IM},
	status = {MEDLINE},
	title = {{DNA damage response inhibitors in cancer therapy: lessons from the past, current status and future implications.}},
	volume = {24},
	year = {2025}}

@article{Shirmanova:2017aa,
	address = {Institute of Biomedical Technologies, Nizhny Novgorod State Medical Academy, 10/1 Minin and Pozharsky Sq., 603005, Nizhny Novgorod, Russia. shirmanovam@mail.ru.; Institute of Biomedical Technologies, Nizhny Novgorod State Medical Academy, 10/1 Minin and Pozharsky Sq., 603005, Nizhny Novgorod, Russia.; Institute of Biomedical Technologies, Nizhny Novgorod State Medical Academy, 10/1 Minin and Pozharsky Sq., 603005, Nizhny Novgorod, Russia.; Institute of Biology and Biomedicine, Lobachevsky State University of Nizhny Novgorod, 23 Gagarin Ave., 603950, Nizhny Novgorod, Russia.; Institute of Biomedical Technologies, Nizhny Novgorod State Medical Academy, 10/1 Minin and Pozharsky Sq., 603005, Nizhny Novgorod, Russia.; Institute of Biology and Biomedicine, Lobachevsky State University of Nizhny Novgorod, 23 Gagarin Ave., 603950, Nizhny Novgorod, Russia.; Institute of Biomedical Technologies, Nizhny Novgorod State Medical Academy, 10/1 Minin and Pozharsky Sq., 603005, Nizhny Novgorod, Russia.; Institute of Biomedical Technologies, Nizhny Novgorod State Medical Academy, 10/1 Minin and Pozharsky Sq., 603005, Nizhny Novgorod, Russia.; Becker \& Hickl GmbH, Nahmitzer Damm 30, 12277, Berlin, Germany.; Molecular technologies laboratory, Shemyakin-Ovchinnikov Institute of Bioorganic Chemistry RAS, 16/10 Miklukho-Maklaya St., 117997, Moscow, Russia.; Institute of Biomedical Technologies, Nizhny Novgorod State Medical Academy, 10/1 Minin and Pozharsky Sq., 603005, Nizhny Novgorod, Russia.},
	author = {Shirmanova, Marina V and Druzhkova, Irina N and Lukina, Maria M and Dudenkova, Varvara V and Ignatova, Nadezhda I and Snopova, Ludmila B and Shcheslavskiy, Vladislav I and Belousov, Vsevolod V and Zagaynova, Elena V},
	cois = {The authors declare that they have no competing interests.},
	crdt = {2017/08/23 06:00},
	date = {2017 Aug 21},
	dcom = {20190315},
	dep = {20170821},
	doi = {10.1038/s41598-017-09426-4},
	edat = {2017/08/23 06:00},
	issn = {2045-2322 (Electronic); 2045-2322 (Linking)},
	jid = {101563288},
	journal = {Sci Rep},
	jt = {Scientific reports},
	language = {eng},
	lid = {10.1038/s41598-017-09426-4 {$[$}doi{$]$}; 8911},
	lr = {20190315},
	mh = {Antineoplastic Agents/*pharmacology; Biomarkers; Cell Line, Tumor; Cell Proliferation/drug effects; Cell Survival/drug effects; Cisplatin/*pharmacology; Energy Metabolism/*drug effects; HeLa Cells; Humans; *Hydrogen-Ion Concentration; Immunohistochemistry; Intracellular Space/drug effects/metabolism; Molecular Imaging},
	mhda = {2019/03/16 06:00},
	month = {Aug},
	number = {1},
	own = {NLM},
	pages = {8911},
	phst = {2016/09/19 00:00 {$[$}received{$]$}; 2017/07/25 00:00 {$[$}accepted{$]$}; 2017/08/23 06:00 {$[$}entrez{$]$}; 2017/08/23 06:00 {$[$}pubmed{$]$}; 2019/03/16 06:00 {$[$}medline{$]$}; 2017/08/21 00:00 {$[$}pmc-release{$]$}},
	pii = {10.1038/s41598-017-09426-4; 9426},
	pl = {England},
	pmc = {PMC5566551},
	pmcr = {2017/08/21},
	pmid = {28827680},
	pst = {epublish},
	pt = {Journal Article; Research Support, Non-U.S. Gov't},
	rn = {0 (Antineoplastic Agents); 0 (Biomarkers); Q20Q21Q62J (Cisplatin)},
	sb = {IM},
	status = {MEDLINE},
	title = {{Chemotherapy with cisplatin: insights into intracellular pH and metabolic landscape of cancer cells in vitro and in vivo.}},
	volume = {7},
	year = {2017}}

@article{Balaban:2019aa,
	address = {Racah Institute of Physics, The Hebrew University, Jerusalem, Israel. nathalie.balaban@mail.huji.ac.il.; MRC Centre for Molecular Bacteriology and Infection, Imperial College London, London, UK.; Department of Biology, Northeastern University, Boston, MA, USA.; Institute of Biogeochemistry and Pollutant Dynamics, ETH Zurich, Zurich, Switzerland.; Department of Environmental Microbiology, Eawag, Dubendorf, Switzerland.; Department of Molecular Biology and Microbiology, Tufts University School of Medicine, Boston, MA, USA.; Department of Medical Biochemistry and Microbiology, Uppsala University, Uppsala, Sweden.; Department of Chemical and Biological Engineering, Princeton University, Princeton, NJ, USA.; Focal Area Infection Biology, Biozentrum of the University of Basel, Basel, Switzerland.; Department of Molecular Biology and Microbiology, Tufts University School of Medicine, Boston, MA, USA.; Institute for Medical Engineering \& Science, Department of Biological Engineering, and Synthetic Biology Center, Massachusetts Institute of Technology, Cambridge, MA, USA.; Wyss Institute for Biologically Inspired Engineering, Harvard University, Boston, MA, USA.; Broad Institute of MIT and Harvard, Cambridge, MA, USA.; Focal Area Infection Biology, Biozentrum of the University of Basel, Basel, Switzerland.; Department of Immunology and Infectious Diseases, Harvard T. H. Chan School of Public Health, Boston, MA, USA.; Institut Pasteur, Genetics of Biofilms Laboratory, Paris, France.; Institute of Microbiology, ETH Zurich, Zurich, Switzerland.; Focal Area Infection Biology, Biozentrum of the University of Basel, Basel, Switzerland.; Molecular Systems Biology, Groningen Biomolecular Sciences and Biotechnology Institute, University of Groningen, Groningen, Netherlands.; Broad Institute of MIT and Harvard, Cambridge, MA, USA.; Focal Area Infection Biology, Biozentrum of the University of Basel, Basel, Switzerland.; Department of Biology, Emory University, Atlanta, GA, USA.; Center for Microbiology, KU Leuven-University of Leuven, Leuven, Belgium.; Division of Molecular and Cellular Biology, Eunice Kennedy Shriver National Institute of Child Health and Human Development, Bethesda, MD, USA.; Infectious Diseases Department, Genentech, South San Francisco, CA, USA.; Institute of Technology, University of Tartu, Tartu, Estonia.; Facult{\'e}des Sciences, Universit{\'e}Libre de Bruxelles, Bruxelles, Belgium.; Division of Infectious Diseases, University Hospital Zurich, University of Zurich, Zurich, Switzerland.},
	auid = {ORCID: 0000-0001-8018-0766; ORCID: 0000-0002-9877-4180; ORCID: 0000-0001-6640-2174; ORCID: 0000-0002-5560-8246; ORCID: 0000-0001-7565-9975; ORCID: 0000-0002-5512-9077},
	author = {Balaban, Nathalie Q and Helaine, Sophie and Lewis, Kim and Ackermann, Martin and Aldridge, Bree and Andersson, Dan I and Brynildsen, Mark P and Bumann, Dirk and Camilli, Andrew and Collins, James J and Dehio, Christoph and Fortune, Sarah and Ghigo, Jean-Marc and Hardt, Wolf-Dietrich and Harms, Alexander and Heinemann, Matthias and Hung, Deborah T and Jenal, Urs and Levin, Bruce R and Michiels, Jan and Storz, Gisela and Tan, Man-Wah and Tenson, Tanel and Van Melderen, Laurence and Zinkernagel, Annelies},
	cois = {The authors declare no competing interests.},
	crdt = {2019/04/14 06:00},
	date = {2019 Jul},
	dcom = {20200303},
	doi = {10.1038/s41579-019-0196-3},
	edat = {2019/04/14 06:00},
	ein = {Nat Rev Microbiol. 2019 Jul;17(7):460. doi: 10.1038/s41579-019-0207-4. PMID: 31036919},
	gr = {MR/P028225/1/MRC{\_}/Medical Research Council/United Kingdom; R01 AI055058/AI/NIAID NIH HHS/United States; R01 GM091875/GM/NIGMS NIH HHS/United States; R37 AI055058/AI/NIAID NIH HHS/United States},
	issn = {1740-1534 (Electronic); 1740-1526 (Print); 1740-1526 (Linking)},
	jid = {101190261},
	journal = {Nat Rev Microbiol},
	jt = {Nature reviews. Microbiology},
	language = {eng},
	lid = {10.1038/s41579-019-0196-3 {$[$}doi{$]$}},
	lr = {20250530},
	mh = {Anti-Bacterial Agents/*pharmacology; Bacteria/*drug effects; Biomedical Research/*methods/*standards; *Drug Tolerance; Guidelines as Topic; Terminology as Topic},
	mhda = {2020/03/04 06:00},
	mid = {NIHMS1576626},
	month = {Jul},
	number = {7},
	own = {NLM},
	pages = {441--448},
	phst = {2019/04/14 06:00 {$[$}pubmed{$]$}; 2020/03/04 06:00 {$[$}medline{$]$}; 2019/04/14 06:00 {$[$}entrez{$]$}; 2019/04/12 00:00 {$[$}pmc-release{$]$}},
	pii = {10.1038/s41579-019-0196-3; 196},
	pl = {England},
	pmc = {PMC7136161},
	pmcr = {2019/04/12},
	pmid = {30980069},
	pst = {ppublish},
	pt = {Journal Article; Research Support, N.I.H., Extramural; Research Support, Non-U.S. Gov't; Review},
	rn = {0 (Anti-Bacterial Agents)},
	sb = {IM},
	status = {MEDLINE},
	title = {{Definitions and guidelines for research on antibiotic persistence.}},
	volume = {17},
	year = {2019}}

@article{KuosmanenPlosCB2025,
	author = {Kuosmanen, Teemu AND Cairns, Johannes AND Noble, Robert AND Beerenwinkel, Niko AND Mononen, Tommi AND Mustonen, Ville},
	journal = {PLoS Comp Biol},
	month = {09},
	number = {9},
	pages = {1-22},
	publisher = {Public Library of Science},
	title = {{Drug-induced resistance evolution necessitates less aggressive treatment}},
	volume = {17},
	year = {2021}}

@article{FischerJTB2024,
	author = {Matthias M. Fischer and Nils Bl{\"u}thgen},
	issn = {0022-5193},
	journal = JTB,
	pages = {111716},
	title = {On minimising tumoural growth under treatment resistance},
	volume = {579},
	year = {2024}}

@article{IyerPlosCB2025,
	author = {Iyer, Anton AND Alva, Adrian AND Granada, Adri{\'a}n E. AND Chakrabarti, Shaon},
	journal = {PLoS Comp Biol},
	month = {09},
	number = {9},
	pages = {1-26},
	publisher = {Public Library of Science},
	title = {{Inheritable cell-states shape drug-persister correlations and population dynamics in cancer cells}},
	volume = {21},
	year = {2025}}

@article{Gevertz2025npjSBA,
	author = {Gevertz, J.L. and Greene, J.M. and Prosperi, S. and et al.},
	journal = {npj Syst Biol Appl},
	number = {30},
	pages = {1-15},
	title = {{Understanding therapeutic tolerance through a mathematical model of drug-induced resistance}},
	volume = {11},
	year = {2025}}

@article{Whiting2025Natcom,
	author = {Whiting, Frederick J. H. and Mossner, Maximilian and Gabbutt, Calum and Kimberley, Christopher and Barnes, Chris P. and Baker, Ann-Marie and Yara-Romero, Erika and Sottoriva, Andrea and Nichols, Richard A. and Graham, Trevor A.},
	journal = {Nat Commun},
	pages = {5282},
	title = {{Quantitative measurement of phenotype dynamics during cancer drug resistance evolution using genetic barcoding}},
	volume = {16},
	year = {2025}}

@article{Gunnarsson2025npjSBA,
	author = {Gunnarsson, E.B. and Magn\'{u}sson B.V. and Foo, J.},
	journal = {npj Syst Biol Appl},
	number = {98},
	pages = {1-15},
	title = {{Optimal dosing of anti-cancer treatment under drug-induced plasticity}},
	volume = {11},
	year = {2025}}

@article{France2024Nature,
	author = {Fran\c{c}a, G.S. and Baron, M. and King, B.R. and et al.},
	journal = {Nature},
	pages = {876-883},
	title = {Cellular adaptation to cancer therapy along a resistance continuum},
	volume = {631},
	year = {2024}}

@article{Gatenby09Adaptive,
	address = {Department of Integrative Mathematical Oncology, Moffitt Cancer Center, Tampa, Florida 33612, USA. Robert.Gatenby@moffitt.org},
	author = {Gatenby, Robert A and Silva, Ariosto S and Gillies, Robert J and Frieden, B Roy},
	cois = {Disclosure of Potential Conflicts of Interest No potential conflicts of interest were disclosed.},
	crdt = {2009/06/03 09:00},
	date = {2009 Jun 1},
	dcom = {20090715},
	edat = {2009/06/03 09:00},
	gr = {R01 CA077575/CA/NCI NIH HHS/United States; U56 CA113004/CA/NCI NIH HHS/United States; R01 CA77575-05/CA/NCI NIH HHS/United States; P20 CA113004-01/CA/NCI NIH HHS/United States},
	issn = {1538-7445 (Electronic); 0008-5472 (Print); 0008-5472 (Linking)},
	jid = {2984705R},
	journal = {Cancer Res},
	jt = {Cancer research},
	language = {eng},
	lid = {10.1158/0008-5472.CAN-08-3658 {$[$}doi{$]$}},
	lr = {20240610},
	mh = {Adaptation, Biological/*physiology; Animals; Antineoplastic Agents/administration \& dosage; *Antineoplastic Protocols; Cell Proliferation/drug effects; Computer Simulation; Drug Dosage Calculations; Drug Resistance, Neoplasm/*physiology; Feasibility Studies; Female; Humans; Maximum Tolerated Dose; Mice; Mice, SCID; Models, Theoretical; Neoplasms/drug therapy/pathology; Tumor Burden; Xenograft Model Antitumor Assays},
	mhda = {2009/07/16 09:00},
	mid = {NIHMS489689},
	month = {Jun},
	number = {11},
	own = {NLM},
	pages = {4894--4903},
	phst = {2009/06/03 09:00 {$[$}entrez{$]$}; 2009/06/03 09:00 {$[$}pubmed{$]$}; 2009/07/16 09:00 {$[$}medline{$]$}; 2013/07/31 00:00 {$[$}pmc-release{$]$}},
	pii = {69/11/4894},
	pl = {United States},
	pmc = {PMC3728826},
	pmcr = {2013/07/31},
	pmid = {19487300},
	pst = {ppublish},
	pt = {Journal Article; Research Support, N.I.H., Extramural},
	rn = {0 (Antineoplastic Agents)},
	sb = {IM},
	status = {MEDLINE},
	title = {{Adaptive therapy}},
	volume = {69},
	year = {2009}}

@article{Marine:2020aa,
	address = {Laboratory for Molecular Cancer Biology, VIB Center for Cancer Biology, KU Leuven, Leuven, Belgium. jeanchristophe.marine@kuleuven.vib.be.; Department of Oncology, KU Leuven, Leuven, Belgium. jeanchristophe.marine@kuleuven.vib.be.; Peter MacCallum Cancer Centre, Melbourne, VIC, Australia. sarah-jane.dawson@petermac.org.; Sir Peter MacCallum Department of Oncology, The University of Melbourne, Melbourne, VIC, Australia. sarah-jane.dawson@petermac.org.; Center for Cancer Research, The University of Melbourne, Melbourne, VIC, Australia. sarah-jane.dawson@petermac.org.; Peter MacCallum Cancer Centre, Melbourne, VIC, Australia. mark.dawson@petermac.org.; Sir Peter MacCallum Department of Oncology, The University of Melbourne, Melbourne, VIC, Australia. mark.dawson@petermac.org.; Center for Cancer Research, The University of Melbourne, Melbourne, VIC, Australia. mark.dawson@petermac.org.},
	auid = {ORCID: 0000-0002-5464-5029},
	author = {Marine, Jean-Christophe and Dawson, Sarah-Jane and Dawson, Mark A},
	crdt = {2020/10/09 05:37},
	date = {2020 Dec},
	dcom = {20210204},
	dep = {20201008},
	doi = {10.1038/s41568-020-00302-4},
	edat = {2020/10/10 06:00},
	gr = {HHMI/Howard Hughes Medical Institute/United States},
	issn = {1474-1768 (Electronic); 1474-175X (Linking)},
	jid = {101124168},
	journal = {Nat Rev Cancer},
	jt = {Nature reviews. Cancer},
	language = {eng},
	lid = {10.1038/s41568-020-00302-4 {$[$}doi{$]$}},
	lr = {20210915},
	mh = {Animals; Cell Lineage; Cell Plasticity; Drug Resistance, Neoplasm/*genetics; Humans; Neoplasm, Residual; Neoplasms/*drug therapy/genetics/pathology; Tumor Escape},
	mhda = {2021/02/05 06:00},
	month = {Dec},
	number = {12},
	own = {NLM},
	pages = {743--756},
	phst = {2020/09/04 00:00 {$[$}accepted{$]$}; 2020/10/10 06:00 {$[$}pubmed{$]$}; 2021/02/05 06:00 {$[$}medline{$]$}; 2020/10/09 05:37 {$[$}entrez{$]$}},
	pii = {10.1038/s41568-020-00302-4},
	pl = {England},
	pmid = {33033407},
	pst = {ppublish},
	pt = {Journal Article; Research Support, Non-U.S. Gov't; Review},
	sb = {IM},
	status = {MEDLINE},
	title = {{Non-genetic mechanisms of therapeutic resistance in cancer.}},
	volume = {20},
	year = {2020}}

@article{Li:2025bd,
	author = {Yakun Li and Xiyin Liang and Jinzhi Lei},
	journal = {CSIAM Trans. Life. Sci.},
	number = {2},
	pages = {320-353},
	title = {{Integrating gene regulatory network dynamics with heterogeneous stem cell regeneration}},
	volume = {1},
	year = {2025}}

@article{Bernard:2003aa,
	address = {D{\'e}partement de Math{\'e}matiques et de Statistique and Centre de recherches math{\'e}matiques, Universit{\'e} de Montr{\'e}al, C.P. 6128, Succ. Centre-Ville, Montr{\'e}al, Qu{\'e}., Canada H3C 3J7. bernard@dms.umontreal.ca},
	author = {Bernard, Samuel and B{\'e}lair, Jacques and Mackey, Michael C},
	crdt = {2003/07/10 05:00},
	date = {2003 Aug 7},
	dcom = {20031106},
	doi = {10.1016/s0022-5193(03)00090-0},
	edat = {2003/07/10 05:00},
	ein = {J Theor Biol. 2004 May 7;228(1):143},
	issn = {0022-5193 (Print); 0022-5193 (Linking)},
	jid = {0376342},
	journal = {J Theor Biol},
	jt = {Journal of theoretical biology},
	language = {eng},
	lr = {20190725},
	mh = {Animals; Apoptosis; Dog Diseases/*immunology/pathology; Dogs; Granulocyte Colony-Stimulating Factor/physiology; Homeostasis; Leukocyte Count; Models, Biological; Neutropenia/*immunology/*veterinary; Neutrophils/*pathology; Stem Cells/*pathology},
	mhda = {2003/11/07 05:00},
	month = {Aug},
	number = {3},
	own = {NLM},
	pages = {283--298},
	phst = {2003/07/10 05:00 {$[$}pubmed{$]$}; 2003/11/07 05:00 {$[$}medline{$]$}; 2003/07/10 05:00 {$[$}entrez{$]$}},
	pii = {S0022519303000900},
	pl = {England},
	pmid = {12850449},
	pst = {ppublish},
	pt = {Journal Article; Research Support, Non-U.S. Gov't},
	rn = {143011-72-7 (Granulocyte Colony-Stimulating Factor)},
	sb = {IM},
	status = {MEDLINE},
	title = {{Oscillations in cyclical neutropenia: new evidence based on mathematical modeling.}},
	volume = {223},
	year = {2003}}

@article{Ma:2023aa,
	address = {Institute for Mathematical Sciences, Renmin University of China, Beijing, 100872, People's Republic of China.; School of Mathematical Science, Tiangong University, Tianjin, 300387, People's Republic of China.; Institute for Mathematical Sciences, Renmin University of China, Beijing, 100872, People's Republic of China. xiulanlai@ruc.edu.cn.},
	auid = {ORCID: 0000-0002-2764-8937},
	author = {Ma, Shizhao and Lei, Jinzhi and Lai, Xiulan},
	copyright = {{\copyright}2023. The Author(s), under exclusive licence to Springer-Verlag GmbH Germany, part of Springer Nature.},
	crdt = {2023/01/25 11:19},
	date = {2023 Jan 25},
	dcom = {20230202},
	dep = {20230125},
	doi = {10.1007/s00285-023-01872-1},
	edat = {2023/01/26 06:00},
	issn = {1432-1416 (Electronic); 0303-6812 (Linking)},
	jid = {7502105},
	journal = {J Math Biol},
	jt = {Journal of mathematical biology},
	language = {eng},
	lid = {10.1007/s00285-023-01872-1 {$[$}doi{$]$}},
	lr = {20230302},
	mh = {Humans; *Neoplasms/drug therapy/genetics; Tumor Microenvironment},
	mhda = {2023/01/28 06:00},
	month = {Jan},
	number = {3},
	oto = {NOTNLM},
	own = {NLM},
	pages = {38},
	phst = {2022/03/18 00:00 {$[$}received{$]$}; 2023/01/09 00:00 {$[$}accepted{$]$}; 2022/12/06 00:00 {$[$}revised{$]$}; 2023/01/25 11:19 {$[$}entrez{$]$}; 2023/01/26 06:00 {$[$}pubmed{$]$}; 2023/01/28 06:00 {$[$}medline{$]$}},
	pii = {10.1007/s00285-023-01872-1},
	pl = {Germany},
	pmid = {36695961},
	pst = {epublish},
	pt = {Journal Article; Research Support, Non-U.S. Gov't},
	rn = {0 (CD274 protein, human)},
	sb = {IM},
	status = {MEDLINE},
	title = {{Modeling tumour heterogeneity of PD-L1 expression in tumour progression and adaptive therapy.}},
	volume = {86},
	year = {2023}}

@article{Wang:2025aa,
	address = {School of Mathematics and Statistics, Wuhan University, Wuhan, China.; School of Mathematical Sciences, Center for Applied Mathematics, Tiangong University, Tianjin, China.; School of Mathematics and Statistics, Wuhan University, Wuhan, China.; School of Mathematics and Statistics, Wuhan University, Wuhan, China.},
	auid = {ORCID: 0000-0003-2694-9559; ORCID: 0000-0001-5294-0764; ORCID: 0000-0002-5131-0215},
	author = {Wang, Shun and Lei, Jinzhi and Zou, Xiufen and Jin, Suoqin},
	cois = {The authors have declared that no competing interests exist.},
	copyright = {Copyright: {\copyright}2025 Wang et al. This is an open access article distributed under the terms of the Creative Commons Attribution License, which permits unrestricted use, distribution, and reproduction in any medium, provided the original author and source are credited.},
	crdt = {2025/02/14 13:43},
	date = {2025 Feb},
	dcom = {20250507},
	dep = {20250214},
	doi = {10.1371/journal.pcbi.1012815},
	edat = {2025/02/14 18:21},
	issn = {1553-7358 (Electronic); 1553-734X (Print); 1553-734X (Linking)},
	jid = {101238922},
	journal = {PLoS Comput Biol},
	jt = {PLoS computational biology},
	language = {eng},
	lid = {10.1371/journal.pcbi.1012815 {$[$}doi{$]$}; e1012815},
	lr = {20250507},
	mh = {*Drug Resistance, Neoplasm/genetics; Humans; *Epigenesis, Genetic/genetics; Cell Line, Tumor; Lung Neoplasms/genetics/drug therapy; Computational Biology; *Neoplasms/genetics/drug therapy; Antineoplastic Agents/pharmacology; Models, Biological; Computer Simulation},
	mhda = {2025/02/18 18:20},
	month = {Feb},
	number = {2},
	own = {NLM},
	pages = {e1012815},
	phst = {2024/05/28 00:00 {$[$}received{$]$}; 2025/01/13 00:00 {$[$}accepted{$]$}; 2025/02/18 00:00 {$[$}revised{$]$}; 2025/02/18 18:20 {$[$}medline{$]$}; 2025/02/14 18:21 {$[$}pubmed{$]$}; 2025/02/14 13:43 {$[$}entrez{$]$}; 2025/02/14 00:00 {$[$}pmc-release{$]$}},
	pii = {PCOMPBIOL-D-24-00898},
	pl = {United States},
	pmc = {PMC11835379},
	pmcr = {2025/02/14},
	pmid = {39951474},
	pst = {epublish},
	pt = {Journal Article},
	rn = {0 (Antineoplastic Agents)},
	sb = {IM},
	status = {MEDLINE},
	title = {{Integrating multiscale mathematical modeling and multidimensional data reveals the effects of epigenetic instability on acquired drug resistance in cancer.}},
	volume = {21},
	year = {2025}}

@article{Zhang:2021gd,
	author = {Can Zhang and Changrong Shao and Xiaopei Jiao and Yue Bai and Miao Li and Hanping Shi and Jinzhi Lei and Xiaosong Zhong},
	journal = {Comput Syst Oncol},
	pages = {e21029},
	title = {{Individual cell-based modeling of tumor cell plasticity-induced immune escape after CAR-T therapy}},
	volume = {1},
	year = {2021}}

@article{Rosell:2012aa,
	address = {Catalan Institute of Oncology, Badalona, Spain. rrosell{\char64}iconcologia.net},
	author = {Rosell, Rafael and Carcereny, Enric and Gervais, Radj and Vergnenegre, Alain and Massuti, Bartomeu and Felip, Enriqueta and Palmero, Ramon and Garcia-Gomez, Ramon and Pallares, Cinta and Sanchez, Jose Miguel and Porta, Rut and Cobo, Manuel and Garrido, Pilar and Longo, Flavia and Moran, Teresa and Insa, Amelia and De Marinis, Filippo and Corre, Romain and Bover, Isabel and Illiano, Alfonso and Dansin, Eric and de Castro, Javier and Milella, Michele and Reguart, Noemi and Altavilla, Giuseppe and Jimenez, Ulpiano and Provencio, Mariano and Moreno, Miguel Angel and Terrasa, Josefa and Mu{\~n}oz-Langa, Jose and Valdivia, Javier and Isla, Dolores and Domine, Manuel and Molinier, Olivier and Mazieres, Julien and Baize, Nathalie and Garcia-Campelo, Rosario and Robinet, Gilles and Rodriguez-Abreu, Delvys and Lopez-Vivanco, Guillermo and Gebbia, Vittorio and Ferrera-Delgado, Lioba and Bombaron, Pierre and Bernabe, Reyes and Bearz, Alessandra and Artal, Angel and Cortesi, Enrico and Rolfo, Christian and Sanchez-Ronco, Maria and Drozdowskyj, Ana and Queralt, Cristina and de Aguirre, Itziar and Ramirez, Jose Luis and Sanchez, Jose Javier and Molina, Miguel Angel and Taron, Miquel and Paz-Ares, Luis},
	cin = {Lancet Oncol. 2012 Mar;13(3):216-7. doi: 10.1016/S1470-2045(12)70037-2. PMID: 22285167},
	cn = {Spanish Lung Cancer Group in collaboration with Groupe Fran{\c c}ais de Pneumo-Canc{\'e}rologie and Associazione Italiana Oncologia, Toracica},
	copyright = {Copyright {\copyright}2012 Elsevier Ltd. All rights reserved.},
	crdt = {2012/01/31 06:00},
	date = {2012 Mar},
	dcom = {20120423},
	dep = {20120126},
	doi = {10.1016/S1470-2045(11)70393-X},
	edat = {2012/01/31 06:00},
	issn = {1474-5488 (Electronic); 1470-2045 (Linking)},
	jid = {100957246},
	journal = {Lancet Oncol},
	jt = {The Lancet. Oncology},
	language = {eng},
	lid = {10.1016/S1470-2045(11)70393-X {$[$}doi{$]$}},
	lr = {20221207},
	mh = {Administration, Oral; Aged; Antineoplastic Combined Chemotherapy Protocols/adverse effects/*therapeutic use; Carboplatin/administration \& dosage; Carcinoma, Non-Small-Cell Lung/*drug therapy/enzymology/*genetics/mortality/pathology; Chi-Square Distribution; Cisplatin/administration \& dosage; Deoxycytidine/administration \& dosage/analogs \& derivatives; Disease-Free Survival; Docetaxel; Drug Administration Schedule; ErbB Receptors/*antagonists \& inhibitors/*genetics; Erlotinib Hydrochloride; Europe; Exons; Female; Humans; Kaplan-Meier Estimate; Lung Neoplasms/*drug therapy/enzymology/*genetics/mortality/pathology; Male; Middle Aged; Molecular Targeted Therapy; *Mutation; Patient Selection; Precision Medicine; Proportional Hazards Models; Prospective Studies; Protein Kinase Inhibitors/administration \& dosage/adverse effects/*therapeutic use; Quinazolines/administration \& dosage/adverse effects/*therapeutic use; Taxoids/administration \& dosage; Time Factors; Treatment Outcome; Gemcitabine},
	mhda = {2012/04/24 06:00},
	month = {Mar},
	number = {3},
	own = {NLM},
	pages = {239--246},
	phst = {2012/01/31 06:00 {$[$}entrez{$]$}; 2012/01/31 06:00 {$[$}pubmed{$]$}; 2012/04/24 06:00 {$[$}medline{$]$}},
	pii = {S1470-2045(11)70393-X},
	pl = {England},
	pmid = {22285168},
	pst = {ppublish},
	pt = {Clinical Trial, Phase III; Comparative Study; Journal Article; Multicenter Study; Randomized Controlled Trial; Research Support, Non-U.S. Gov't},
	rn = {0 (Protein Kinase Inhibitors); 0 (Quinazolines); 0 (Taxoids); 0W860991D6 (Deoxycytidine); 15H5577CQD (Docetaxel); BG3F62OND5 (Carboplatin); DA87705X9K (Erlotinib Hydrochloride); EC 2.7.10.1 (EGFR protein, human); EC 2.7.10.1 (ErbB Receptors); Q20Q21Q62J (Cisplatin); 0 (Gemcitabine)},
	sb = {IM},
	si = {ClinicalTrials.gov/NCT00446225},
	status = {MEDLINE},
	title = {{Erlotinib versus standard chemotherapy as first-line treatment for European patients with advanced EGFR mutation-positive non-small-cell lung cancer (EURTAC): a multicentre, open-label, randomised phase 3 trial.}},
	volume = {13},
	year = {2012}}

@article{Reyes:2025aa,
	address = {Pathology and Oral Medicine Department, Faculty of Odontology, Universidad de Chile, Santiago, Chile.; School of Odontology, Faculty of Odontology, Universidad San Sebasti{\'a}n, Santiago, Chile.; School of Odontology, Faculty of Odontology, Universidad San Sebasti{\'a}n, Santiago, Chile.},
	author = {Reyes, Montserrat and Urra, Hery and Pe\~{n}a-Oyarz\'{u}n, Daniel},
	cois = {The authors declare that the research was conducted in the absence of any commercial or financial relationships that could be construed as a potential conflict of interest.},
	copyright = {{\copyright}2025 Reyes, Urra and Pe{\~n}a-Oyarz{\'u}n.},
	crdt = {2025/05/27 04:30},
	date = {2025},
	dep = {20250512},
	doi = {10.3389/froh.2025.1575721},
	edat = {2025/05/27 06:27},
	issn = {2673-4842 (Electronic); 2673-4842 (Linking)},
	jid = {9918227262706676},
	journal = {Front Oral Health},
	jt = {Frontiers in oral health},
	language = {eng},
	lid = {10.3389/froh.2025.1575721 {$[$}doi{$]$}; 1575721},
	lr = {20250530},
	mhda = {2025/05/27 06:28},
	oto = {NOTNLM},
	own = {NLM},
	pages = {1575721},
	phst = {2025/02/12 00:00 {$[$}received{$]$}; 2025/04/28 00:00 {$[$}accepted{$]$}; 2025/05/27 06:28 {$[$}medline{$]$}; 2025/05/27 06:27 {$[$}pubmed{$]$}; 2025/05/27 04:30 {$[$}entrez{$]$}; 2025/05/12 00:00 {$[$}pmc-release{$]$}},
	pl = {Switzerland},
	pmc = {PMC12104182},
	pmcr = {2025/05/12},
	pmid = {40421179},
	pst = {epublish},
	pt = {Journal Article; Review},
	status = {PubMed-not-MEDLINE},
	title = {{Evaluating the link between periodontitis and oral squamous cell carcinoma through Wnt/$\beta$-catenin pathway: a critical review.}},
	volume = {6},
	year = {2025}}

@article{Rambow:2018aa,
	address = {Laboratory for Molecular Cancer Biology, VIB Center for Cancer Biology, KU Leuven, Leuven, Belgium; Department of Oncology, KU Leuven, Leuven, Belgium.; Laboratory for Molecular Cancer Biology, VIB Center for Cancer Biology, KU Leuven, Leuven, Belgium; Department of Oncology, KU Leuven, Leuven, Belgium.; Laboratory for Molecular Cancer Biology, VIB Center for Cancer Biology, KU Leuven, Leuven, Belgium; Department of Oncology, KU Leuven, Leuven, Belgium.; Laboratory of Computational Biology, VIB Center for Brain \& Disease Research, KU Leuven, Leuven, Belgium; Department of Human Genetics, KU Leuven, Leuven, Belgium.; Department of Cell, Developmental, and Cancer Biology, Knight Cancer Institute, Oregon Health and Science University, Portland, OR, USA.; Laboratory for Molecular Cancer Biology, VIB Center for Cancer Biology, KU Leuven, Leuven, Belgium; Department of Oncology, KU Leuven, Leuven, Belgium.; Laboratory for Molecular Cancer Biology, VIB Center for Cancer Biology, KU Leuven, Leuven, Belgium; Department of Oncology, KU Leuven, Leuven, Belgium.; Laboratory of reproductive genomics, Department of Human Genetics, KU Leuven, Leuven, Belgium.; Department of Biomedical Engineering, Oregon Center for Spatial Systems Biomedicine, Oregon Health and Science University, Portland, OR, USA.; Laboratory for Genetics of Malignant Disorders, Department of Human Genetics, KU Leuven, Leuven, Belgium.; Laboratory for Molecular Cancer Biology, VIB Center for Cancer Biology, KU Leuven, Leuven, Belgium; Department of Oncology, KU Leuven, Leuven, Belgium.; Comparative Pathology Core, University of Pennsylvania, Department of Pathobiology, Philadelphia, PA, USA.; Department of General Medical Oncology, UZ Leuven, Leuven, Belgium.; Department of General Medical Oncology, UZ Leuven, Leuven, Belgium.; Department of Dermatology, University of Z{\"u}rich Hospital, Z{\"u}rich, Switzerland.; Department of Dermatology, University of Z{\"u}rich Hospital, Z{\"u}rich, Switzerland.; Department of Surgical Oncology, Massachusetts General Hospital, Boston, MA, USA.; Department of Surgical Oncology, Massachusetts General Hospital, Boston, MA, USA.; Department of Surgical Oncology, Massachusetts General Hospital, Boston, MA, USA.; Department of Medical Oncology, Massachusetts General Hospital, Boston, MA, USA.; Laboratory of Translational Cell and Tissue Research, Department of Pathology, UZ Leuven, Leuven, Belgium.; Laboratory of reproductive genomics, Department of Human Genetics, KU Leuven, Leuven, Belgium.; Laboratory of Computational Biology, VIB Center for Brain \& Disease Research, KU Leuven, Leuven, Belgium; Department of Human Genetics, KU Leuven, Leuven, Belgium.; Department of Cell, Developmental, and Cancer Biology, Knight Cancer Institute, Oregon Health and Science University, Portland, OR, USA.; Laboratory for Molecular Cancer Biology, VIB Center for Cancer Biology, KU Leuven, Leuven, Belgium; Department of Oncology, KU Leuven, Leuven, Belgium. Electronic address: jeanchristophe.marine@kuleuven.vib.be.},
	author = {Rambow, Florian and Rogiers, Aljosja and Marin-Bejar, Oskar and Aibar, Sara and Femel, Julia and Dewaele, Michael and Karras, Panagiotis and Brown, Daniel and Chang, Young Hwan and Debiec-Rychter, Maria and Adriaens, Carmen and Radaelli, Enrico and Wolter, Pascal and Bechter, Oliver and Dummer, Reinhard and Levesque, Mitchell and Piris, Adriano and Frederick, Dennie T and Boland, Genevieve and Flaherty, Keith T and van den Oord, Joost and Voet, Thierry and Aerts, Stein and Lund, Amanda W and Marine, Jean-Christophe},
	cin = {Pigment Cell Melanoma Res. 2019 Jan;32(1):6-8. doi: 10.1111/pcmr.12744. PMID: 30339326},
	copyright = {Copyright {\copyright}2018 Elsevier Inc. All rights reserved.},
	crdt = {2018/07/19 06:00},
	date = {2018 Aug 9},
	dcom = {20190520},
	dep = {20180712},
	doi = {10.1016/j.cell.2018.06.025},
	edat = {2018/07/19 06:00},
	gr = {P30 CA069533/CA/NCI NIH HHS/United States},
	issn = {1097-4172 (Electronic); 0092-8674 (Linking)},
	jid = {0413066},
	journal = {Cell},
	jt = {Cell},
	language = {eng},
	lid = {S0092-8674(18)30793-1 {$[$}pii{$]$}; 10.1016/j.cell.2018.06.025 {$[$}doi{$]$}},
	lr = {20190520},
	mh = {Animals; Biomarkers, Tumor; Drug Resistance, Neoplasm/drug effects; Female; Gene Expression Regulation, Neoplastic/*drug effects; Humans; MAP Kinase Kinase 1/antagonists \& inhibitors/genetics; Male; Melanoma/*drug therapy/metabolism/pathology; Mice, SCID; Mutation; Neoplasm, Residual/*drug therapy/metabolism/pathology; Neoplastic Stem Cells/*drug effects/metabolism/pathology; Neural Stem Cells/*drug effects/metabolism/pathology; Protein Kinase Inhibitors/*pharmacology; Proto-Oncogene Proteins B-raf/antagonists \& inhibitors/genetics; Retinoid X Receptor gamma/*antagonists \& inhibitors; Tumor Cells, Cultured; Xenograft Model Antitumor Assays},
	mhda = {2019/05/21 06:00},
	month = {Aug},
	number = {4},
	oto = {NOTNLM},
	own = {NLM},
	pages = {843--855},
	phst = {2017/10/09 00:00 {$[$}received{$]$}; 2018/04/13 00:00 {$[$}revised{$]$}; 2018/06/12 00:00 {$[$}accepted{$]$}; 2018/07/19 06:00 {$[$}pubmed{$]$}; 2019/05/21 06:00 {$[$}medline{$]$}; 2018/07/19 06:00 {$[$}entrez{$]$}},
	pii = {S0092-8674(18)30793-1},
	pl = {United States},
	pmid = {30017245},
	pst = {ppublish},
	pt = {Journal Article; Research Support, N.I.H., Extramural; Research Support, Non-U.S. Gov't; Research Support, U.S. Gov't, Non-P.H.S.},
	rn = {0 (Biomarkers, Tumor); 0 (Protein Kinase Inhibitors); 0 (Retinoid X Receptor gamma); EC 2.7.11.1 (BRAF protein, human); EC 2.7.11.1 (Proto-Oncogene Proteins B-raf); EC 2.7.12.2 (MAP Kinase Kinase 1); EC 2.7.12.2 (MAP2K1 protein, human)},
	sb = {IM},
	status = {MEDLINE},
	title = {{Toward minimal residual disease-directed therapy in melanoma}},
	volume = {174},
	year = {2018}}

@article{Sharma:2010aa,
	address = {Massachusetts General Hospital Cancer Center, 149 13th Street, Charlestown, MA 02129, USA.},
	author = {Sharma, Sreenath V and Lee, Diana Y and Li, Bihua and Quinlan, Margaret P and Takahashi, Fumiyuki and Maheswaran, Shyamala and McDermott, Ultan and Azizian, Nancy and Zou, Lee and Fischbach, Michael A and Wong, Kwok-Kin and Brandstetter, Kathleyn and Wittner, Ben and Ramaswamy, Sridhar and Classon, Marie and Settleman, Jeff},
	cin = {Cell. 2010 Apr 2;141(1):18-20. doi: 10.1016/j.cell.2010.03.020. PMID: 20371339},
	copyright = {Copyright 2010 Elsevier Inc. All rights reserved.},
	crdt = {2010/04/08 06:00},
	date = {2010 Apr 2},
	dcom = {20100422},
	doi = {10.1016/j.cell.2010.02.027},
	edat = {2010/04/08 06:00},
	gr = {P20 CA090578/CA/NCI NIH HHS/United States; R01 CA115830/CA/NCI NIH HHS/United States; R01CA115830/CA/NCI NIH HHS/United States; R01 CA142825/CA/NCI NIH HHS/United States; P50 CA090578/CA/NCI NIH HHS/United States},
	issn = {1097-4172 (Electronic); 0092-8674 (Print); 0092-8674 (Linking)},
	jid = {0413066},
	journal = {Cell},
	jt = {Cell},
	language = {eng},
	lid = {10.1016/j.cell.2010.02.027 {$[$}doi{$]$}},
	lr = {20250529},
	mh = {Cell Line, Tumor; Chromatin/metabolism/pathology; DNA Damage; *Drug Resistance, Neoplasm; Histone Deacetylase Inhibitors/pharmacology; Histone Demethylases/metabolism; Humans; Jumonji Domain-Containing Histone Demethylases/antagonists \& inhibitors/genetics/metabolism; Neoplasms/*drug therapy/metabolism/*pathology; Receptor, IGF Type 1/metabolism},
	mhda = {2010/04/23 06:00},
	mid = {NIHMS182868},
	month = {Apr},
	number = {1},
	own = {NLM},
	pages = {69--80},
	phst = {2009/07/15 00:00 {$[$}received{$]$}; 2009/11/26 00:00 {$[$}revised{$]$}; 2010/02/08 00:00 {$[$}accepted{$]$}; 2010/04/08 06:00 {$[$}entrez{$]$}; 2010/04/08 06:00 {$[$}pubmed{$]$}; 2010/04/23 06:00 {$[$}medline{$]$}; 2011/04/02 00:00 {$[$}pmc-release{$]$}},
	pii = {S0092-8674(10)00180-7},
	pl = {United States},
	pmc = {PMC2851638},
	pmcr = {2011/04/02},
	pmid = {20371346},
	pst = {ppublish},
	pt = {Journal Article; Research Support, N.I.H., Extramural; Research Support, Non-U.S. Gov't},
	rn = {0 (Chromatin); 0 (Histone Deacetylase Inhibitors); EC 1.14.11.- (Histone Demethylases); EC 1.14.11.- (Jumonji Domain-Containing Histone Demethylases); EC 1.5.- (KDM4A protein, human); EC 2.7.10.1 (Receptor, IGF Type 1)},
	sb = {IM},
	status = {MEDLINE},
	title = {{A chromatin-mediated reversible drug-tolerant state in cancer cell subpopulations.}},
	volume = {141},
	year = {2010}}

@article{Calderon:2024aa,
	address = {The Jackson Laboratory for Genomic Medicine, Farmington, CT, USA.; The Jackson Laboratory for Genomic Medicine, Farmington, CT, USA.; The Jackson Laboratory for Genomic Medicine, Farmington, CT, USA; Graduate Program in Genetics and Development, UConn Health, Farmington, CT, USA.; The Jackson Laboratory for Genomic Medicine, Farmington, CT, USA; Department of Genetics and Genome Sciences, UConn Health, Farmington, CT, USA; Institute for Systems Genomics, University of Connecticut Health Center, Farmington, CT, USA. Electronic address: eric.wang@jax.org.},
	author = {Calderon, Alexander and Han, Cuijuan and Karma, Sadik and Wang, Eric},
	cois = {Declaration of interests No interests are declared by the authors.},
	copyright = {Copyright {\copyright}2023 Elsevier Inc. All rights reserved.},
	crdt = {2023/10/15 22:00},
	date = {2024 Jan},
	dcom = {20240112},
	dep = {20231013},
	doi = {10.1016/j.trecan.2023.09.003},
	edat = {2023/10/16 00:42},
	issn = {2405-8025 (Electronic); 2405-8025 (Linking)},
	jid = {101665956},
	journal = {Trends Cancer},
	jt = {Trends in cancer},
	language = {eng},
	lid = {S2405-8033(23)00187-5 {$[$}pii{$]$}; 10.1016/j.trecan.2023.09.003 {$[$}doi{$]$}},
	lr = {20240313},
	mh = {Humans; *Leukemia, Myeloid, Acute/drug therapy/genetics; Recurrence; Drug Resistance},
	mhda = {2024/01/12 06:42},
	month = {Jan},
	number = {1},
	oto = {NOTNLM},
	own = {NLM},
	pages = {38--51},
	phst = {2023/06/20 00:00 {$[$}received{$]$}; 2023/09/13 00:00 {$[$}revised{$]$}; 2023/09/14 00:00 {$[$}accepted{$]$}; 2024/01/12 06:42 {$[$}medline{$]$}; 2023/10/16 00:42 {$[$}pubmed{$]$}; 2023/10/15 22:00 {$[$}entrez{$]$}},
	pii = {S2405-8033(23)00187-5},
	pl = {United States},
	pmid = {37839973},
	pst = {ppublish},
	pt = {Journal Article; Review},
	sb = {IM},
	status = {MEDLINE},
	title = {{Non-genetic mechanisms of drug resistance in acute leukemias.}},
	volume = {10},
	year = {2024}}

@article{McDonald:2024aa,
	address = {Department of Integrative Oncology, BC Cancer Research Institute, Vancouver, BC V5Z 1L3, Canada.; Department of Integrative Oncology, BC Cancer Research Institute, Vancouver, BC V5Z 1L3, Canada; Department of Biochemistry and Molecular Biology, University of British Columbia, Vancouver, BC V6T 1Z3, Canada. Electronic address: sdedhar@bccrc.ca.},
	author = {McDonald, Paul C and Dedhar, Shoukat},
	cois = {Declaration of Competing Interest The authors declare that they have no known competing financial interests or personal relationships that could have appeared to influence the work reported in this paper.},
	copyright = {Copyright {\copyright}2023 Elsevier Ltd. All rights reserved.},
	crdt = {2023/11/17 18:24},
	date = {2024 Mar 15},
	dcom = {20231205},
	dep = {20231116},
	doi = {10.1016/j.semcdb.2023.11.003},
	edat = {2023/11/18 11:41},
	issn = {1096-3634 (Electronic); 1084-9521 (Linking)},
	jid = {9607332},
	journal = {Semin Cell Dev Biol},
	jt = {Seminars in cell \& developmental biology},
	language = {eng},
	lid = {S1084-9521(23)00229-X {$[$}pii{$]$}; 10.1016/j.semcdb.2023.11.003 {$[$}doi{$]$}},
	lr = {20240103},
	mh = {Humans; *Drug Resistance, Neoplasm/genetics; Cell Plasticity; *Neoplasms/pathology; Apoptosis; Cell Death},
	mhda = {2023/12/05 12:43},
	month = {Mar},
	oto = {NOTNLM},
	own = {NLM},
	pages = {1--10},
	phst = {2023/08/04 00:00 {$[$}received{$]$}; 2023/11/08 00:00 {$[$}revised{$]$}; 2023/11/08 00:00 {$[$}accepted{$]$}; 2023/12/05 12:43 {$[$}medline{$]$}; 2023/11/18 11:41 {$[$}pubmed{$]$}; 2023/11/17 18:24 {$[$}entrez{$]$}},
	pii = {S1084-9521(23)00229-X},
	pl = {England},
	pmid = {37977107},
	pst = {ppublish},
	pt = {Journal Article; Review},
	sb = {IM},
	status = {MEDLINE},
	title = {{Persister cell plasticity in tumour drug resistance.}},
	volume = {156},
	year = {2024}}

@article{LeiJTB20framework,
	author = {Jinzhi Lei},
	journal = JTB,
	pages = {110196},
	title = {A general mathematical framework for understanding the behavior of heterogeneous stem cell regeneration},
	volume = {492},
	year = {2020}}

@article{Liau17Glioblastoma,
	author = {Liau, Brian B. and Sievers, Cem and Donohue, Laura K. and Gillespie, Shawn M. and Flavahan, William A. and Miller, Tyler E. and Venteicher, Andrew S. and Hebert, Christine H. and Carey, Christopher D. and Rodig, Scott J.},
	journal = {Cell Stem Cell},
	number = {2},
	pages = {233--246},
	title = {{Adaptive chromatin remodeling drives stem cell plasticity and drug tolerance}},
	volume = {20},
	year = {2017}}

@article{Shahrezaei08PNAS,
	author = {Vahid Shahrezaei and Peter S Swain},
	journal = {Proc Natl Acad Sci USA},
	number = {45},
	pages = {17256-61},
	title = {{Analytical distributions for stochastic gene expression}},
	volume = {105},
	year = {2008}}

@article{SunX18MCT,
	author = {Yongjiang Zheng and Jiguang Bao and Qiyi Zhao and Ttianshou Zhou and Xaioqiang Sun},
	journal = {Mol Cancer Ther},
	month = {04},
	number = {4},
	pages = {814-824},
	title = {{A spatio-temporal model of macrophage-mediated drug resistance in glioma immunotherapy}},
	volume = {17},
	year = {2018}}

@article{Lima2021Bayesian,
	author = {Lima, Ernesto A. B. F. AND Faghihi, Danial AND Philley, Russell AND Yang, Jianchen AND Virostko, John AND Phillips, Caleb M. AND Yankeelov, Thomas E.},
	journal = {PLoS Comp Biol},
	month = {11},
	number = {11},
	pages = {1-35},
	publisher = {Public Library of Science},
	title = {Bayesian calibration of a stochastic, multiscale agent-based model for predicting in vitro tumor growth},
	volume = {17},
	year = {2021}}

@article{Cell21Colorectal,
	author = {Sumaiyah K. Rehman and Jennifer Haynes and Evelyne Collignon and Kevin R. Brown and Yadong Wang and Allison M.L. Nixon and Jeffrey P. Bruce and Jeffrey A. Wintersinger and Arvind {Singh Mer} and Edwyn B.L. Lo and Cherry Leung and Evelyne Lima-Fernandes and Nicholas M. Pedley and Fraser Soares and Sophie McGibbon and Housheng Hansen He and Aaron Pollet and Trevor J. Pugh and Benjamin Haibe-Kains and Quaid Morris and Miguel Ramalho-Santos and Sidhartha Goyal and Jason Moffat and Catherine A. OBrien},
	journal = {Cell},
	number = {1},
	pages = {226-242.e21},
	title = {{Colorectal cancer cells enter a diapause-like DTP state to survive chemotherapy}},
	volume = {184},
	year = {2021}}

@article{Ramirez16Diverse,
	author = {Ramirez, M. and Rajaram, S. and Steininger, R. and others},
	journal = {Nat. Commun.},
	pages = {10690},
	title = {Diverse drug-resistance mechanisms can emerge from drug-tolerant cancer persister cells.},
	volume = {7},
	year = {2016}}

@article{CR15Emergence,
	author = {Chisholm, Rebecca H. and Lorenzi, Tommaso and Lorz, Alexander and Larsen, Annette K. and Almeida, Luis Neves de and Escargueil, Alexandre and Clairambault, Jean},
	journal = {Cancer Research},
	month = {03},
	number = {6},
	pages = {930-939},
	title = {{Emergence of drug tolerance in cancer cell populations: An evolutionary outcome of selection, nongenetic instability, and stress-induced adaptation}},
	volume = {75},
	year = {2015}}

@article{Lei20evolutionary,
	author = {Jinzhi Lei},
	journal = {Sci. China. Math},
	number = {3},
	pages = {411--424},
	title = {Evolutionary dynamics of cancer: From epigenetic regulation to cell population dynamics--mathematical model framework, applications, and open problems},
	volume = {63},
	year = {2020}}

@article{Probst2009Epigenetic,
	author = {Probst, A.V and Dunleavy, E. and Almouzni, G.},
	journal = {Nat Rev Mol Cell Biol.},
	number = {3},
	pages = {192--206},
	title = {{Epigenetic inheritance during the cell cycle}},
	volume = {10},
	year = {2009}}

@article{PlosCB21Inferring,
	author = {Jagiella, Nick AND Muller, Benedikt AND Muller, Margareta AND Vignon-Clementel, Irene E. AND Drasdo, Dirk},
	journal = {PLoS Comp Biol},
	month = {02},
	number = {2},
	pages = {1-39},
	publisher = {Public Library of Science},
	title = {Inferring Growth Control Mechanisms in Growing Multi-cellular Spheroids of NSCLC Cells from Spatial-Temporal Image Data},
	volume = {12},
	year = {2016}}

@article{Friedman06linking,
	author = {Nir Friedman and Long Cai and X Sunney Xie},
	journal = {Phys Rev Lett},
	number = {16},
	pages = {168302},
	title = {{Linking stochastic dynamics to population distribution: An analytical framework of gene expression}},
	volume = {97},
	year = {2006}}

@article{McClatchy20Modeling,
	author = {McClatchy, David M. and Willers, Henning and Hata, Aaron N. and Piotrowska, Zofia and Sequist, Lecia V. and Paganetti, Harald and Grassberger, Clemens},
	journal = {Cancer Research},
	month = {11},
	number = {22},
	pages = {5121-5133},
	title = {{Modeling resistance and recurrence patterns of combined targeted-hemoradiotherapy predicts benefit of shorter induction period}},
	volume = {80},
	year = {2020}}

@article{Natcom18Notch,
	author = {Arasada, RR. and Shilo, K. and Yamada, T. and others},
	journal = {Nat Commun.},
	number = {1},
	pages = {3198},
	title = {Notch3-dependent $\beta$-catenin signaling mediates EGFR TKI drug persistence in EGFR mutant NSCLC},
	volume = {9},
	year = {2018}}

@article{Burns70onthe,
	author = {F J Burns and I F Tannock},
	journal = {Cell Prolif.},
	number = {4},
	pages = {321--334},
	title = {{On the existence of a G0-phase in the cell cycle}},
	volume = {3},
	year = {1970}}

@article{Bell20Principles,
	author = {Charles C Bell and Omer Gilan},
	journal = {Br J Cancer},
	number = {4},
	pages = {465-472},
	title = {{Principles and mechanisms of non-genetic resistance in cancer}},
	volume = {122},
	year = {2020}}

@article{Huz22JCI,
	author = {Meng Nie AND Na Chen AND Huanhuan Pang AND Tao Jiang AND Wei Jiang AND Panwen Tian AND LiAng Yao AND Yangzi Chen AND Ralph J. DeBerardinis AND Weimin Li AND Qitao Yu AND Caicun Zhou AND Zeping Hu},
	journal = {J Clin Invest},
	number = {20},
	pages = {e160152},
	title = {{Targeting acetylcholine signaling modulates persistent drug tolerance in EGFR-mutant lung cancer and impedes tumor relapse}},
	volume = {132},
	year = {2022}}

@article{Boumahdi20greatescape,
	author = {S. Boumahdi and F.J. de Sauvage},
	journal = {Nat Rev Drug Discov.},
	number = {1},
	pages = {39--56},
	title = {{The great escape: tumour cell plasticity in resistance to targeted therapy}},
	volume = {19},
	year = {2020}}

@article{Straussman12CAF,
	author = {Straussman, R. and Morikawa, T. and Shee, K. and others},
	journal = {Nature},
	pages = {500-504},
	title = {Tumour micro-environment elicits innate resistance to RAF inhibitors through HGF secretion.},
	volume = {487},
	year = {2012}}

@article{Aissa20Natcom,
	author = {Aissa, A.F. and Islam, A.B.M.M.K. and Ariss, M.M. and et al.},
	journal = {Nat Commun},
	pages = {110162},
	title = {Single-cell transcriptional changes associated with drug tolerance and response to combination therapies in cancer},
	volume = {12},
	year = {2021}}

@article{Feinberg2023Science,
	author = {Andrew P. Feinberg and Andre Levchenko},
	journal = {Science},
	number = {6632},
	pages = {eaaw3835},
	title = {Epigenetics as a mediator of plasticity in cancer},
	volume = {379},
	year = {2023}}

@article{Kolokotroni2016PlosCB,
	author = {Kolokotroni, Eleni AND Dionysiou, Dimitra AND Veith, Christian AND Kim, Yoo-Jin AND Sabczynski, J{\"o}rg AND Franz, Astrid AND Grgic, Aleksandar AND Palm, Jan AND Bohle, Rainer M. AND Stamatakos, Georgios},
	doi = {10.1371/journal.pcbi.1005093},
	journal = {PLoS Comp Biol},
	month = {09},
	number = {9},
	pages = {1-43},
	publisher = {Public Library of Science},
	title = {In Silico Oncology: Quantification of the In Vivo Antitumor Efficacy of Cisplatin-Based Doublet Therapy in Non-Small Cell Lung Cancer (NSCLC) through a Multiscale Mechanistic Model},
	volume = {12},
	year = {2016}}

\end{document}